\documentclass[traditabstract,longauth]{aa} 

\usepackage{graphicx}
\usepackage{txfonts}
\usepackage{longtable}
\usepackage{lscape}
\usepackage{hyperref}
\usepackage{soul}
\hypersetup{colorlinks=true, citecolor=blue}
\newcommand{\hi}{{\sc H\,i}}
\newcommand{\hii}{{\sc H\,ii}}
\newcommand{\nii}{{\sc N\,ii}}
\newcommand{\mhi}{{$M$(\sc H\,i})}
\newcommand{\nhi}{{$N$(\sc H\,i})}
\newcommand{\shi}{{$\Sigma$(\sc H\,i})}
\newcommand{\kms}{$\,$km$\,$s$^{-1}$}

\begin{document}

   \title{A Virgo Environmental Survey Tracing Ionised Gas Emission (VESTIGE)\\ XX. Star formation in the tidal tail of NGC 4254}
   
   \subtitle{}
  \author{A. Boselli\inst{1,2},         
    A. Lupi\inst{3,4},
    P. Serra\inst{2},
    P. Andreani\inst{5,6},
    F. Calura\inst{3},
    M-A. Miville-Desch\^enes\inst{7}, 
    G. Hensler\inst{8},  
          M. Boquien\inst{9}, 
          M. Fossati\inst{10,11},
          S. Boissier\inst{1},
    J. Braine\inst{12},
    P. C{\^o}t{\'e}\inst{13},
    J.C. Cuillandre\inst{14},
          F. de Gasperin\inst{15},
          H. Edler\inst{16},
    L. Ferrarese\inst{13},
          G. Gavazzi\inst{10},
    S. Gwyn\inst{13},
          J. Hutchings\inst{13},
          K. Kianfar\inst{17,5},
          A. Longobardi\inst{10},
    E.S. Mangola\inst{1},
    S. Martocchia\inst{1},
          E. Peng\inst{18},
    H. Plana\inst{19},
          J. Postma\inst{20},  
          J. Roediger\inst{13},
    Y. Roehlly\inst{1},
    M. Sun\inst{21}
       }

\institute{     
        Aix Marseille Univ, CNRS, CNES, LAM, Marseille, France
    \and
        INAF - Osservatorio Astronomico di Cagliari, Via della Scienza 5, 09047 Selargius (CA), Italy
        \email{alessandro.boselli@lam.fr}
    \and
        Como Lake Center for Astrophysiscs, DiSAT, Universit\`a degli Studi dell'Insubria, via Valleggio 11, 22100, Como, Italy
    \and
        INAF - Osservatorio Astronomico di Bologna, via Gobetti 93/3, I-40129 Bologna, Italy 
    \and
        European Southern Observatory, Karl-Schwarzschild-Strasse 2, 85748, Garching, Germany
    \and
        INAF - Osservatorio Astronomico di Capodimonte, Salita Moiariello 16, 80131 Napoli, Italy
    \and
        Laboratoire de Physique de l'Ecole Normale Sup\'erieure, ENS, Universit\'e PSL, CNRS, Sorbonne Universit\'e, Universit\'e de Paris, Observatoire de Paris, 24 rue Lhomond, 75005, Paris Cedex 05, France
    \and
        Department of Astrophysics, University of Vienna, T\"urkenschanzstrasse 17, 1180 Vienna, Austria
    \and
        Universit\'e C\^ote d'Azur, Observatoire de la C\^ote d'Azur, CNRS, Laboratoire Lagrange, 06000 Nice, France 
    \and
        Universit\`a di Milano-Bicocca, piazza della scienza 3, 20100 Milano, Italy
    \and
        INAF - Osservatorio Astronomico di Brera, via Brera 28, 20121 Milano, Italy
    \and
        Laboratoire d'Astrophysique de Bordeaux, Univ. Bordeaux, CNRS, B18N, allee Geoffroy Saint-Hilaire, 33615 Pessac, France
    \and
        National Research Council of Canada, Herzberg Astronomy and Astrophysics, 5071 West Saanich Road, Victoria, BC, V9E 2E7, Canada
    \and
        AIM, CEA, CNRS, Universit\'e Paris-Saclay, Universit\'e Paris Diderot, Sorbonne Paris Cit\'e, Observatoire de Paris, PSL University, F-91191 Gif-sur-Yvette Cedex, France
    \and
        INAF - Istituto di Radioastronomia, Via P. Gobetti 101, 40129 Bologna, Italy   
    \and
        ASTRON, Oude Hoogeveensedijk 4, 7991 PD Dwingeloo, The Nederlands
    \and
        Physics Department, Aeronautics Institute of Technology - ITA, Pra\c{c}a Marechal Eduardo Gomes, 50, S\~ao Jos\'e dos Campos, 12228-900, S\~ao Paulo, Brazil
    \and
        Department of Astronomy, Peking University, Beijing 100871, PR China
    \and
        Laborat{\'o}rio de Astrof{\'i}sica Te{\'o}rica e Observacional, Dept. de Ci\^encias Exatas, Universidade Estadual de Santa Cruz, Bahia, Brazil 
    \and
        University of Calgary, 2500 University Drive NW, Calgary, Alberta, Canada
    \and
        Department of Physics and Astronomy, University of Alabama in Huntsville, Huntsville, AL 35899, USA
    }
         
\authorrunning{Boselli et al.}
\titlerunning{VESTIGE: SFR in the tail of NGC 4254 }

   \date{}

 
  \abstract  
{ALMA $^{12}$CO(1-0) observations of 42 star-forming regions located outside the disc of the Virgo Cluster galaxy NGC 4254 within an \hi\ gas tail produced during the galaxy's 
interaction with another cluster member have revealed the presence of ten giant molecular clouds (GMCs) in four of these regions. All of the GMCs were resolved at the angular 
resolution of the observations ($\simeq$ 160 pc) and have molecular gas masses of $M(\mathrm{H_2}) \simeq (0.8-2.0) \times 10^{6}$ M$_\odot$.
These ten clouds are characterised by gas column densities [$\Sigma\mathrm{(H_2)}$ $\simeq$ 10 M$_{\odot}$ pc$^{-2}$] and velocity dispersions [$\sigma_{v}({\rm CO})$ $\simeq$ 3--12 km s$^{-1}$] 
respectively lower and comparable to those encountered in similar GMCs in the Milky Way. They follow the relation between the gas column density and the star formation activity 
(Schmidt law) derived using similar data over the stellar disc of NGC 4254 and other local and Virgo cluster galaxies. With analytic calculations and tuned simulations, we show 
that these clouds are unstable and thus expected to dissolve on relatively short timescales ($\sim$10--30 Myr). We show that they probably formed after the collapse of dense gas 
clouds in the \hi\ gas tail stripped during the gravitational interaction that the galaxy suffered several hundreds millions of years ago. The clouds are short-lived and 
isolated given the low density of the surrounding intracluster medium, which cannot confine the gas expelled by stellar feedback. We discuss the implications of these 
results in the general context of the fate of stripped gas in hostile cluster environments.}


   \maketitle
%

\section{Introduction}
Galaxies located in high-density regions such as groups and clusters are subject to different kinds of gravitational or hydrodynamic
interactions with their surrounding environment (e.g. Boselli \& Gavazzi 2006). These interactions are able to remove part 
of their interstellar medium (ISM), thus reducing (on different timescales) the activity of star formation of the perturbed systems.
Gravitational perturbations indifferently affect all the different galaxy components (stars, gas, dust, dark matter),
while hydrodynamic interactions act only on the different components of the ISM.
Regardless of the dominant perturbing mechanism, the cold atomic gas phase, which is the one more loosely bound to 
the gravitational potential well of the galaxy, is the gas phase of the ISM more easily removed in any kind of interaction.
Indeed, it is now well established that star-forming galaxies in rich environments are characterised by a lower atomic gas
content than their counterparts in the field (e.g. Haynes \& Giovanelli 1984, Cayatte et al. 1990, Gavazzi et al. 2005, Cortese et al. 2021).
The molecular gas phase, which is mainly concentrated within dense giant molecular clouds (GMCs) located within the stellar disc, is 
hardly removed in hydrodynamic interactions. Only its diffuse component can be stripped along with the cold atomic phase, explaining the 
moderate molecular gas deficiency observed in cluster systems (e.g. Fumagalli et al. 2009, Boselli et al. 2014a, Zabel et al. 2022).

If the different stripping mechanisms are now well identified and their effects on galaxy evolution well understood, what is still 
totally unclear is the fate of the stripped material. Once removed from the galaxy disc, the cold gas becomes mixed with the hot and 
tenuous surrounding intracluster medium (ICM), which strongly emits in X-rays, and it is trapped within the gravitational potential 
well of the group or cluster massive halo. Tuned hydrodynamic simulations (Kronberger et al. 2008; Vollmer et al. 2008; Tonnesen \& Bryan 2009, 2010, 2012; Roediger et al. 2014; 
Steyrleithner et al. 2020; Boselli et al. 2021) and multi-frequency observations of representative objects
(e.g. Jachym et al. 2013, 2014, 2017, 2019, 2022; Moretti et al. 2018, 2020a, 2020b) suggest that part of 
this gas can collapse into GMCs to form new stars. There are indeed several examples
of perturbed galaxies with star formation in their tails (IC 3418, Hester et al. 2010, Fumagalli et al. 2011, Kenney et al. 2014;
ESO137-001, Sun et al. 2007, Fossati et al. 2016; D100, Cramer et al. 2019; the GASP sample, Poggianti et al. 2019a). There are, however,
several other examples of cluster systems with extended tails of stripped gas without any associated star formation activity
(e.g. Gavazzi et al. 2001, Yagi et al. 2010, Boissier et al. 2012, Boselli et al. 2016). This observed unpredictable behaviour of the stripped 
gas in the tail has been tentatively explained with several arguments. They include the observations of different epochs of the stripping process 
(Moretti et al. 2023) and the variations in the formation of turbulent vortices in galaxies entering in the ICM with different impact parameters 
(face-on versus edge-on; Roediger \& Br\"uggen 2006), 
possibly combined with a relative difference in the physical properties of the gas (density, temperature) 
in the surrounding medium (Roediger \& Hensler 2005; Boselli et al. 2016). Furthermore, the presence of magnetic fields can contribute 
to confining matter, inhibiting the energy and momentum exchange between the ISM and the ICM 
(Tonnesen \& Stone 2014; Ruszkowski et al. 2014; M\"uller et al. 2021). However, these are all tentative explications that still cannot be 
used to predict what will happen to the gas once stripped from a galaxy in a rich environment.
There are several reasons behind this uncertainty. First of all, tails of stripped material have very low column densities
or surface brightnesses and have only recently been mapped at different wavelengths thanks to the advent of 
very sensitive wide field instruments coupled with large telescopes. Their statistics are thus still very limited.
Secondly, although large efforts have been 
made in the past years, simulations of the star formation process in the tail of stripped gas are still challenging.
This is due to the fact that they should consider a complex multi-phase medium (cold atomic and molecular, ionised, hot gas, 
e.g. Fossati et al. 2016; Boselli et al. 2016, 2021, 2022; Poggianti et al. 2019b; Sun et al. 2022) 
possibly coupled with dust (e.g. Longobardi et al. 2020) on different scales, 
from the whole cluster ($\sim$ 1--2 Mpc scales) down to GMCs and individual \hii\ regions (10--100 pc scales; see Roediger 2009), with magnetic fields possibly having an important role.
Finally, the stripped gas is subject to different competitive mechanisms, some of which induce its change of phase 
(heat conduction, mixing, evaporation, photoionisation, stellar feedback; e.g. Calura et al. 2020), while 
others favour its collapse into GMC (cooling), and all are possibly modulated by the presence of
weak magnetic fields (Tonnesen \& Stone 2014; Ruszkowski et al. 2014).

We carried out a deep H$\alpha$ narrow-band imaging survey of the Virgo cluster with MegaCam at the Canada 
French Hawaii Telescope (CFHT). This survey, called VESTIGE (A Virgo Environmental Survey Tracing Ionised Gas Emission;
Boselli et al. 2018a) covers the cluster up to its virial radius ($\simeq$ 104 deg$^2$). Thanks to its blind nature,
VESTIGE allowed us to identify in a complete and unbiased way several tails of stripped gas in its ionised phase associated with
perturbed galaxies in the cluster. One spectacular example is the massive spiral NGC 4254 (M99) located at the north-western periphery
of Virgo. A detailed analysis of the multi-frequency data in hand allowed us to identify $\sim$ 60 star-forming regions outside
the stellar disc of the galaxy located within a long tail of diffuse \hi\ gas (Haynes et al. 2007; Minchin et al. 2007) that probably formed during the gravitational interaction
with another cluster member (Vollmer et al. 2005, Duc \& Bournaud 2008). The properties of these star-forming regions
have been studied in detail in Boselli et al. (2018b). We observed these regions with the Atacama Large 
Millimeter/submillimeter Array (ALMA) radio telescope 
with the purpose of understanding the relationship between the gas column density and the star-formation process 
in this extreme environment. The results of this study are presented in this work.
This paper is structured as follows: In Sec. 2 we describe the new ALMA observations along with unpublished ASTROSAT/UVIT 
and MeerKAT data, gathered during the ViCTORIA (Virgo
Cluster multi Telescope Observations in Radio of Interacting galaxies and AGN) survey of the cluster (de Gasperin et al. 2025),
and the VESTIGE H$\alpha$ data already presented in Boselli et al. (2018b). In Sec. 3 we use these new data to derive several physical 
parameters useful for the analysis in Sec. 4. We also present two sets of different simulations expressly constructed to study the fate of a GMC 
in a hot surrounding gas once subject to an external wind and discuss the results in Sec. 5. 
Consistent with previous VESTIGE works, we assumed NGC 4254 at 16.5 Mpc, the mean distance of the cluster substructure to which the galaxy 
belongs (Gavazzi et al. 1999; Mei et al. 2007; Cantiello et al. 2024). We recall that at the assumed distance of the galaxy, 1\arcsec = 80 pc. 
The main parameters of NGC 4254 are given in Table \ref{gal}.

\begin{table}
\caption{Properties of the galaxy NGC 4254 (M99).}
\label{gal}
{
\[
\begin{tabular}{ccc}
\hline
\noalign{\smallskip}
\hline
Variable                          & Value                                                 & Ref.          \\      
\hline
\rm{Type}                         & SA(s)c                                                & 1       \\
$cz_{\rm Hel}$                    & 2406 km s$^{-1}$                                      & 1       \\
$M_\mathrm{star}^a$               & 3.3$\times$10$^{10}$ M$_{\odot}$                      & 2       \\ 
\mhi                              & 5.5$\times$10$^{9}$ M$_{\odot}$                       & TW      \\ 
$M\mathrm{(H_2)}^b$               & 7.6$\times$10$^{9}$ M$_{\odot}$                       & 3       \\ 
${\rm SFR}^a$                     & 4.35 M$_{\odot}$ yr$^{-1}$                            & 2       \\              
Distance                          & 16.5 Mpc                                              & 4,5,6,7,8 \\
Proj.~ distance~from~M87          & 0.98 Mpc, 0.63 $r/r_\mathrm{200}^c$                   & TW       \\
\noalign{\smallskip}
\hline
\end{tabular}
\]
References: (1) NED; (2) Boselli et al. (2023); (3) Brown et al. (2021), (4) Mei et al. (2007); (5) Gavazzi et al. (1999); 
(6) Blakeslee et al. (2009); (7) Cantiello et al. (2018); (8) Cantiello et al. (2024); (TW) this work.\\
(a) $M_\mathrm{star}$ and SFR are derived assuming a Chabrier (2003) IMF and the Calzetti et al. (2010) calibration.
(b) derived assuming $X_{\rm CO}$ = 2.0 $\times$ 10$^{20}$ cm$^{-2}$ (K km s$^{-1}$)$^{-1}$ and $R_{21}$ = 0.8, Leroy et al. (2009).
(c) assuming $r_\mathrm{200}$ = 1.55 Mpc (Ferrarese et al. 2012). }
\end{table}

\section{Observations and data reduction}

\subsection{ALMA 114 GHz $^{12}$CO(1-0) line }
\label{sec:almadata}

\begin{figure*}
\centering
\includegraphics[width=0.99\textwidth]{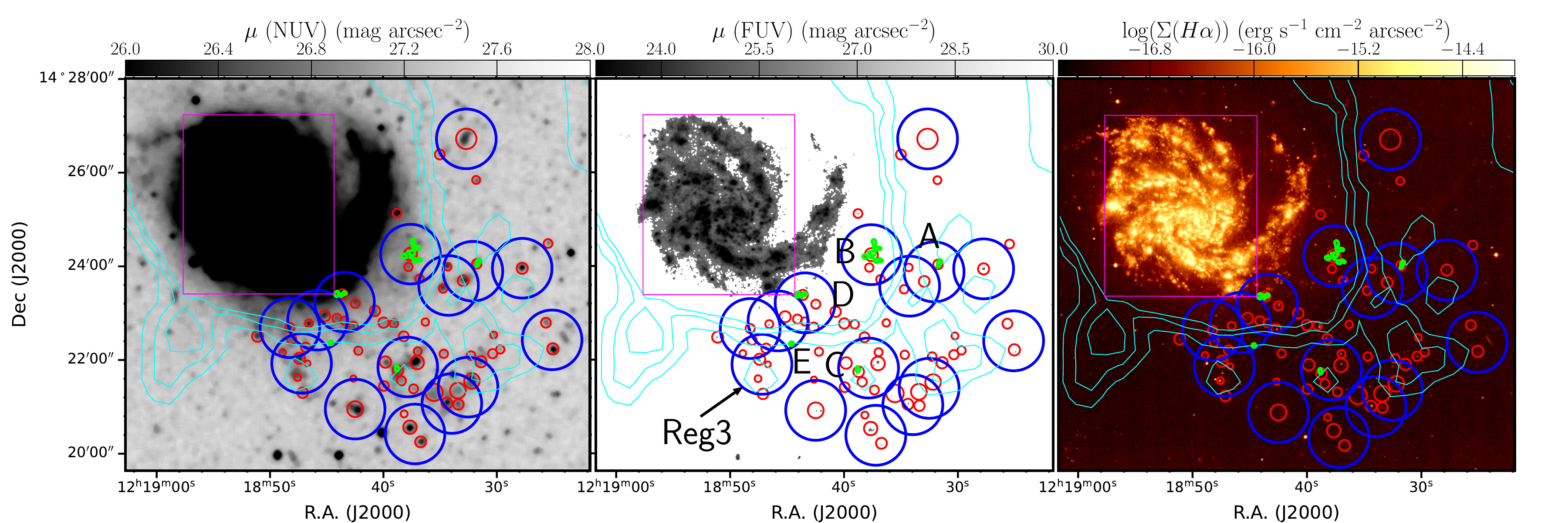}\\
\caption{GALEX NUV (\textit{left}), FUV ASTROSAT/UVIT (\textit{centre}), and continuum-subtracted H$\alpha$ (\textit{right}) images of the galaxy NGC 4254. 
The cyan contour shows the \hi\ column density at \shi\ $=$ 0.1, 0.5, 1 M$_{\odot}$ pc$^{-2}$ at 27\arcsec\ angular resolution. 
The star-forming regions located outside the stellar disc and identified in Boselli et al. (2018b) and studied in this work are indicated with red circles (see Fig. \ref{idregions} 
for their identification). The green contours show the detected molecular clouds at a column density of $\Sigma\mathrm{(H_2)}$ $=$ 1 M$_{\odot}$ pc$^{-2}$. Labels A, B, C, D, and E 
indicate the different molecular gas complexes (see Fig. \ref{zoom_region} for details), and Reg3 is the region detected in \hi.
ALMA has a beam size of 2.2\arcsec $\times$ 1.9\arcsec\, and the position angle is $PA$ = $-$34.5 degrees. MeerKAT has a beam size of 30.5\arcsec $\times$ 24.5\arcsec\, and the 
position angle is $PA$ = 165 degrees. The coverage of the 15 different ALMA pointings are indicated with blue circles of diameter 75\arcsec, the size of the cube, while the magenta 
box shows the footprint of ALMA/PHANGS.}
\label{ALMA_pallini}%
\end{figure*}

The ALMA coverage of the tail of NGC4254 was achieved with 15 different pointings. These were chosen to cover most of the star-forming regions analysed in Boselli et al. (2018b) 
and located outside the star-forming disc (see Fig. \ref{ALMA_pallini}). The observations were carried out with the 12m array during ALMA Cycle 7, between 23 and 29 March 2019 
(Project 2018.1.00982.S, PI: M. Boquien). The on-source integration was 544 seconds for each of the 15 fields. The array configuration had a minimum and a maximum baseline of 
15\,m and 500\,m, respectively. The spectral setup was optimised for the $^{12}$CO(1-0) transition line in band 3, at 114.352 GHz (2.6 mm) observing frequency, with a channel 
width of $\sim1.3$ km s$^{-1}$ (0.488 MHz). At this frequency, the primary beam full width at  half maximum (FWHM) is $\sim$ 55\arcsec (corresponding to 4.4 kpc at the distance of the galaxy).

We downloaded the 15 $^{12}$CO(1-0) cubes from the ALMA archive. The cubes have pixel size 0.38\arcsec and channel width 1.27 \kms. The typical noise level of the cubes 
is $\sim5.2$ mJy beam$^{-1}$ [0.12 K]. The typical beam size is 2.2\arcsec $\times$ 1.9\arcsec [176$\times$152 pc], with a position angle $PA\sim-35$ degrees measured 
from North counter-clockwise. We searched the cubes for line emission using \texttt{SoFiA} (Serra et al. 2015, Westmeier et al. 2021). In fact, we noticed that all cubes 
exhibit faint and spectrally broad stripes -- both positive and negative -- visible on RA-frequency and Dec-frequency projections. These artefacts are likely caused by imperfect 
continuum subtraction and, since they cover a significant fraction of each cube's volume, affect the quality of the source finding and parameterisation. Therefore, we first 
used \texttt{SoFiA} to identify voxels with likely line emission, masked them out while fitting and subtracting the continuum with the \texttt{imcontsub} task of the \texttt{mowjsub} 
package\footnote{\url{https://github.com/laduma-dev/mowjsub}} to improve the continuum subtraction (Fig.~\ref{fig:contsub}), and finally run \texttt{SoFiA} again on the improved 
cubes for the final source finding. The key steps of \texttt{SoFiA} are: the normalisation of the cube to the local noise; the detection of voxels with possible emission with a 
smoothing+clipping algorithm; the friends-of-friends linking of detected voxels to form sources; the rejection of unreliable sources (see Serra et al. 2015 for more details). 
Our settings for these key steps were: local noise measurement in 3D windows with a width of 25 pixel and 101 channels, followed by interpolation to map the smooth noise gradients 
across the cube; Gaussian spatial smoothing $= 0$, 3, 6 pixel FWHM, boxcar spectral smoothing  $=0$, 3, 7, 15 channels, and detection threshold $=4\sigma$ for each combination of 
spatial and spectral smoothing kernels; linking radius of 3 pixels and 3 channels; minimum source reliability of 99\% and minimum integrated S/N of 5. We also dilated the final 
3D detection masks by 1 pixel and 1 channel in order to include faint emission at the edge of each source.

For comparison with the inner disc of NGC 4254, we also use the combined 12m+7m+TP ALMA data gathered during 
the Physics at High Angular Resolution in Nearby Galaxies (PHANGS)-ALMA survey (Leroy et al. 2021) and available on the 
PHANGS database\footnote{\url{https://sites.google.com/view/phangs/home/data}}. These data have been taken at 
230 GHz (1.3 mm) to map the $^{12}$CO(2-1) transition with a typical integration time of 20-30 seconds
per pointing and a fairly symmetric beam of size FWHM $\simeq$ 2\arcsec. The available map covers the inner 
regions of the disc. For comparison with our data, taken in the 
$^{12}$CO(1-0) transition line, we convert the observed moment 0 map (in K km s$^{-1}$) of the disc assuming
a $^{12}$CO(1-0)/$^{12}$CO(2-1) = 0.8 line ratio (Leroy et al. 2009). 
We stress that the proximity of the galaxy (16.5 Mpc) allowed us to study the relation between gas distribution and star formation at an angular resolution ($\simeq$ 2\arcsec\ [$\simeq$ 160 pc]) 
sufficient to resolve individual GMCs. This resolution has never been reached in other perturbed galaxies in further clusters such as Coma, Norma, or in the GASP sample. It also exceeds the 
resolution reached during the VERTICO survey of the Virgo cluster ($\simeq$ 720 pc; Brown et al. 2021).

\subsection{UVIT imaging}

NGC 4254 has been observed with ASTROSAT/UVIT (Agrawal et al. 2006; Tandon et al. 2020) during targeted 
observations of a dozen of galaxies in the Virgo cluster (Proposal A08003, PI: J. Hutchings). The galaxy has been observed using the far-ultraviolet (FUV) filter BaF2 
($\lambda_c$ = 1541 \AA; $\Delta\lambda$ = 380 \AA) with an integration time of 8055 s, sufficient to reach a typical surface brightness of $\mu$(FUV) $\simeq$ 26.6 AB mag arcsec$^{-2}$.
The data were reduced as described in Tandon et al. (2020) using a zero point of $z_p$ = 17.771 mag. 
The astrometry of the field was checked against the accurate Next Generation Virgo Survey (NGVS) imaging data (Ferrarese et al. 2012).
All the extraplanar \hii\ regions identified in Boselli et al. (2018b) are included within the 
large field of view ($\simeq$ 28$\arcmin$ in diameter). Most of them (53/60) 
are detected with a signal-to-noise $S/N>3$ and resolved thanks to 
the angular resolution ($\simeq$ 1.5$\arcsec$) of the instrument. The FUV UVIT image of the galaxy is shown in Fig. \ref{ALMA_pallini}, while magnitudes are given in Table \ref{HIIUVIT}.

\subsection{VESTIGE narrow-band H$\alpha$ imaging}

The galaxy NGC 4254 and its surrounding regions have been observed during VESTIGE, a narrow-band (NB) H$\alpha$ imaging survey of 
the Virgo cluster (Boselli et al. 2018a). The NB H$\alpha$ imaging data of the galaxy, along with those in the optical $ugiz$ and NUV and FUV bands gathered 
during the NGVS (Ferrarese et al. 2012) and the GALEX UltraViolet Virgo Cluster Survey (GUViCS, Boselli et al. 2011) of the cluster, have been extensively analysed in Boselli et al. (2018b). 
The reader can refer to these works for a detailed description of the observing strategy and data reduction
procedure. Here we just summarise the most relevant points necessary for the following analysis. NGC 4254
was observed using MegaCam at the CFHT in the NB filter MP9603 ($\lambda_c$ = 6590 \AA; $\Delta\lambda$ = 104 \AA;
$T$ = 93\%) optimally designed to include the redshifted ($v$ = 2404 km s$^{-1}$) H$\alpha$+[\nii] emission of 
the galaxy and of its \hi\ stripped tail. MegaCam is composed of 40 CCD with a pixel scale of 
0.187 arcsec pixel$^{-1}$. The galaxy has several exposures gathered during the complete survey 
of the cluster but also from a previous targeted observation carried out during a pilot project.
The resulting integration time is 15\,000 s in the NB filter and 1500 s in the broad $r$-band filter 
necessary for the stellar continuum subtraction. The data are of excellent quality, with a typical seeing of 
0.66\arcsec ~in the stacked image. The typical sensitivity of the VESTIGE survey is of
$f\mathrm{(H\alpha)}$ $\simeq$ 4 $\times$ 10$^{-17}$ erg s$^{-1}$ cm$^{-2}$ for point sources 
(5$\sigma$) and $\Sigma\mathrm{(H\alpha)}$ $\simeq$ 2 $\times$ 10$^{-18}$ erg s$^{-1}$ cm$^{-2}$ 
arcsec$^{-2}$ for extended sources (1$\sigma$ detection limit at $\sim$ 3\arcsec angular resolution),
while those in the $r$-band are 24.5 AB mag (5$\sigma$) for point sources
and 25.8 AB mag arcsec$^{-2}$ (1$\sigma$) for extended sources. Given that the integration time here is
about a factor of two longer, the depth of the NB H$\alpha$ and $r$-band data is a factor of $\sqrt{2}$ 
better than those gathered elsewhere in the cluster. The subtraction of the stellar continuum is done 
using a combination of the $g$ and $r$ broad-band images, as extensively described in Boselli et al. (2019, see also Boselli et al. 2025). 
The continuum subtracted image of the galaxy and of the star-forming regions located along the tail of stripped material analysed in this work are shown in Fig. \ref{ALMA_pallini}.

\subsection{MeerKAT \ion{H}{i} data}

\begin{figure}
\centering
\includegraphics[width=0.49\textwidth]{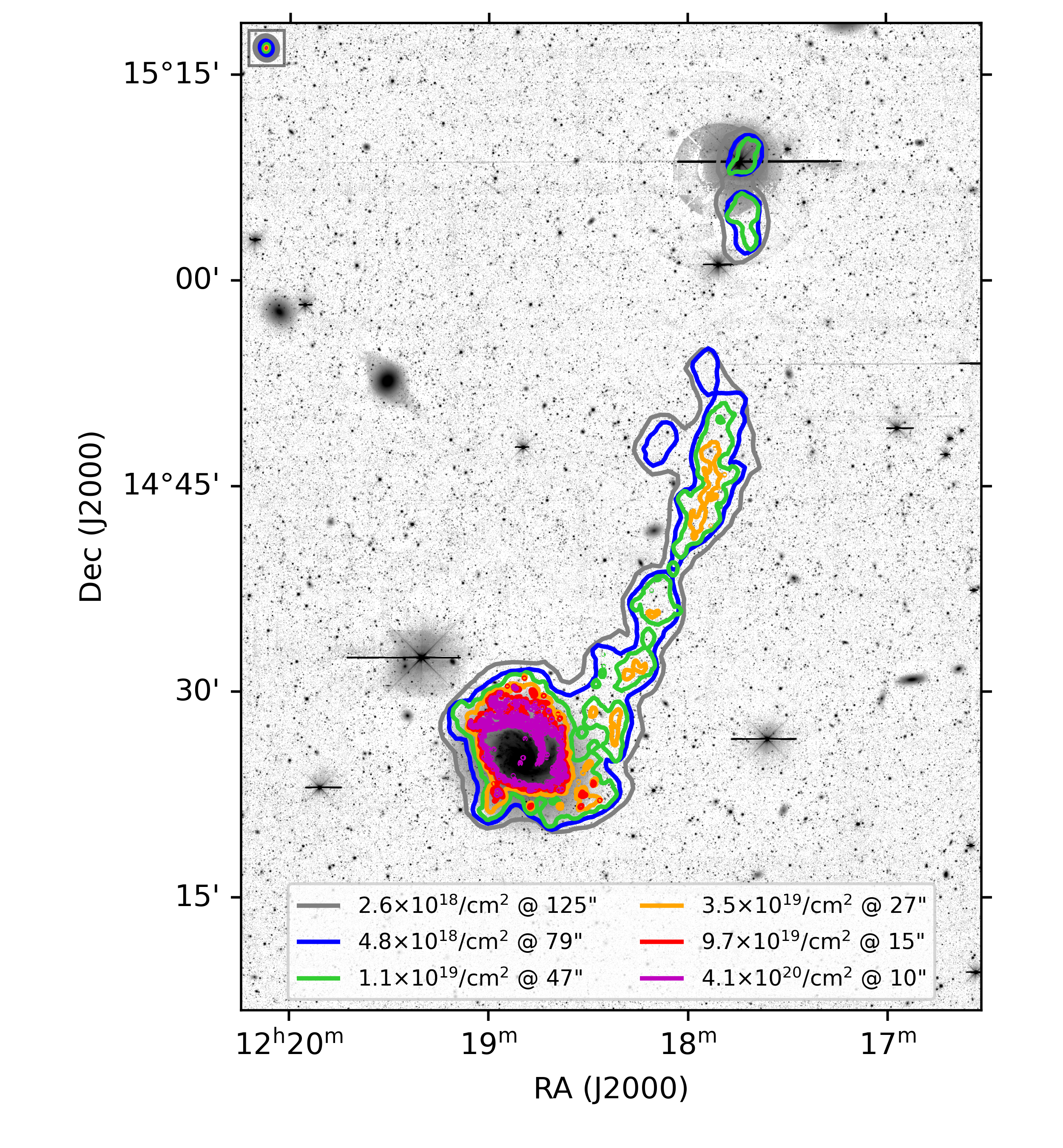}\\
\caption{\hi\ gas distribution in NGC 4254 and in its tail. For each resolution, we show the lowest reliable contour (3$\sigma$ over a line width of 25 km s$^{-1}$). 
The contours are overlaid on an $r$-band image. The contour levels and resolutions are listed in the legend at the bottom and shown in the top-left corner.}
\label{MeerKAT_full}%
\end{figure}

\begin{figure}
\centering
\includegraphics[width=0.48\textwidth]{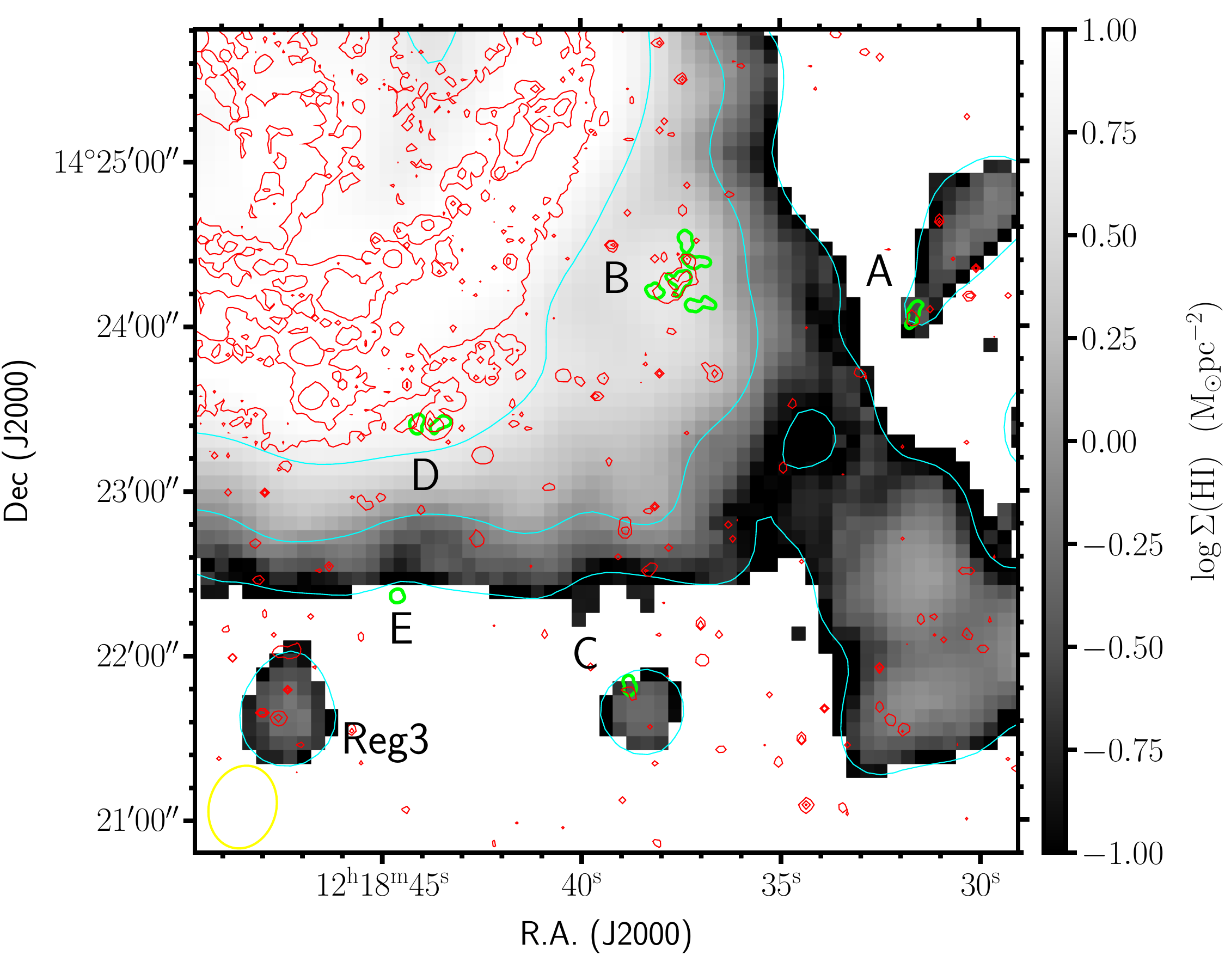}\\
\caption{\hi\ gas distribution around the CO detected regions. The beam size of the MeerKAT observations (yellow ellipse) is 30.5\arcsec $\times$ 24.5\arcsec\, and the position angle is 
$PA$ = 165 degrees, while the beam size of ALMA is $\simeq$ 2\arcsec. The green contours show the detected molecular gas at column density $\Sigma\mathrm{(H_2)}$ = 1 M$_{\odot}$ pc$^{-2}$; 
the cyan contours show the \hi\ at column densities \shi\ $=$ 0.1, 1, 5 M$_{\odot}$ pc$^{-2}$.
Red contours show the H$\alpha$ surface brightness $\Sigma\mathrm{(H\alpha)}$ = 2, 10, 40 $\times$ 10$^{-17}$
erg s$^{-1}$ cm$^{-2}$ arcsec$^{-2}$. Labels A, B, C, D, and E indicate the different molecular gas complexes detected with ALMA, and Reg3 is the region detected in \hi.}
\label{MeerKAT}%
\end{figure}

In the following analysis we also use the \hi\ data obtained with MeerKAT as part of the ViCTORIA project, which is described in de Gasperin et al. (2025). We refer to that paper 
for all details on the \hi\ observations and data processing. For the purpose of this study we remind the reader that, thanks to the excellent $uv$ coverage of MeerKAT, we were able 
to make sensitive \hi\ cubes and images at different angular resolutions from 10\arcsec\ to 125\arcsec (the channel width is always 5.5 \kms). The \hi\ column density sensitivity ranges 
between \nhi\ = $4.1 \times 10^{20}$ cm$^{-2}$ [\shi\ $=$ 3.3 M$_{\odot}$ pc$^{-2}$] at the highest resolution and \nhi\ = $2.6 \times 10^{18}$ cm$^{-2}$ [\shi\ $=$ 0.02 M$_{\odot}$ pc$^{-2}$] 
at the lowest resolution (at each resolution, the sensitivity is defined as the median column density of pixels whose \hi\ signal-to-noise ratio is between 2 and 3 in the NGC~4254 field).
Figures \ref{MeerKAT_full} and \ref{MeerKAT_color} show the \hi\ gas distribution (moment-0 map) derived using the MeerKAT data of the whole galaxy including the extended \hi\ tail, 
while Fig. \ref{MeerKAT} shows the atomic gas distribution at $\simeq$ 27\arcsec\ resolution around the regions detected in CO outside the stellar disc.

\subsection{MUSE/PHANGS IFU spectroscopy}

The galaxy NGC 4254 has been observed using the Multi Unit Spectroscopic Explorer (MUSE) at the Very Large Telescope (VLT) in spectroscopic mode during the PHANGS survey (Emsellem et al. 2022). 
The galaxy, which has an optical diameter of 7.59\arcmin ~(Binggeli et al. 1985), has been mapped using twelve independent pointings, with a typical integration time of 43 min/pointing. 
Given its angular extension, the MUSE pointings do not cover the external disc of the galaxy and the outer \hii\ regions observed with ALMA. The combined datacube has $FWHM=0.89$\arcsec, 
an angular resolution that is close to the one reached in the deep NB H$\alpha$ image. The spectral range covered during the observations is 4750--9350\AA, with a typical spectral 
resolution $\sigma$ $\simeq$ 80 km s$^{-1}$ at the blue end up to $\sigma$ $\simeq$ 35 km s$^{-1}$ at the red end. For a detailed description of the observations and data reduction we 
refer the reader to Emsellem et al. (2022).

\section{Derived parameters}

\subsection{Atomic hydrogen}

The new MeerKAT data allowed us to measure the atomic hydrogen mass of the galaxy disc, \mhi\ $= (5.5 \pm 0.6) \times 10^9 $ M$_{\odot}$, and of the extended tail, 
\mhi\ $= (3.4 \pm 0.4) \times 10^8 $ M$_{\odot}$, where the uncertainties include the statistical error and a 10\% flux-scale error. To make these measurements, we define the tail 
as the region including all the \hi\ that deviates clearly from the rotating gas disc when viewing the cubes (at all resolutions) in virtual reality with \texttt{iDaVIE} 
(Jarrett et al. 2021; Sivitilli et al. 2026). The resulting mass values can be compared with those gathered for the VIVA and ALFALFA surveys (once scaled to our assumed distance of 16.5 Mpc). 
The former gives \mhi\ $= (4.7 \pm 0.4) \times 10^9$ M$_{\odot}$ for the disc (Chung et al. 2009). The latter \mhi\ $= 4.2 \times 10^9$ M$_{\odot}$ and $4.9 \times 10^9$ M$_{\odot}$ 
for the disc (Giovanelli et al. 2007 and Haynes et al. 2018, respectively), and \mhi\ $= (4.2\pm 0.6) \times 10^8$ M$_{\odot}$ for the tail (Haynes et al. 2007).

The \hi\ gas not associated with the stellar disc has a patchy distribution with a typical column density ranging from
\nhi\ $\simeq$ 1--5 $\times 10^{19}$ cm$^{-2}$ [\shi\ $=$ 0.08--0.4 M$_{\odot}$ pc$^{-2}$] in its diffuse component (Fig. \ref{MeerKAT}) to 
\nhi\ $\simeq$ 5--10 $\times 10^{20}$ cm$^{-2}$ [\shi\ $=$ 4--8 M$_{\odot}$ pc$^{-2}$] where the molecular gas has been detected. Region 3 [RA(J2000)=12:18:47.554; Dec=+14:21:37.30] 
is the only one among the external star-forming regions which has been detected as an \hi\ cloud physically detached in space and velocity from the main galaxy disc 
[$v$(\hi\ ) = 2285 km s$^{-1}$ for the cloud versus $v$(\hi\ ) = 2360 km s$^{-1}$ for the disc at this position when measured at 50\arcsec -- 80\arcsec\ resolution; see 
Table \ref{HIIUVIT} and Fig. \ref{MeerKAT}]. Its spectrum does not change when measured at 15\arcsec\ and 27\arcsec\ angular resolution, suggesting that the cloud is compact. 
Its mass is \mhi\ $= (1.7 \pm 0.4) \times 10^6 $ M$_{\odot}$, $v$(\hi\ ) = 2285 km s$^{-1}$ and a line width $W_{20}$(\hi) $= 18$ km s$^{-1}$. The remaining regions located outside 
the diffuse \hi\ disc of the galaxy are all undetected with an upper limit of \mhi\ $= 10^6 $ M$_{\odot}$ (3$\sigma$ over 25 km s$^{-1}$). For those located within the extended, 
diffuse \hi\ disc, confusion prevented us from estimating a reliable upper limit.

These column densities are measured here with a resolution of $\simeq$ 15\arcsec.
With the star-forming complex and the molecular gas cloud being totally unresolved in the MeerKAT data, 
these column densities should be considered only as mean values. If the gas is distributed within clumpy
structures, locally its column density might be significantly higher.
Further out along the tail, which extends by $\simeq$ 240 kpc in projected distance, the \hi\ column density detected by MeerKAT is in the range 
2$\times$10$^{18}$ $\lesssim$ \nhi\ $\lesssim 5\times10^{19}$ cm$^{-2}$, with lower column densities measured at 80\arcsec--120\arcsec\ angular resolution in the diffuse 
component while the highest values in the clumpy regions at 27\arcsec--50\arcsec resolution.

\subsection{Molecular hydrogen}

\begin{figure*}
\centering
\includegraphics[width=0.33\textwidth]{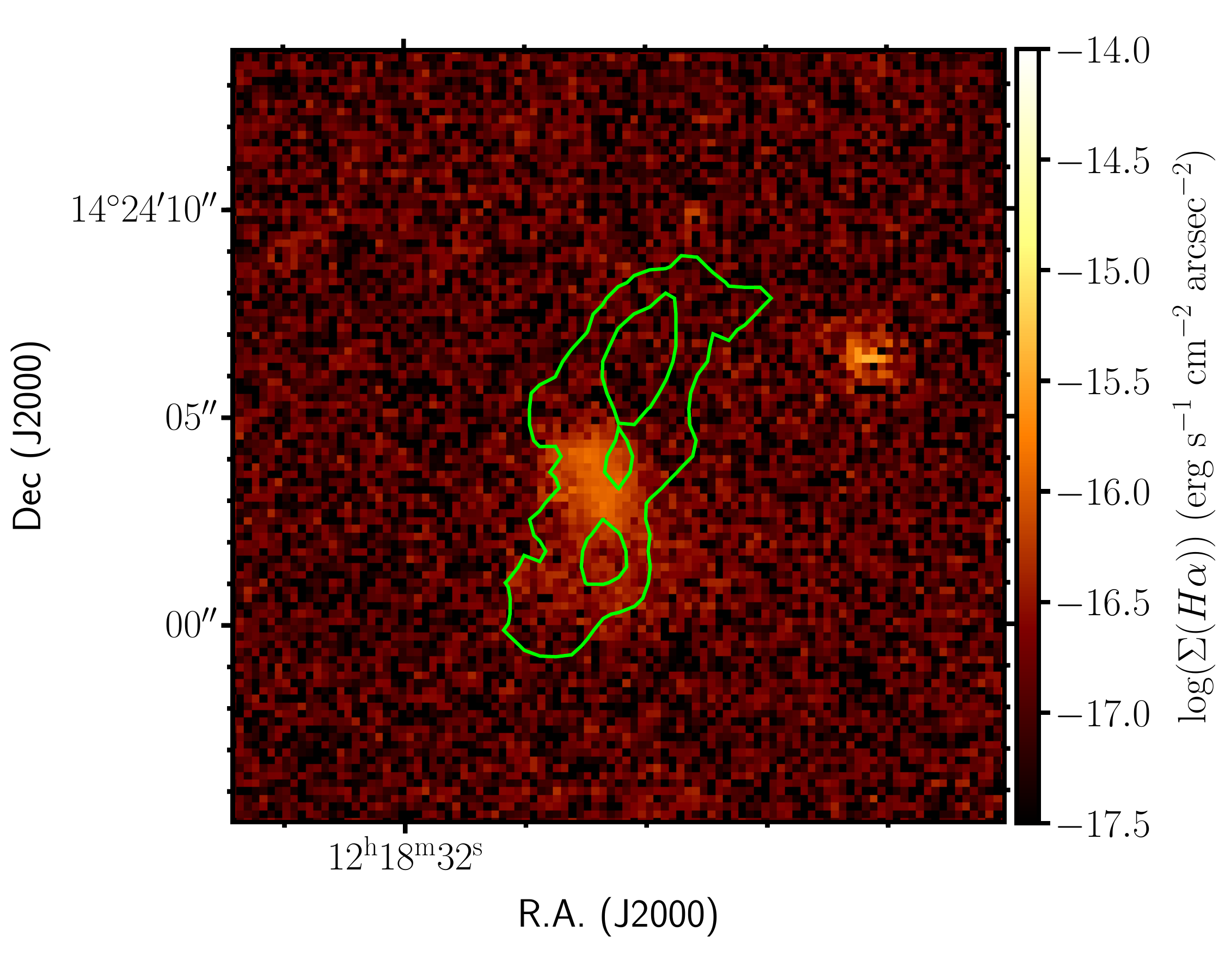}
\includegraphics[width=0.33\textwidth]{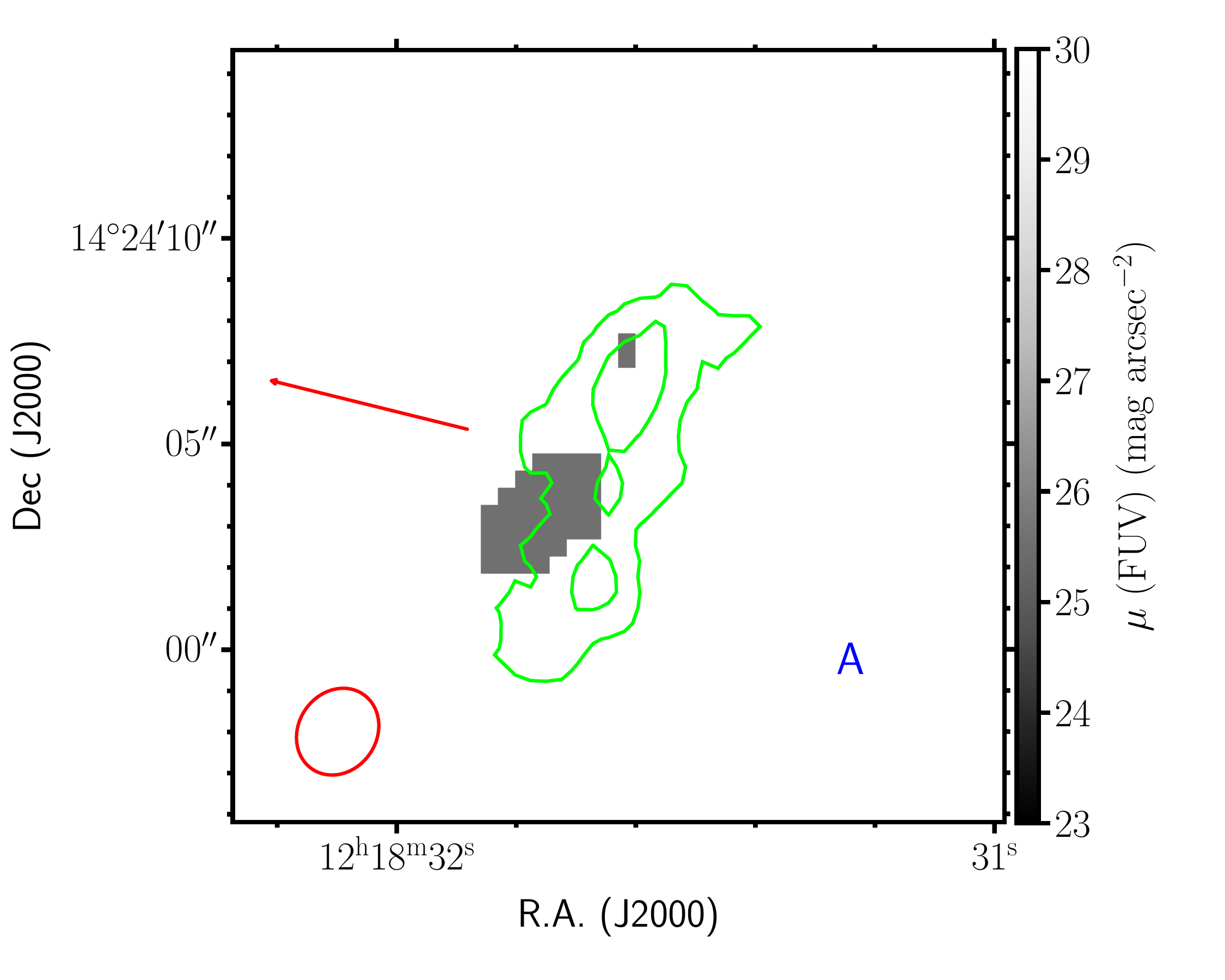}
\includegraphics[width=0.33\textwidth]{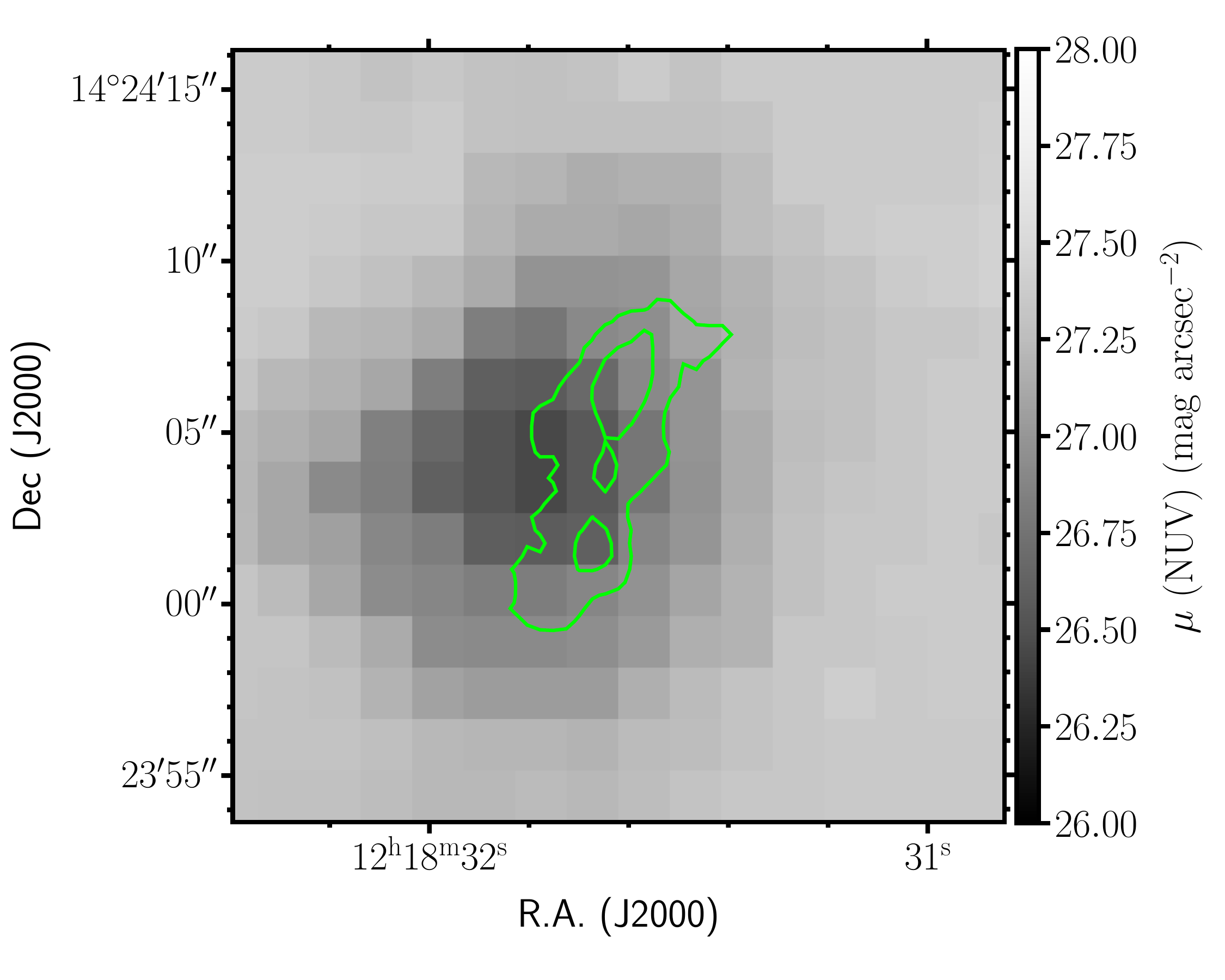}\\
\includegraphics[width=0.33\textwidth]{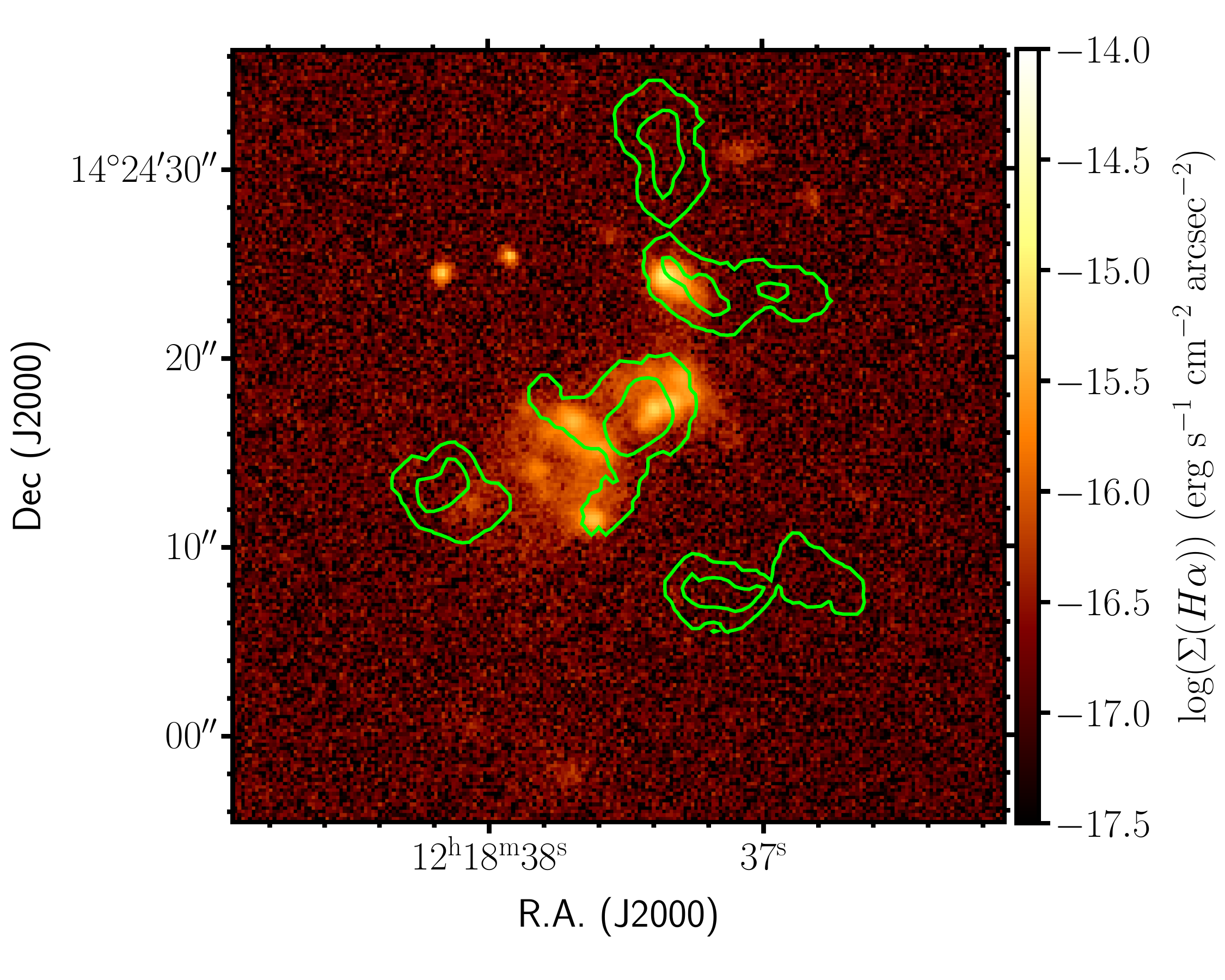}
\includegraphics[width=0.33\textwidth]{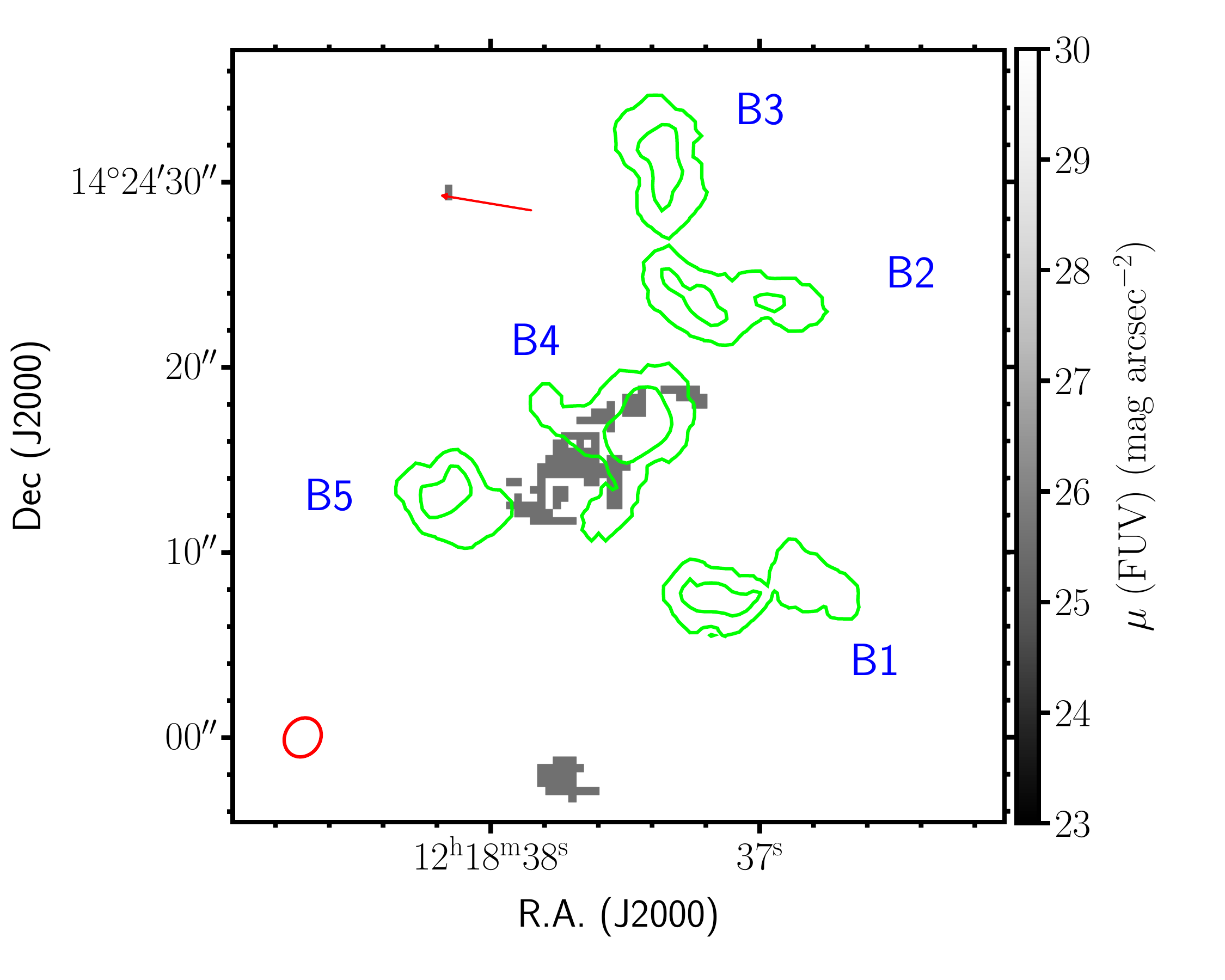}
\includegraphics[width=0.33\textwidth]{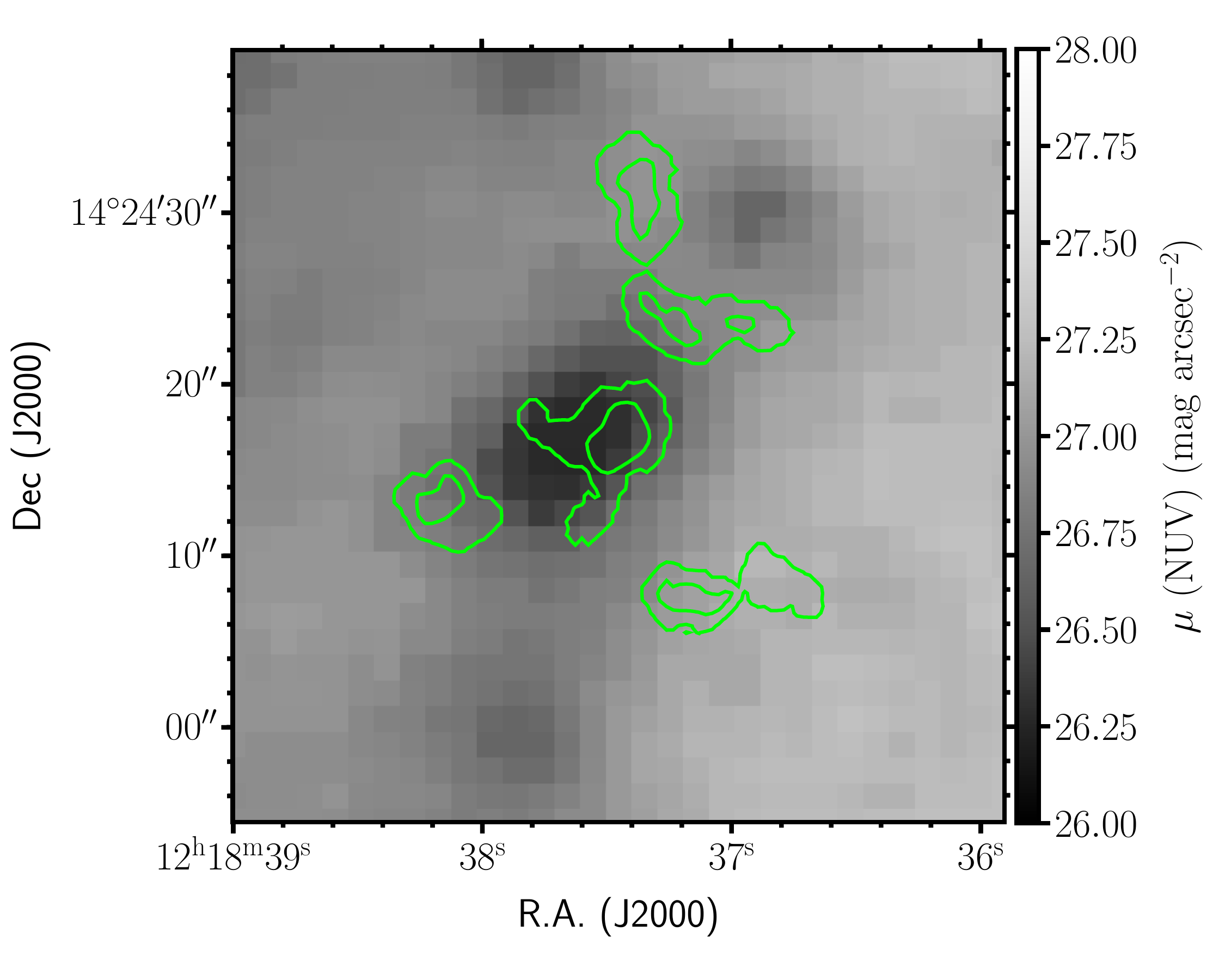}\\
\includegraphics[width=0.33\textwidth]{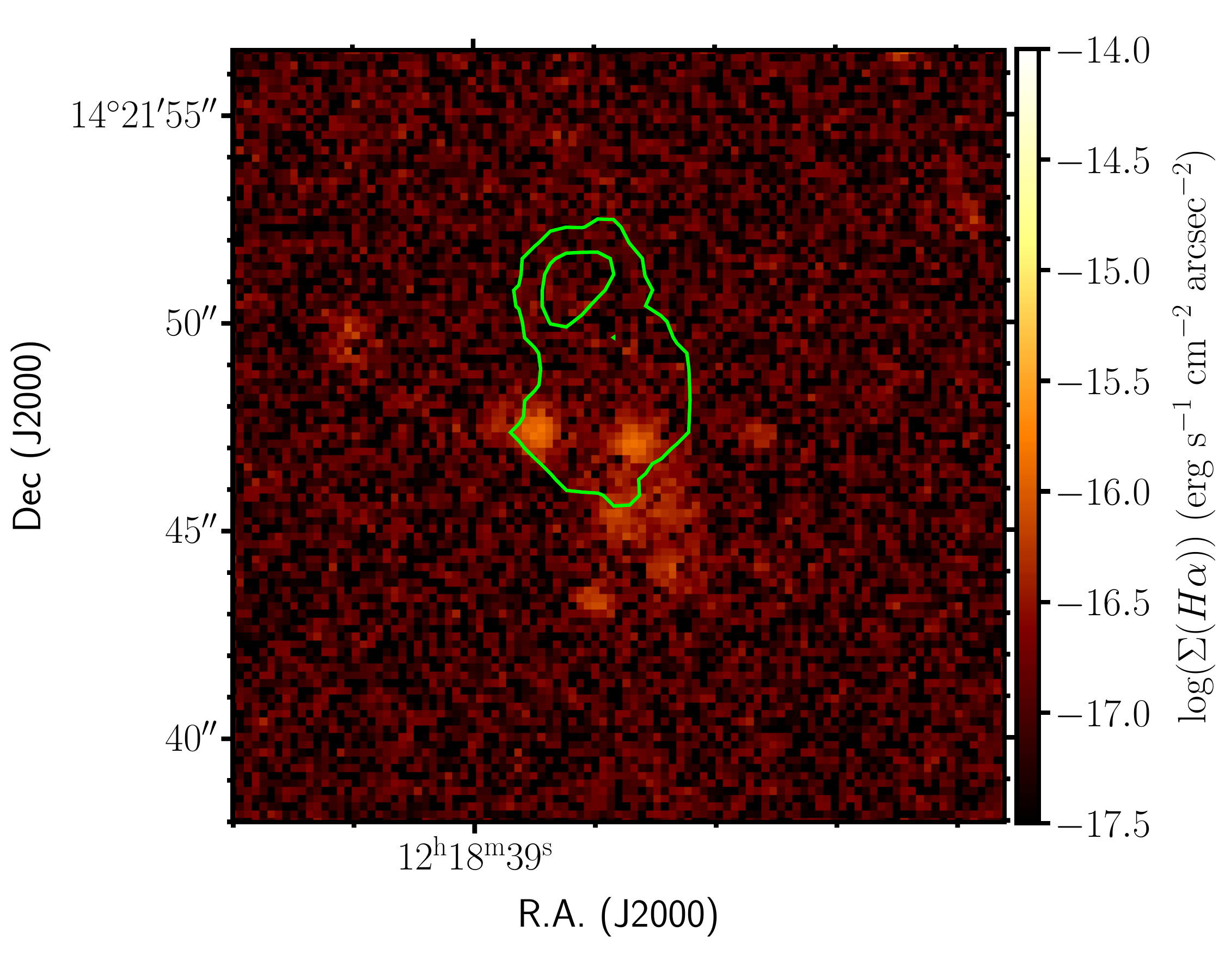}
\includegraphics[width=0.33\textwidth]{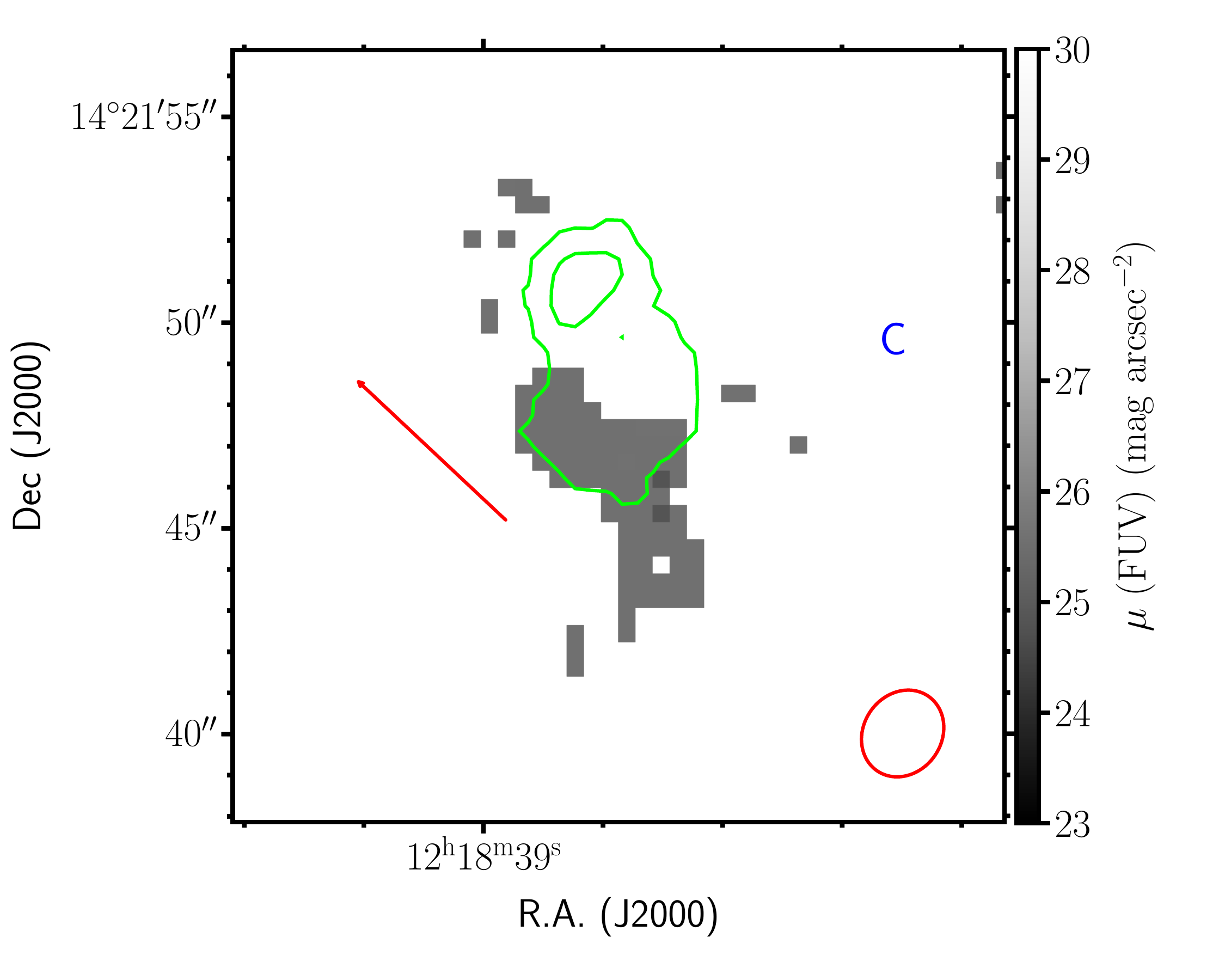}
\includegraphics[width=0.33\textwidth]{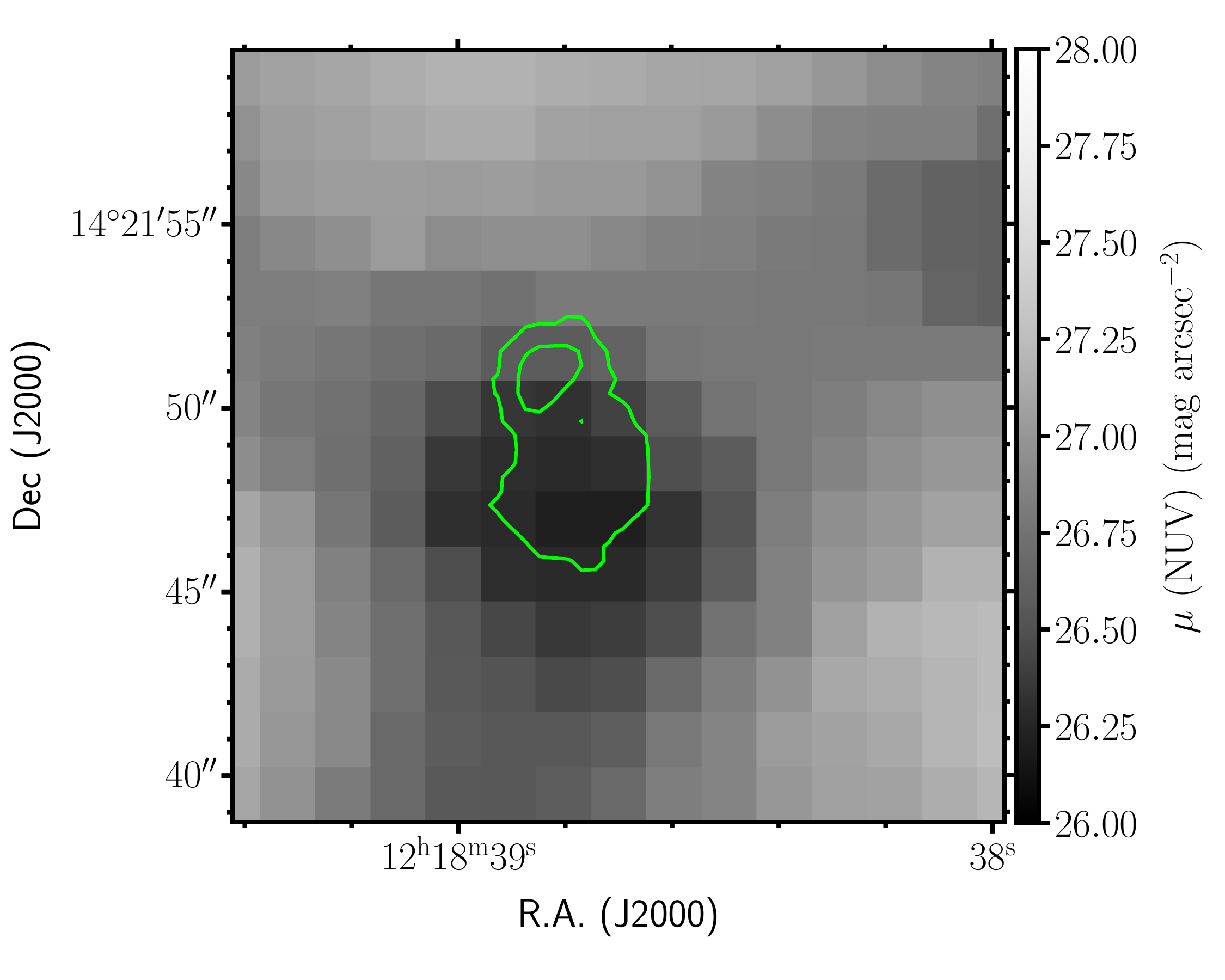}\\\includegraphics[width=0.33\textwidth]{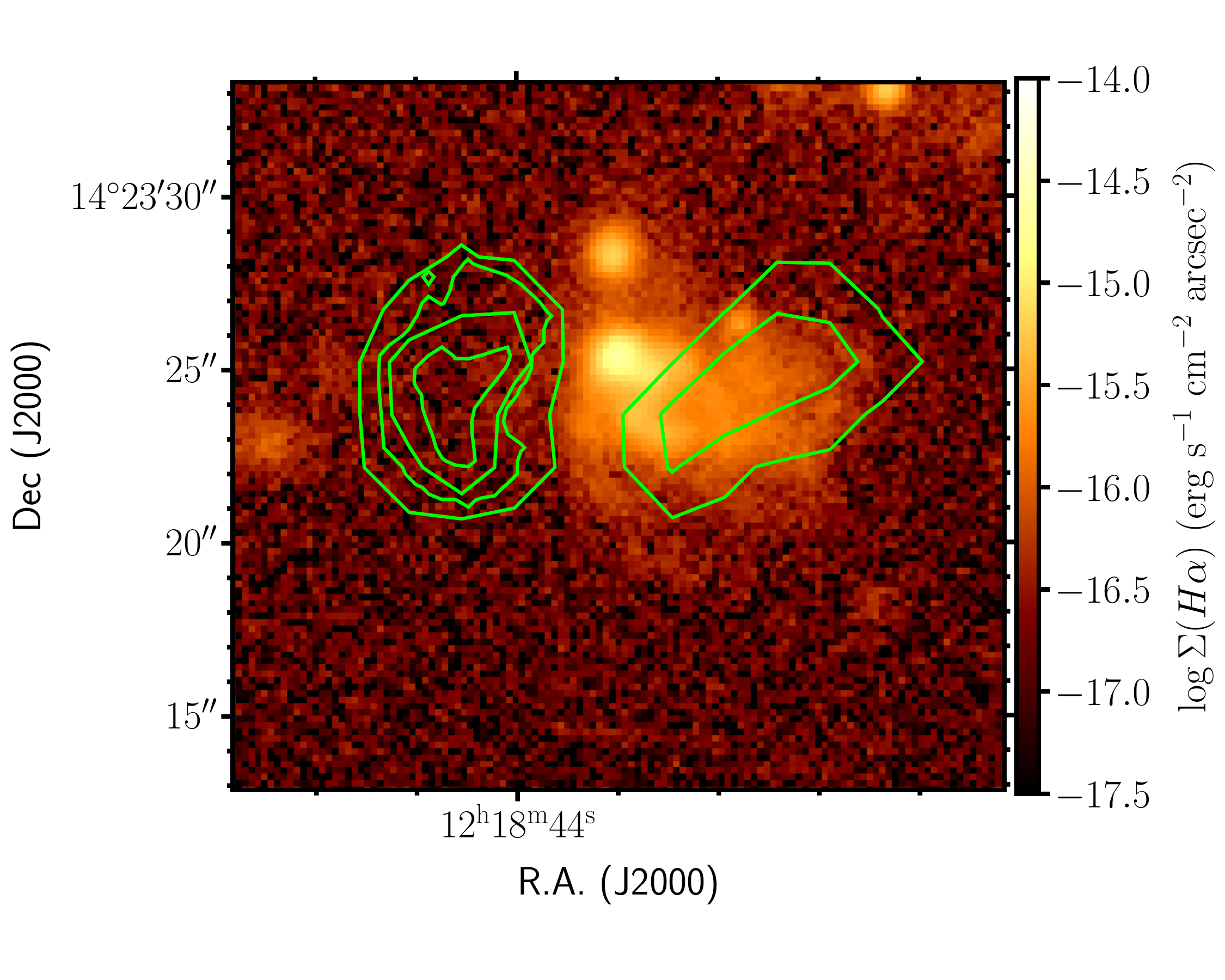}
\includegraphics[width=0.33\textwidth]{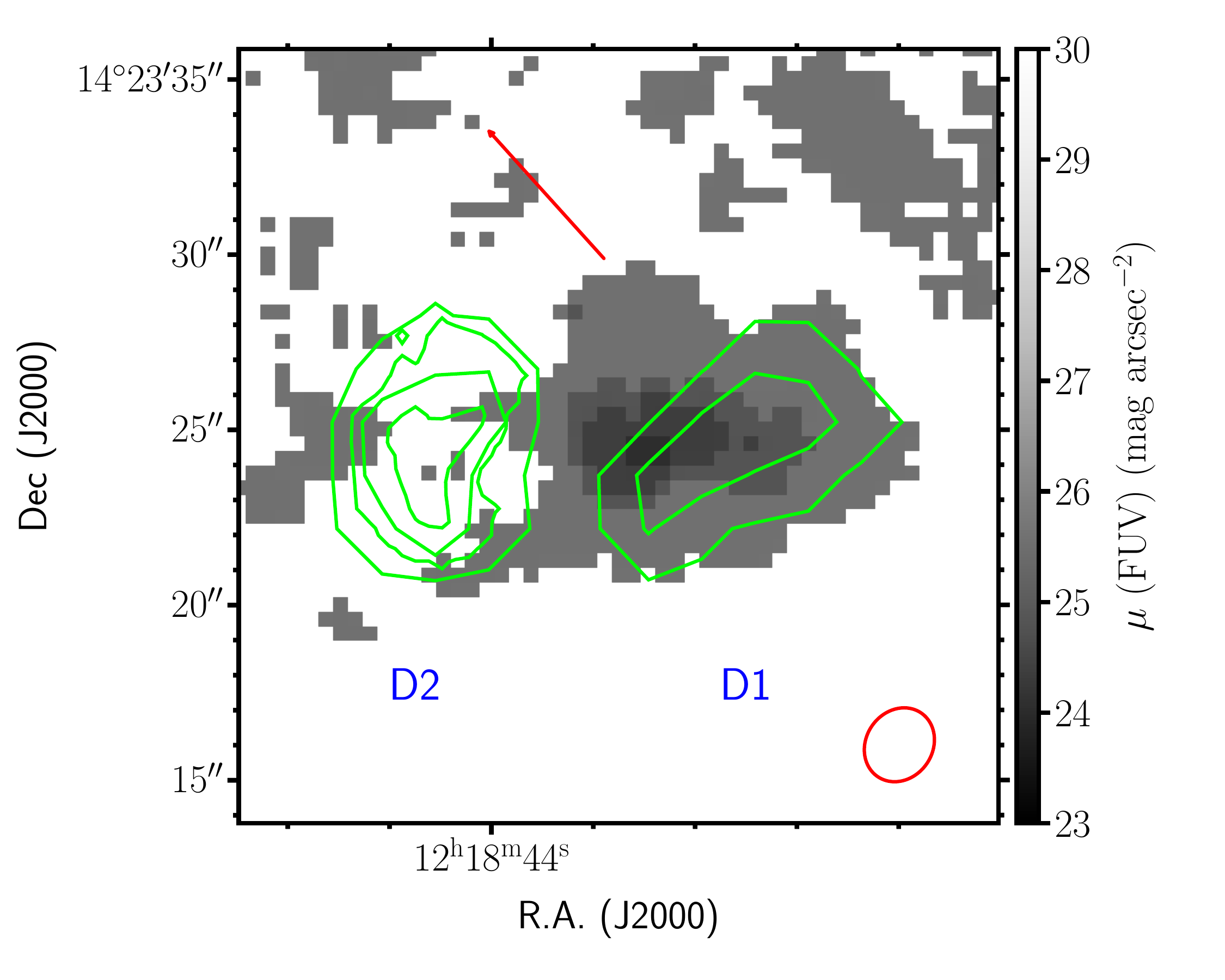}
\includegraphics[width=0.33\textwidth]{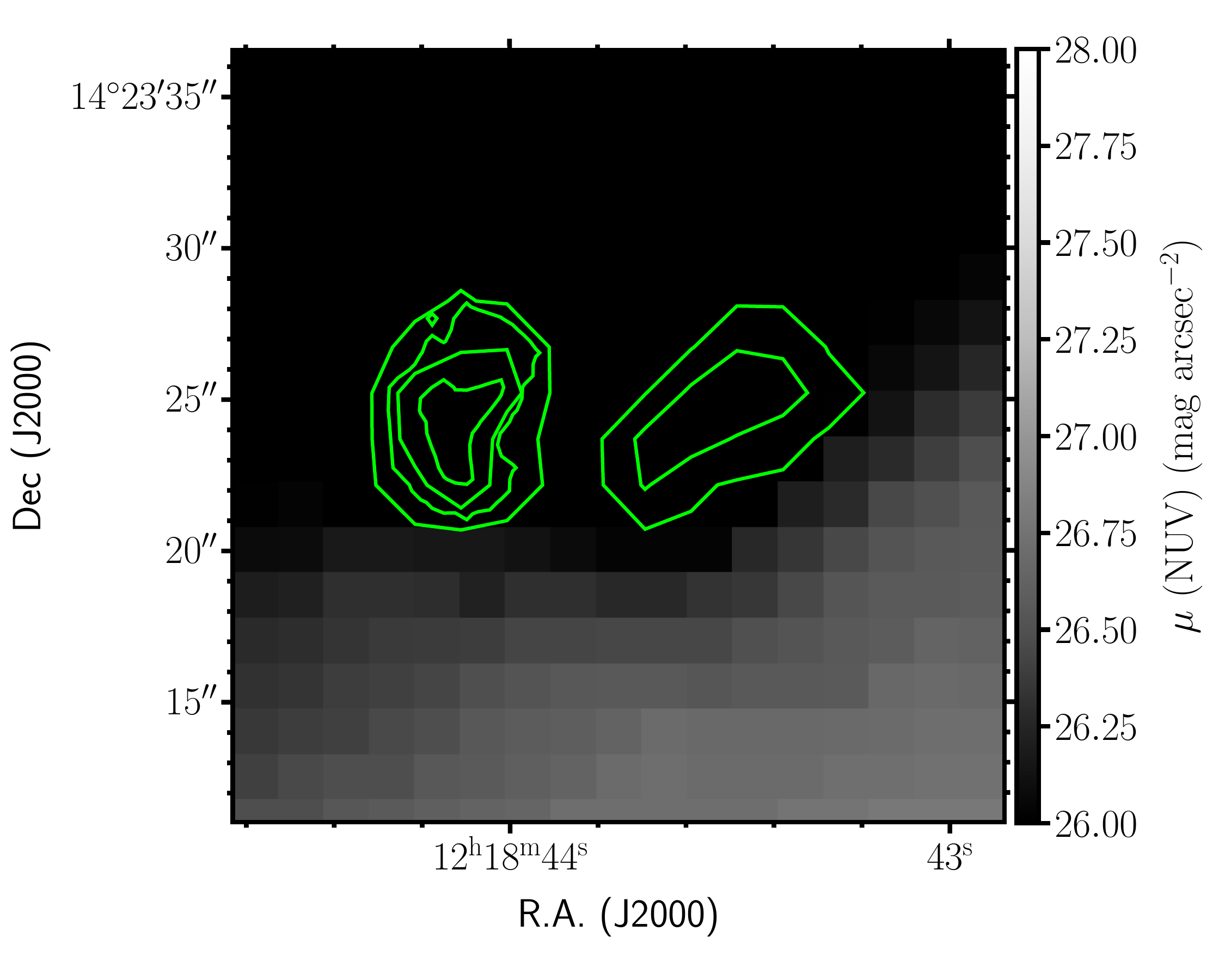}\\
\includegraphics[width=0.33\textwidth]{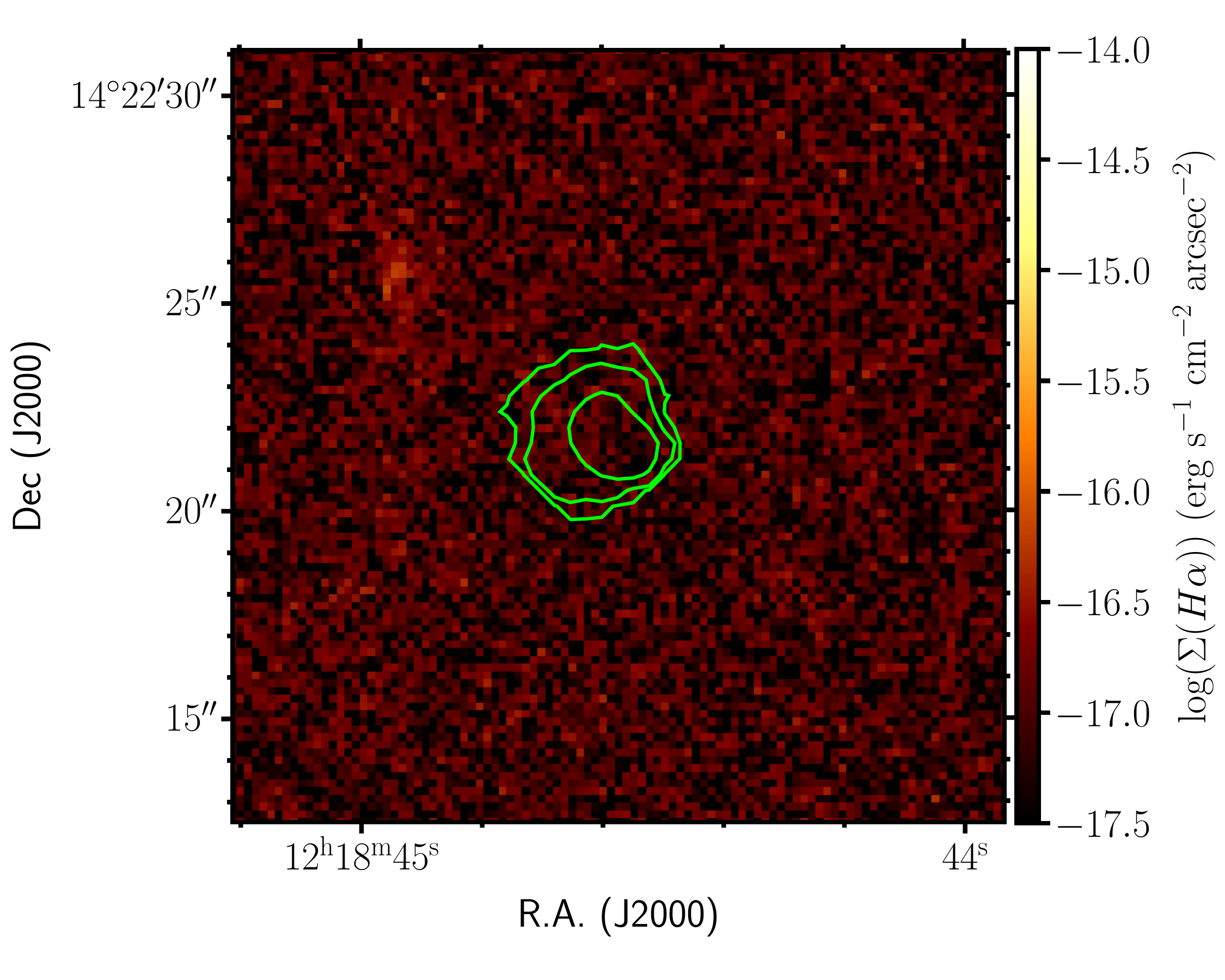}
\includegraphics[width=0.33\textwidth]{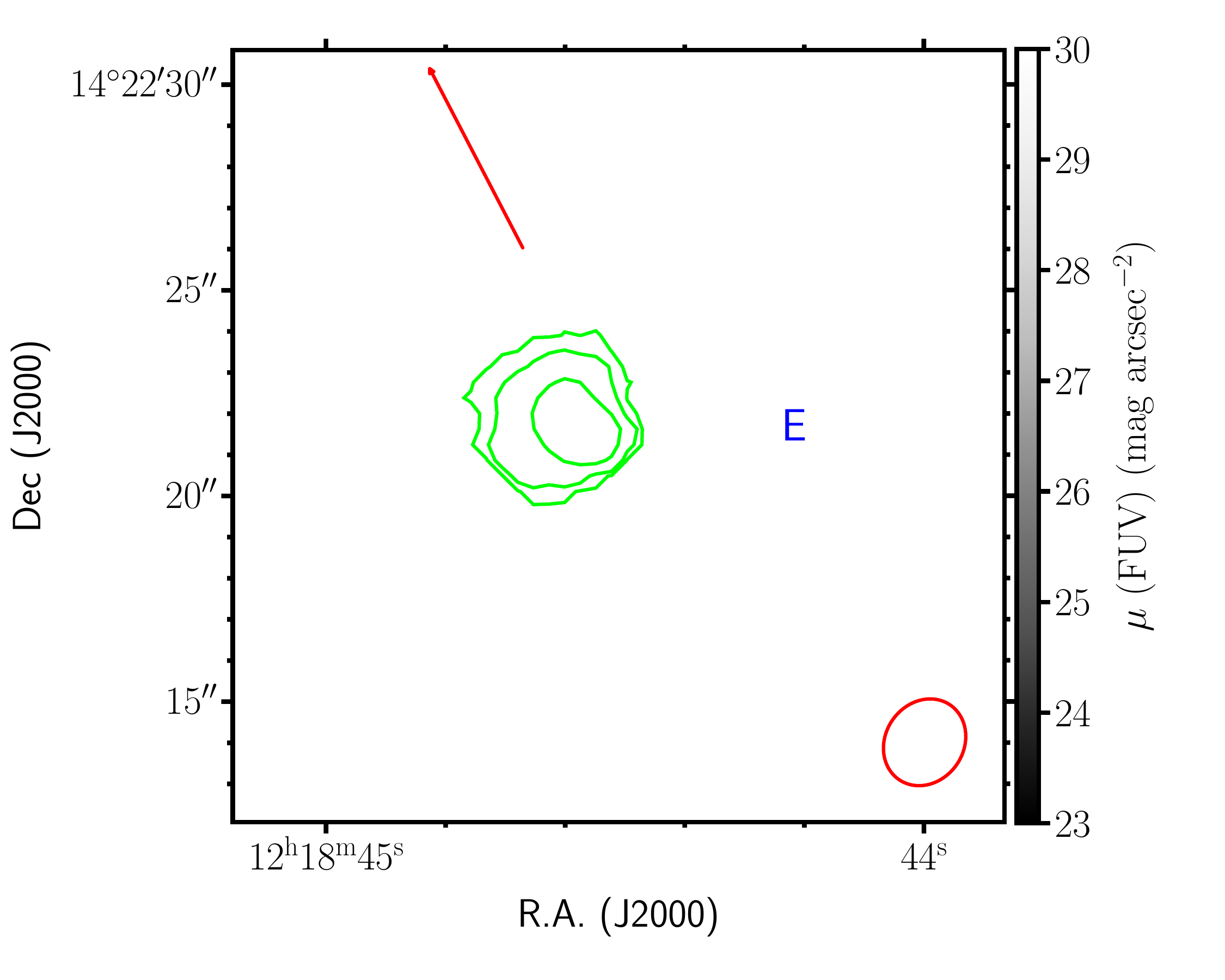}
\includegraphics[width=0.33\textwidth]{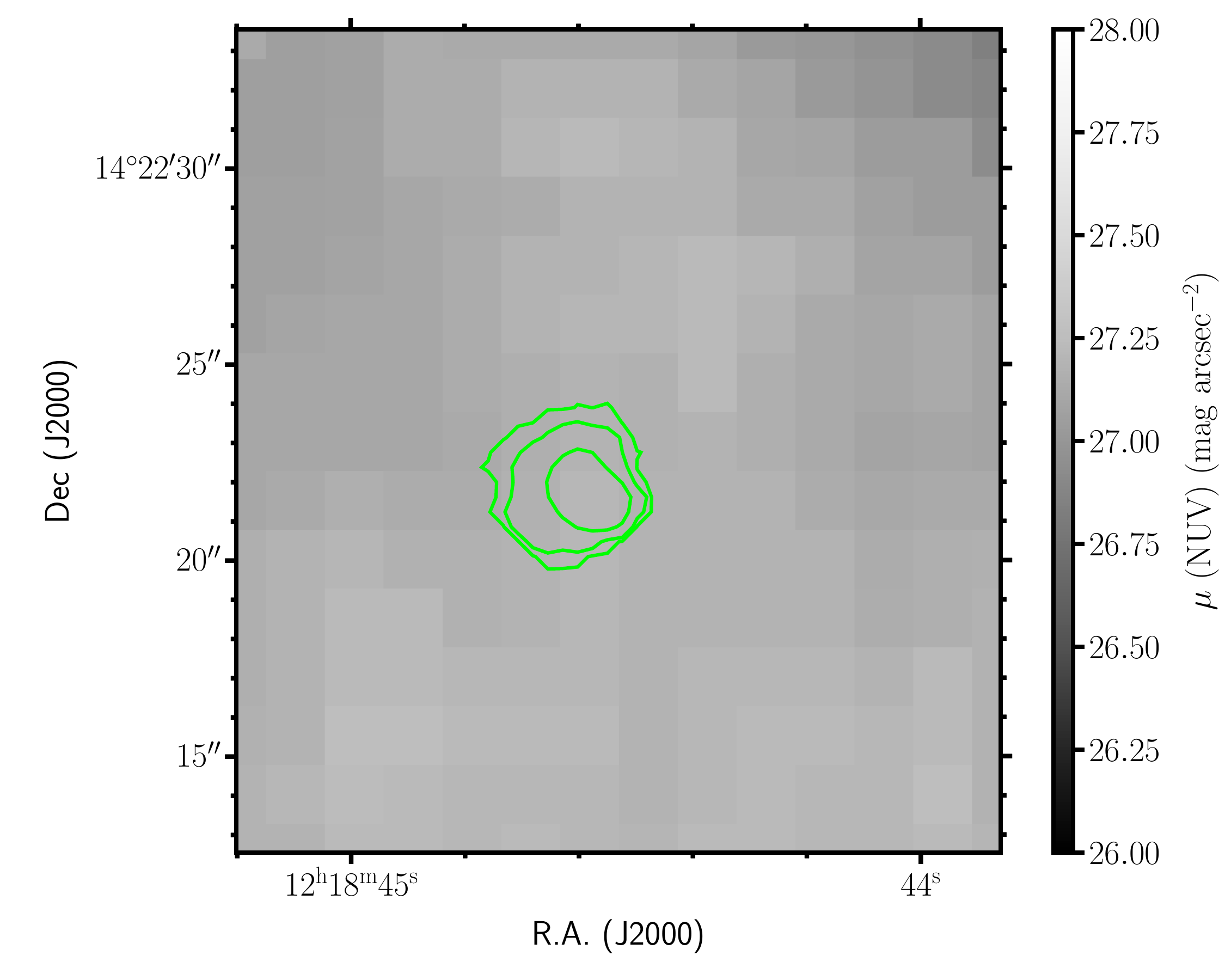}\\
\caption{\textit{Left column}: Continuum-subtracted H$\alpha$ images of the molecular clouds detected by ALMA in the CO(1-0) emission line. 
Assuming a Chabrier IMF, [\nii]$\lambda$6583\AA/H$\alpha$=0.2, and $A\mathrm{(H\alpha)}$=0.7 mag as in Boselli et al. (2018b). An observed H$\alpha$ surface brightness 
of $\Sigma\mathrm{(H\alpha)}$=10$^{-16}$ erg s$^{-1}$ cm$^{-2}$ arcsec$^{-2}$ corresponds to $\Sigma({\rm SFR})$ = 3.8$\times$10$^{-3}$ M$_{\odot}$ yr$^{-1}$ kpc$^{-2}$. 
\textit{Central column}: UVIT BaF2 images. \textit{Right column}: GALEX NUV images. The green contours show the molecular gas distribution at column densities of 
$\Sigma\mathrm{(H_2)}$ $=$ 1, 10, 40 M$_{\odot}$ pc$^{-2}$. The red arrow in the central column indicates the direction of the centre of the galaxy. The red ellipse shows the beam size of the
ALMA observations (2.2\arcsec\ $\times$ 1.9\arcsec\, $PA$ = $-$34.5 degrees), while blue labels identify the different GMCs.
}
\label{zoom_region}%
\end{figure*}

The 15 ALMA pointings cover only 49 out of the 60 external star-forming regions discovered by Boselli et al. (2018b; Fig. \ref{ALMA_pallini}). Within these regions, we detected ten 
molecular complexes, all spatially resolved in the ALMA data (Fig. \ref{zoom_region}). Some of them have a Gaussian line profile (clouds B3, B4, B5, C, D1, D2) suggesting that they 
are also resolved in the velocity space. Others have multiple components, suggesting that the clouds are not fully resolved in the velocity space (see Fig. \ref{COvelprofile}). \footnote{Given 
that GMCs have typical sizes of 10-100 pc in radius, these complexes might be composed by more than a single cloud as assumed in this work. The results presented in the following analysis 
which depend on this assumption should be taken with caution. We notice, however, that their CO brightness temperature is in the range 4 $<$ $T_{\rm B}$ $<$ 12 K, similar to the one observed 
in resolved Galactic GMCs ($T_{\rm B}$ $\simeq$ 4-20 K; Solomon et al. 1987), suggesting minimal beam dilution and thus that the clouds are resolved.}
They all have line intensities of $S$(CO) $= 250-700$ mJy km s$^{-1}$ (see Table \ref{TabCO}). These line intensities were converted into CO luminosities using the relation 
(Solomon \& Vanden Bout 2005)

\begin{equation}
  L^{'}(\mathrm{CO}) = 3.25\times 10^7 \left[\frac{S(\mathrm{CO})}{\rm{Jy~km~s^{-1}}}\left(\frac{D}{\rm{Mpc}}\right)^2\right]  \frac{(1+z)^{-3}}{(\nu/\mathrm{GHz})^2}
\end{equation}

\noindent
and into molecular hydrogen masses using

\begin{equation}
    M\mathrm{(H_2)} = \alpha_{\mathrm{CO}} L^{'}(\mathrm{CO})
,\end{equation}

\noindent
which assuming a Galactic conversion factor $\alpha_{\rm CO}$ = 4.3 M$_{\odot}$ pc$^{-2}$ (K km s$^{-1}$)$^{-1}$ [$X_{\rm CO}$ = 2.0 $\times$ 10$^{20}$ cm$^{-2}$ (K km s$^{-1}$)$^{-1}$]
give molecular hydrogen masses of $M\mathrm{(H_2)}$ $\simeq$ 0.75--2.08 $\times$ 10$^6$ M$_{\odot}$ including He
(see Table \ref{TabCO}) and mean column densities 
$\Sigma\mathrm{(H_2)}$ $\simeq$ 10 M$_{\odot}$ pc$^{-2}$, with only one cloud (GMC E) with $\Sigma\mathrm{(H_2)}$ $\simeq$ 45 M$_{\odot}$ pc$^{-2}$. 

For the other undetected regions, we derived 3$\sigma$ upper limits on their molecular gas content. We assumed a line width of 25 km s$^{-1}$ [the typical $W_\mathrm{20}$(CO) value for 
the detections] and integrated the CO emission spatially within a box of size 2$R$, where $R$ is the radius of the aperture used to identify the selected \hii\ region (see Table \ref{TabCO}). 
We converted the resulting upper limits on $S({\rm CO})$ into upper limits on $M\mathrm{(H_2)}$ with the same conversions given above.

\section{Analysis}
\label{sec:analysis}

The detection rate of molecular gas in the star-forming regions located outside the stellar disc of NGC 4254 and 
discovered by Boselli et al. (2018b) is extremely low, with only ten GMCs associated with four star-forming complexes out of the 42 identified so far, plus one (J121844.60+142221.8) 
located far from any star-forming region (Fig. \ref{zoom_region}). We describe their properties in this section.

\subsection{General properties}

\begin{figure}
\centering
\includegraphics[width=0.48\textwidth]{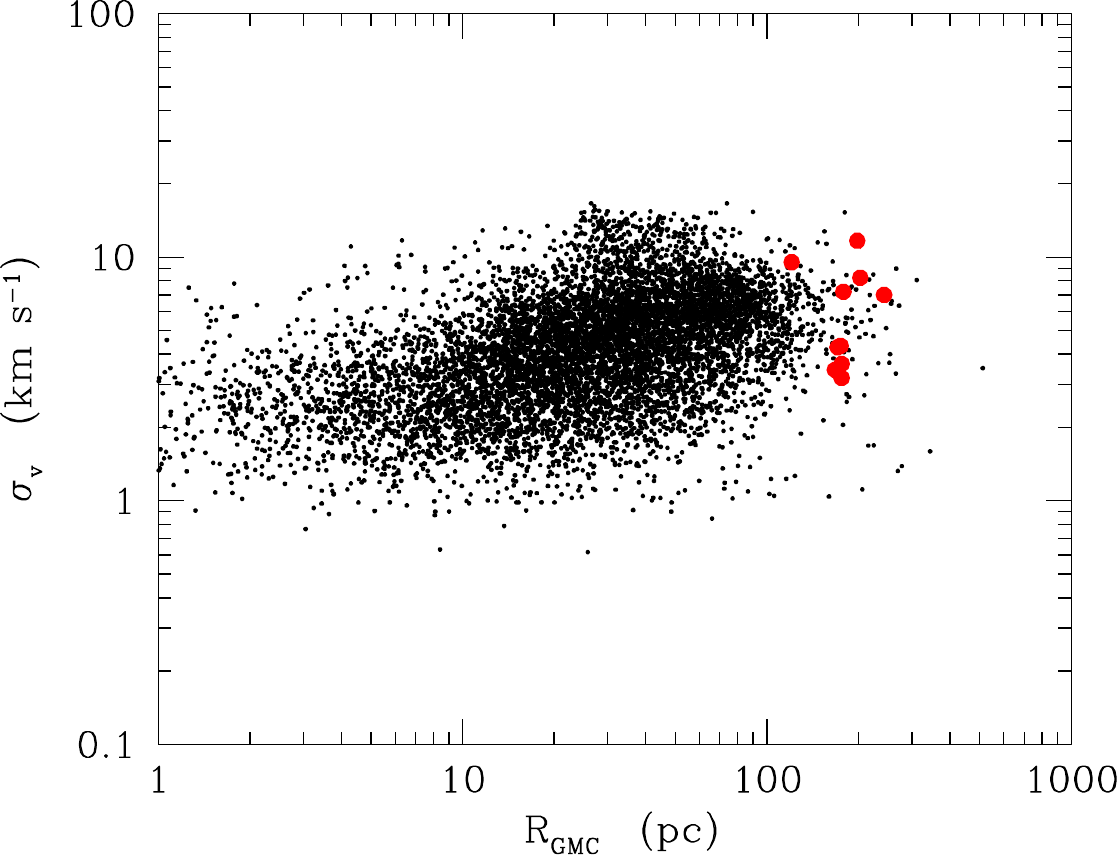}\\
\caption{Relationship between the CO line width and the radius of the GMCs detected outside the stellar disc of NGC 4254 (red dots) compared to those in the Milky Way (black dots, 
from Miville-Desch\^enes et al. 2017). } 
\label{Solomon}%
\end{figure}

The molecular hydrogen properties of these regions can be compared to those observed in the disc of normal, nearby star-forming systems in terms of estimated total mass, velocity width, 
and column density. In Fig. \ref{Solomon} we compare the size and velocity dispersion of the ten GMCs of NGC 4254 (see Table \ref{TabCO}) with those measured in the Milky Way by 
Miville-Desch\^enes et al. (2017). Our measurements are made in the same way as in that paper, namely: we estimate major and minor axis of each cloud through the spatial moment 
analysis performed by \texttt{SoFiA} (see \texttt{SoFiA} user manual v2.6.53), and use the resulting values to calculate the cloud size with Eq.~18 of Miville-Desch\^enes et al. 
(2017); and we estimate the velocity dispersion as the second moment of the integrated spectrum. We caution here that, as mentioned above, we consider these molecular gas complexes 
as individual GMCs, an assumption which should be confirmed with higher resolution data. Figure \ref{Solomon} shows that our regions are populating the large size-velocity dispersion 
range observed in the Milky Way, but following the mean relation between these two quantities first observed by Larson (1981) and Solomon et al. (1987). They also follow the 
$\sigma_{v}$ versus $\Sigma(\mathrm{H}_2)$$\times$$R$ relation of Milky Way GMCs (Miville-Desch\^enes et al. 2017). We recall that, in general, GMCs located in the outer disc of 
the Milky Way or of other galaxies such as M33 are less turbulent and are thus characterised by a lower velocity dispersion than those located in the inner regions (Braine et al. 2018).

\begin{figure}
\centering
\includegraphics[width=0.48\textwidth]{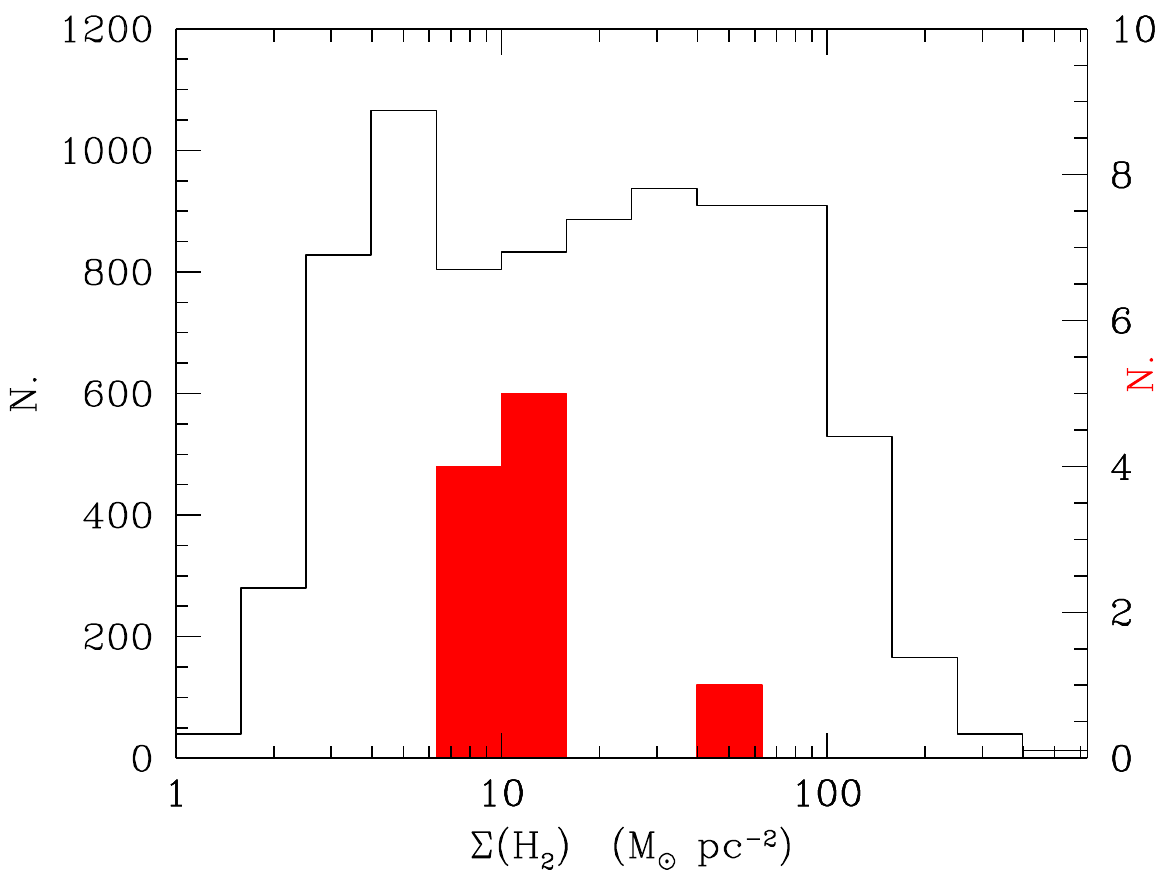}\\
\caption{Density distribution of the GMCs detected outside the stellar disc of NGC 4254 (red shaded histogram) compared to the one of the GMC in the Milky Way (black histogram, 
from Miville-Desch\^enes et al. 2017). The left Y-axis gives the number of Galactic GMCs; the right Y-axis  shows the number of GMCs in the outskirts of NGC 4254.}
\label{density}%
\end{figure}

Their mean molecular hydrogen column densities 
are, on average, lower than those measured within GMCs in the Milky Way ($\langle\Sigma\mathrm{(H_2)}\rangle$ = 41 $\pm$ 5 M$_{\odot}$ pc$^{-2}$; Lada \& Dame 2020), as depicted in 
Fig. \ref{density}, or in other nearby galaxies observed during the PHANGS survey (Rosolowsky et al. 2021). They rather correspond to those measured in the outer disc of the Milky 
Way (Miville-Desch\^enes et al. 2017).
These values can be compared to the mean cold atomic gas column density measured with MeerKAT on a spatial scale of a few kpc$^2$ in Fig. \ref{MeerKAT}.

Considering the molecular hydrogen mass of the detected GMCs [$M({\rm H_2})$ = 0.75--2.08 $\times$ 10$^6$ M$_{\odot}$], their corresponding star-formation rates derived using the SFR 
versus $M({\rm H_2})$ scaling relation measured within the Milky Way (Lada et al. 2012)
should be SFR $\simeq$ 4--10 $\times$ 10$^{-4}$ M$_{\odot}$ yr$^{-1}$ (derived assuming a fraction of dense gas of only 1\%)\footnote{The dense gas fraction generally decreases 
in the outer disc of star-forming galaxies (Braine et al. 2023)}, 
while the one measured using our H$\alpha$ imaging data is more than  an order of magnitude lower. Dust attenuation cannot
be at the origin of this large difference since it would imply an attenuation of $A\mathrm{(H\alpha)}$ $\simeq$ 3 mag. This number is inconsistent with the mean dust attenuation 
measured using the Balmer decrement derived from the MUSE data in the outer disc of the galaxy, where $A\mathrm{(H\alpha)}$ $\simeq$ 0.8 mag (Fig. A.1 in Boselli et al. 2025).
This suggests that the activity of star formation within these regions is reduced with respect to similar GMCs in the Milky Way, or that here the depletion time is $\simeq$ a 
factor of five longer than within the disc of star-forming galaxies (e.g. Leroy et al. 2008).
Figure \ref{zoom_region}, however, suggests that these GMCs belong to larger complexes of star-forming regions, composed of several individual \hii\ regions. Their integrated 
H$\alpha$ emission is much higher, suggesting that at these scales (160 pc), the star-forming complexes are locally decoupled by $\simeq$ 100--200 pc from the associated GMCs, 
as predicted by simulations (e.g. Steyrleithner et al. 2020). We do not observe any evident preferential direction in the relative position of GMCs and associated star-forming 
complexes versus the centre of the galaxy or of the cluster, or the direction of the tidal \hi\ gas tail.

\subsection{Schmidt relation}

\begin{figure}
\centering
\includegraphics[width=0.49\textwidth]{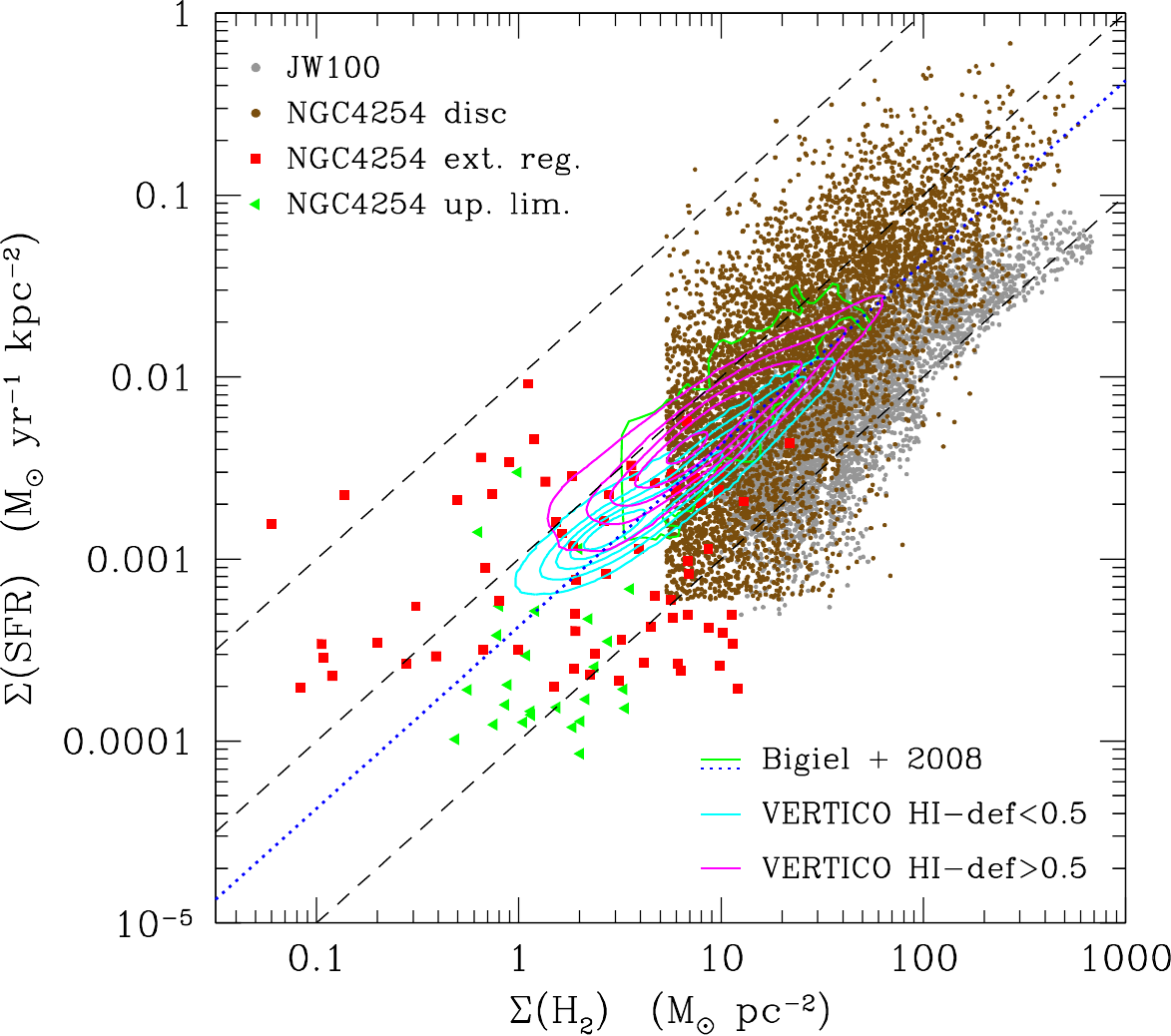}\\
\caption{Pixel-by-pixel molecular gas column density versus SFR surface density relation derived after smoothing the H$\alpha$ imaging data to the angular resolution of ALMA 
($\sim$ 2\arcsec, corresponding to $\sim$\,160\,pc). Only 3$\sigma$ and 2$\sigma$ detections are considered in the CO and H$\alpha$ line, respectively. Red filled squares represent pixels 
representing external star-forming regions of NGC 4254. Green triangles indicate 3$\sigma$ upper limits. These values are compared to those derived for the disc of NGC 4254 (brown dots) 
at a similar angular resolution, to the jellyfish galaxy JW100 (grey dots, Moretti et al. 2020b, at 1 kpc scale), to the THINGS sample of star-forming discs (Bigiel et al. 2008; green 
contour and dotted blue line, best fit; 750 pc scale), and to the VERTICO sample of \hi-rich (cyan contours) and \hi-poor (magenta contours) 
gaalxies in the Virgo cluster (Brown et al. 2023; 720 pc scale). The three parallel dashed black lines show lines of constant star formation efficiency, indicating the 
level of $\Sigma({\rm SFR})$ needed to consume 100\%, 10\%, and 1\%~ (from top to bottom) of the gas reservoir in 10$^8$ years.}
\label{schmidt}%
\end{figure}

The data gathered in this work can be used to see whether the detected regions follow the Schmidt diagram which relates the mean star formation surface density to the mean molecular gas 
column density (Fig. \ref{schmidt}) at 160 pc scale. To do this, we convolve the H$\alpha$ imaging data of the individual star-forming clouds and of the full galaxy to the resolution of 
ALMA and pixelise the convolved images on pixels of size 2\arcsec $\times$ 2\arcsec. We then measure molecular gas column densities and star formation rate (SFR) surface densities in 
each pixel whenever the signal-to-noise ratio $S/N$ is $S/N>3$ and $S/N>2$ in the CO and H$\alpha$ data, respectively. For comparison with other extreme or representative environments, Fig.
\ref{schmidt} also shows the Schmidt relation derived for the main disc of NGC 4254 using PHANGS data of similar angular resolution
and slightly lower sensitivity, the main relation for star-forming discs derived by Bigiel et al. (2008) for the THINGS sample where star formation is derived using H$\alpha$ data corrected 
for dust attenuation with 24~$\mu$m data, the VERTICO sample of Brown et al. (2021) for \hi-normal and \hi-deficient galaxies (Brown et al. 2023, Jimenez-Donaire 2023), and to the jellyfish 
galaxy JW100 observed by Moretti et al. (2020b). 
The MUSE data available for NGC 4254 and gathered during the PHANGS survey allowed us to measure the H$\alpha$ emission uncontaminated by the two [\nii] lines, and to measure pixel by pixel 
the dust attenuation with the Balmer decrement. We do not apply any dust attenuation nor [\nii] contamination correction for the external \hii\ regions, where the metallicity is expected to 
be low. The different sets of data have angular resolutions going from 160 pc (NGC 4254) to $\simeq$ 750 pc (VERTICO, THINGS) and 1 kpc (JW100).
Figure \ref{schmidt} shows that within the detected regions, gas is transformed into stars in a way comparable to that observed within the disc of NGC 4254 at similar angular scales, or on 
larger scales in the THINGS and in the VERTICO samples of Virgo cluster galaxies.

\section{Discussion}

The most interesting results of the analysis are summarised as follows:\\
1) A very low detection rate (4/49) of molecular gas associated with the star-forming regions located in the tail of stripped gas 
formed during the high-velocity encounter of NGC 4254 with another Virgo cluster member (gravitational perturbation).
These four star-forming complexes host nine GMCs, while another cloud was detected far from any star-forming region. \\
2) All of these GMCs are spatially resolved in the ALMA data, have sizes $R_{\rm GMC}$ $\gtrsim$ 120--250 pc, and have velocity dispersions
$\sigma_{v}(\rm CO)$ $\gtrsim$ 3 km s$^{-1}$. Compared to GMCs in the Milky Way, these newly detected clouds have, on average,
lower molecular gas column densities.\\
3) These molecular gas clouds generally belong to relatively large star-forming complexes, but they are not always physically associated with individual \hii\ regions, which are 
located at $\sim$ 100--200 pc in projected separation.\\
4) The clouds detected outside the stellar disc follow the same Schmidt relation measured at similar angular scales ($\simeq$ 160 pc) within the disc of NGC 4254, or at larger scales 
($\simeq$ 750 pc) in other local or Virgo cluster galaxies.\\

An interesting question is to understand how these results can be explained in a picture of galaxy evolution in a cluster environment, where different and competing physical 
processes are acting at the same time on the gas stripped from the galaxy disc after the interaction.

\subsection{CO-to-H$_2$ conversion factor}

The very low detection rate of molecular clouds could be partly explained by a very high CO-to-H$_2$ conversion factor
if the stripped gas is metal-poor (Boselli et al. 2002, Bolatto et al. 2013, Bisbas et al. 2025). Being a massive galaxy, NGC 4254 is metal rich (Groves et al. 2023).
The gas in the tail, being removed from the outer disc, is expected to have a lower metallicity than the one measured within the disc. The observed radial metallicity gradient of 
NGC 4254 measured with the PHANGS/MUSE data is relatively flat compared to other galaxies. It goes from 12+log(O/H) $\simeq$ 8.6 in the centre to 12+log(O/H) $\simeq$ 8.5 at 
$R/R_{\rm eff}$ $\simeq$ 4, for $R_{\rm eff}$ = 0.6\arcmin\ (Groves et al. 2023), where $R_{\rm eff}$ is the effective radius of the galaxy. We recall that the \hii\ regions 
analysed in this work are located at a distance from the galaxy nucleus ranging from $\simeq$ 2.3\arcmin\ for the closest up to $\simeq$ 7.1\arcmin\ for the farthest: 
$R/R_{\rm eff}$ $\simeq$ 3.8--11.8, or equivalently 0.6 $\lesssim$ $R/R_{\rm B25.5}$ $\lesssim$ 1.9, where $R_{\rm B25.5}$ is the isophotal diameter in the $B$ band as 
measured by Binggeli et al. (1985; $R_{\rm B25.5}$ = 3.8\arcmin). Thus, it is very likely that the metallicity here drops by another 0.1--0.2 dex, reaching 12+log(O/H) 
$\simeq$ 8.3--8.4 at $R/R_{\rm eff}$ $\simeq$ 10.  Given the limited metallicity gradient, and the known $X_{\rm CO}$-metallicity relation (Boselli et al. 2002; Sandsrtom et al. 
2013; Bolatto et al. 2013, Bisbas et al. 2025), we expect a variation of the conversion factor of only 10--20\%\ at most due to the decrease of metallicity, insufficient to 
explain the low detection rate.

The CO-to-H$_2$ conversion factor, however, could change for other reasons. Indeed, the physical condition of the gas in the tail might significantly differ from those 
observed in the disc of a massive spiral, where the dominant heating source is the interstellar radiation field
mainly produced by recently formed stars. In the tail of NGC 4254, star formation is very limited and sporadic, and the gas,
at least its diffuse component, can be heated after mixing with the surrounding hot intracluster medium (X-ray heating). $X_{\rm CO}$ is known to change with gas density and dust 
attenuation in a different way once subject to different ionising sources (stellar UV radiation, X-rays, cosmic rays, e.g. Wolfire et al. 1995, 2010; Kaufman et al. 1999; Shetty 
et al. 2011; Bolatto et al. 2013, Bisbas et al 2025) The signal-to-noise in the detected regions, however, is fairly high ($S/N$ $\gtrsim$ 6), and the column density sensitivity 
reached in the observations is low [$\Sigma\mathrm{(H_2)}$ $\simeq$ 1 M$_{\odot}$ pc$^{-2}$]. It
is thus surprising that none of the remaining regions, given their recent or ongoing star formation activity, has been 
detected even at a lower signal-to-noise level. The low detection rate suggests a genuine lack of molecular gas, a low gas temperature, or a CO-destruction mechanism that renders 
these clouds CO-dark, as discussed in Bisbas et al. (2025). The latter does not seem to result from inefficient CO self-shielding from FUV and cosmic-ray ionisation rate (see 
Papadopoulos et al. 2004; Bisbas et al. 2025) given that these two variables are low in these outer disc environments.

\subsection{GMC formation}

Simulations consistently indicate 
that the elongated tail of atomic gas was formed by a gravitational perturbation that occurred $\simeq$ 280--750 Myr ago (Vollmer et al. 2005; Duc \& Bournaud 2008). 
Thus, GMCs might have been removed during the gravitational interaction, which is able to affect all the different components of the perturbed system.
However, the age of these star-forming complexes ($\simeq$ 10--30 Myr) is significantly younger than the age of the perturbation ($\simeq$ 280--750 Myr), 
and it is thus plausible that the molecular clouds where these star-forming regions were born formed within the stripped material. This implies that the diffuse stripped gas, 
whenever the density was sufficiently high to self-shield it from the external hot ICM, cooled and collapsed to form GMCs. But several questions remain open. 
Indeed, it is still unclear why cooling was possible in the tail only $\lesssim$ 100 Myr ago (the age of the star-forming complexes), only close to the galaxy disc and not 
further out in the tail, and why the formation of GMCs is not possible any more even close to the galaxy.

The cooling time in spiral arms of normal, star-forming systems is generally short, $\simeq$10--20 Myr (Dobbs et al. 2008). The conditions of the gas in the stripped tail 
are significantly different from those encountered in the ISM of unperturbed galaxies. Cooling agents such as dust and carbon lines (Wolfire et al. 1995) might be less 
abundant than in the inner disc since most of the stripped gas is the one located in the outer disc which is metal-poor. The nature of the heating sources also changes given 
the lack of stars (UV-radiation, cosmic rays). The stripped gas, on the other hand, is embedded in a hot ICM which might contribute to the heating of the gas, although the 
densest regions should be shielded by the diffuse \hi\ gas component of the tail. Quantifying the typical gas cooling time under these conditions, where cooling is probably 
dominated by local instabilities, is thus very difficult. 
The GMCs where these star-forming regions have been created must have been formed recently. We recall that the galaxy is also suffering a ram pressure stripping event: the 
direction of the \hi\ gas tail and the observed compression of the \hi\ gas on the leading edge of the disc suggest that the galaxy is in its first infall into the cluster 
and that it is moving from the NNW periphery towards the cluster centre (Phookun et al. 1993). Because of this infalling trajectory, the external pressure on the gas disc 
is increasing with time, and the densest gas regions of the disc, where cooling becomes more and more efficient, might have been stripped only recently.

The formation of molecular clouds in the ISM of galaxies is due to several mechanisms, as summarised in K\"ortgen et al. (2017) and Beuther et al. (2020). 
It is related to gas compression due to converging flows driven by stellar feedback and turbulence, agglomeration of smaller clouds, gravitational and magneto-gravitational 
instabilities, and instabilities due to differential buoyancy. Converging flows can be formed during the expansion of \hii\ regions and by supernova winds.
The molecular clouds are formed whenever the pressure of the flow exceeds that of the surrounding medium. Although this 
mechanism generally produces small molecular clouds of mass $\sim$ 10$^4$ M$_{\odot}$, the agglomeration of several clouds in an overdensity region such as the prominent 
extended arm formed during the gravitational perturbation that affected NGC 4254
is possible. The presence of magnetic fields can help confine the matter and increase the density of the diffuse gas component, thus favouring the formation of molecular 
clouds (Tonnesen \& Stone 2014; Ruszkowski et al. 2014). Furthermore, the transition between \hi\ and H$_2$ is shifted to higher
column densities in low metallicity, low dust attenuation environments as those expected in the tail of stripped material. 
This is due to the fact that H$_2$ molecules are principally formed on dust grains (Hollenbach \& Salpeter 1971; Bolatto et al. 2011; Wong et al. 2013; Wakelam et al. 2017; Bisbas et al. 2025).

Using the data gathered during this work, we can derive the mean molecular gas column density fraction, defined as $R_{\mathrm{mol}}$ =  $\frac{\Sigma(\mathrm{H_2})}{\Sigma(\mathrm{\ion{H}{i}})}$, 
in the gas located in the CO detected regions. These are 1 $\lesssim$ $R_{\mathrm{mol}}$ $\lesssim$ 20 when measured using the {\hi} moment-0 map at 27\arcsec\ angular resolution. 
These values, which are probably overestimated because of the limited resolution of the {\hi} data, can be compared to the one derived using the relation between the ratio of the 
H$_2$-to-\hi\ gas column density and the external pressure $P_{\rm ext}$ derived by Blitz \& Rosolowsky (2006) for the diffuse gas in the tail:

\begin{equation}
{R_{\mathrm{mol}} = \frac{\Sigma(\mathrm{H_2})}{\Sigma(\mathrm{\ion{H}{i}})} = \left[\frac{P_{\mathrm{ext}}/k}{(4.6\pm 0.6)\times 10^4}\right]^{0.92\pm 0.07}}
\label{rmol}.
\end{equation}

\noindent
In the diffuse tail, where \shi\ $\simeq$ 0.02 M$_{\odot}$ pc$^{-2}$, the external pressure derived assuming Eq. 5 is $P_{\rm ext}$ $\simeq$ 1.5$\times$ 10$^{-15}$ g cm$^{-1}$ s$^{-2}$. 
In the densest regions the \hi\ column density reaches \shi\ $\simeq$ 0.3 M$_{\odot}$ pc$^{-2}$ (see Fig. \ref{MeerKAT_full}), and thus 
$P_{\rm ext}$ $\simeq$ 2.25$\times$ 10$^{-14}$ g cm$^{-1}$ s$^{-2}$. By extrapolating Eq.~\ref{rmol} (which has been derived for pressures
$P$ $\gtrsim$ 10$^{-12}$ g cm$^{-1}$ s$^{-2}$), we can estimate the molecular gas column density to be $\Sigma\mathrm{(H_2)}$ $\simeq$ 0.5-2 $\times$ 10$^{-4}$ M$_{\odot}$ pc$^{-2}$. 
We can estimate whether the gas can collapse at these low column densities. 
The atomic gas in the tail is embedded in a hot ($T$ $\simeq$ 10$^7$ K, e.g. Simionescu et al. 2017) ICM emitting in X-rays. As discussed in Burkhart \&  Loeb (2016), the stripped gas 
is supported against collapse by turbulence, thermal pressure, and rotation. It can become Toomre unstable and collapse into GMCs if it cools efficiently.  This can happen on dust grains, 
as previously mentioned, or via molecular line cooling (Burkhart \& Loeb 2016). If the column density of the gas does not exceed a critical density sufficient to self-shield the gas cloud, 
the formed molecular hydrogen is efficiently photodissociated by the harsh X-rays radiation. Taylor \& Webster (2005), using theoretical arguments, indicate that the critical density of 
the gas embedded in a hot X-ray emitting medium necessary to make molecular hydrogen formation possible is $\Sigma_{\rm crit}$ $\gtrsim$ 4 M$_{\odot}$ pc$^{-2}$, a column density well 
above the \hi\ column density measured with MeerKAT within the extended tail 
or the one derived above for the molecular gas phase. As expected, this critical density is exceeded in all the CO detected regions (see Fig. \ref{density}).
These simple considerations suggest that the formation of GMCs in the stripped gas, if possible close to the galaxy disc where the density of the stripped medium can become sufficiently 
high because of cloud-cloud collision due to the turbulent motion of the gas induced by the external perturbation, it is now very unlikely elsewhere in the tail where the density of the 
gas is significantly lower. Nine of the ten detected molecular clouds are indeed located in relatively high \hi\ density regions, have a low molecular gas density, and might thus represent 
the first phase of gas collapse. This would also explain their limited star-forming activity. They are thus quite similar to GMCs in the Milky Way, which are rarely associated with 
star-forming regions, suggesting that these GMCs are in a precursor phase of star formation (Miville-Desch\^enes et al. 2017).

\subsection{GMC destruction}

\begin{figure}
\centering
\includegraphics[width=0.48\textwidth]{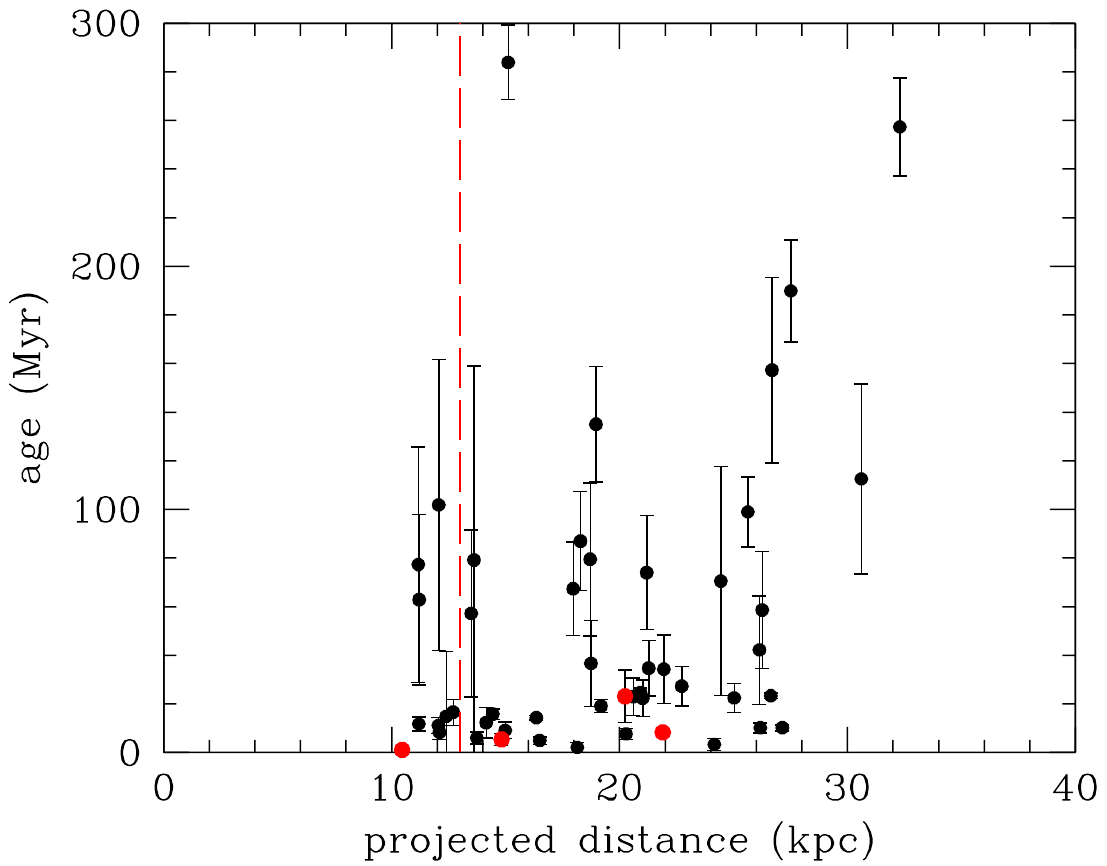}\\
\caption{Relation between the age of the observed \hii\ regions derived with SED fitting 
in Boselli et al. (2018b) and their projected distance from the nucleus of NGC 4254 (in kpc). Red dots are for the regions detected in CO. 
The long-dashed red vertical line indicates the 23.5 mag arcsec$^{-2}$ $i$-band isophotal radius of the galaxy.}
\label{distance}%
\end{figure}

\begin{figure}
\centering
\includegraphics[width=0.48\textwidth]{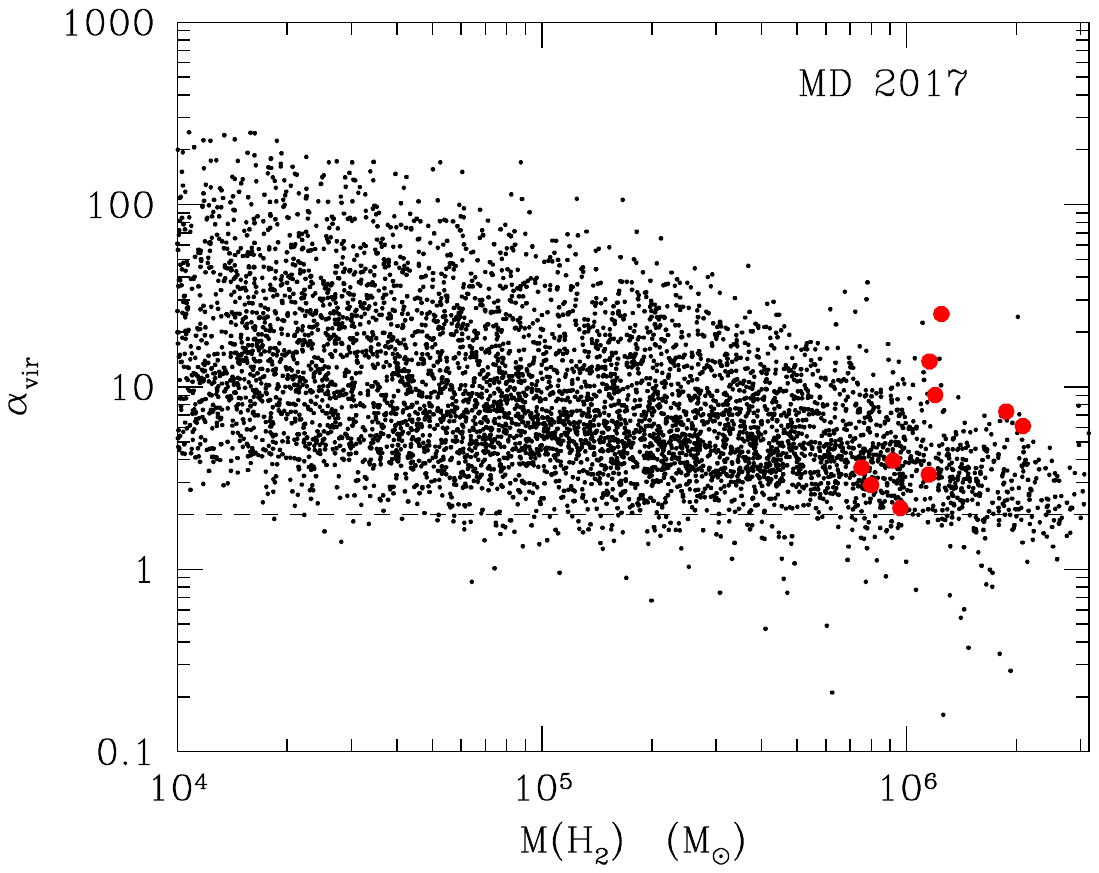}\\
\caption{Relationship between the virial ratio $\alpha_{\rm vir}$ and the molecular gas mass of the GMCs in the outskirts of NGC 4254 (red dots) and in the Milky Way (black dots, 
from Miville-Desch\^enes et al. 2017). The horizontal dashed black line gives the critical limit of $\alpha_{\rm vir}$ = 2, above which GMCs are expected to dissolve.}
\label{avirial}%
\end{figure}

Most of the observed star-forming complexes located outside the disc of the galaxy do not contain a significant amount of molecular gas, as also observed at small scales in star-forming discs. 
The four external star-forming regions detected in CO have stellar populations with typical ages 1 $\lesssim$ age $\lesssim$ 23 Myr, and are among the youngest one discovered by Boselli et al. 
(2108b), which rather span a much wider range in age (1 $\lesssim$ age $\lesssim$ 200 Myr; see Fig. \ref{distance}). Two of them also have the highest H$\alpha$/NUV flux ratio, consistent with
a young age, and H$\alpha$ surface brightness \footnote{Only one of the detected regions have an accurate estimate of the H$\alpha$ diameter derived using \textsc{HIIphot} 
(see Boselli et al. 2018b). The surface brightness estimated here is defined as the ratio of the H$\alpha$ flux divided by the surface of the aperture used to extract the 
flux and is thus only a rough estimate of the surface brightness that is available for all the external \hii\ regions of NGC 4254.} among the 60 external \hii\ regions. This evidence might 
suggest an age effect and could be explained 
if the GMCs are destroyed by the first generation of supernovae
($\sim$ 10 Myr; notice that a similar age effect is seen in ESO 137-001, Waldron et al. 2023). To test this scenario, we calculated the virial ratio $\alpha_{\rm vir}$ for a spherical 
gas cloud as done by Jachym et al. (2019) using the relation (Bertoldi \& McKee 1992)

\begin{equation}
{\alpha_{\mathrm{vir}} = \frac{5\sigma^2R}{GM}}
,\end{equation}

\noindent
where $G$ is the gravitational constant, 
$\sigma$ is the velocity dispersion of the cloud, and $M$ and $R$ are its mass and radius. If $\alpha_{\rm vir}$ is subcritical ($\alpha_{\rm vir}$ $>$ 2
for a non-magnetised cloud, Kauffmann et al. 2013),
the cloud will dissolve with time. If we take the values of $\sigma$ and $M$ derived from the ALMA observations 
and we assume that the molecular cloud complexes are formed by an individual GMC with mass dominated by molecular gas, $\alpha_{\rm vir}$ $\gtrsim$ 2 for all the detected 
clouds (see Fig. \ref{avirial}). 
Under these assumptions, which depend on the limited angular resolution of the ALMA data, it is expected that all the clouds will dissolve with time.

A similar conclusion can be reached considering the pressure exerted by the same clouds on its surrounding medium.
We estimated the external pressure exerted by the \hi\ gas tail $P_{\ion{H}{i}}$ on the GMCs following the Burkhart \& Loeb (2016) recipe:

\begin{equation}
    {P_{\ion{H}{i}} = \frac{1}{2}\sigma_{\ion{H}{i}}^2\frac{M({\rm \ion{H}{i}})}{f_{\ion{H}{i}}\pi r^2 h}}
,\end{equation}

\noindent
where $\sigma_{\rm \ion{H}{i}}$ is the velocity dispersion including the turbulent and thermal components of the diffuse gas, \mhi\ its total mass, $\pi r^2 h$ its volume whenever 
the tail is approximated with a cylinder of radius $r$ and height $h$, and $f_{\rm \ion{H}{i}}$ a filling factor for the \hi\ gas within the tail. Assuming $\sigma_{\rm \ion{H}{i}}$ $\simeq$ 
10 km s$^{-1}$ as roughly derived from the moment2 map of the \hi\ gas in the tail and consistently with typical values in the ISM of local galaxies, \mhi\ $\simeq 3.4\times10^8$ M$_{\odot}$, 
$h$ $\simeq$ 240 kpc, and $r$ $\simeq$ 10 kpc, we obtain $P_{\rm \ion{H}{i}}$ $\simeq$ 1.5 $\times$ 10$^{-15}$ g cm$^{-1}$ s$^{-2}$ for $f_{\rm \ion{H}{i}}$ =0.1. The pressure exerted by the 
GMCs on the surrounding medium ($P_{\rm H_2}$ $\gtrsim$ 1.5 $\times$ 10$^{-13}$ g cm$^{-1}$ s$^{-2}$ derived assuming spherical clouds, $f_{\rm H_2}$=1, and the parameters given in Table 
\ref{TabCO}) is significantly higher, and will naturally make the clouds dissolve with time.

The kinetic energy injected by the stellar winds, and in particular those of supernovae, 
is sufficient to dissolve the GMCs, which are not kept bound by the gravitational potential well of the disc, by the external pressure of the ISM, and by that of the expanding shells of the 
nearby \hii\ regions (self-propagating star formation). 
We can estimate the expected number of supernovae formed within the four star-forming complexes detected in CO from their measured SFR (3 $\times$ 10$^{-4}$ $\lesssim$ SFR $\lesssim$ 10$^{-2}$  
M$_{\odot}$ yr$^{-1}$, Boselli et al. 2018b).\footnote{These are the star formations measured within the whole star-forming complexes identified in Fig. \ref{ALMA_pallini} with red circles.} 
This number is $\simeq$ 20--200 on the lifetime of the star-forming complex. They could provide an energy of 2 $\times$ 10$^{52}$ $\lesssim$ $E_{\rm tot}$ $\lesssim$ 2 $\times$ 10$^{53}$ ergs.
Considering that only 1--10\% of this energy can be transferred to the gas as kinetic energy, we could estimate the 
mass of gas that can be expelled by a wind of velocity $v_{\rm wind}$ $\sim$ 10 km s$^{-1}$ as (Boselli et al. 2016)
\begin{equation}
{M_{\mathrm{out}} = \frac{2E_{\mathrm{kin}}}{v_{\mathrm{wind}}^2}}
.\end{equation}

\noindent 
We obtained $M_{\rm out}$ $\simeq$ 10$^6$--10$^7$ M$_{\odot}$, a value comparable to the mass of molecular gas observed in the detected GMCs [$M\mathrm{(H_2)}$ $\simeq$ 10$^6$ M$_{\odot}$].

The typical lifetime of a cloud embedded in a diffuse medium is determined by two competing mechanisms, the
external accretion rate of gas feeding the giant molecular cloud and star formation, and the gas ejected by stellar feedback (Jeffreson et al. 2024).  
Gas accretion is driven by large scale flows, while gas ejection depends on stellar winds.
The lifetimes of GMCs have been derived using models and simulations, and ranges between 1 and 100 Myr within the ISM 
of disc galaxies (e.g. Goldbaum et al. 2011; Inutsuka et al. 2015; 
Burkert 2017; Kobayashi et al. 2017; Jeffreson et al. 2024). Goldbaum et al. (2011) derived the typical lifetime of GMCs embedded
in a diffuse \hi\ medium of column density \shi\ = 8 M$_{\odot}$ pc$^{-2}$ to be $t_{\rm life}$ $\simeq$ 26 Myr and  
$t_{\rm life}$ $\simeq$ 52 Myr when \shi\ = 16 M$_{\odot}$ pc$^{-2}$. The mean \hi\ column density around the detected GMCs is \shi\ = 4--8 M$_{\odot}$ pc$^{-2}$, and rapidly drops below 
this value moving towards the tail.
Although the relation between $t_{\rm life}$ and \shi\ is non-linear, it is conceivable that in these regions $t_{\rm life}$ $\lesssim$ 20Myr, but significantly shorter in the outer regions 
where the density of the diffuse gas is significantly lower [\shi\ = 0.02--0.3 M$_{\odot}$ pc$^{-2}$].

What looks different from star-forming complexes within galaxy discs is the fact that these star-forming regions
are not located close to each other, but at a rather large distance (kpc scales). This might suggest that self-propagating star formation
does not occur any more in the tail of stripped gas where spiraling density waves are absent and where the density of 
the stripped medium is significantly lower than within the stellar disc. 
The fact that these star-forming regions have been observed in the tail only close to the galaxy disc ($\lesssim$ 20 kpc) and not further out
is probably related to the observed gradient in the diffuse gas column density.
The decrease in gas column density might be due to the fact that with time the cold ISM stripped from the galaxy warms up once mixed with the surrounding hot 
ICM emitting in X-rays, as observed in other Virgo cluster galaxies (NGC 4569, Boselli et al. 2016, Sun et al. 2025).
Cooling becomes more unlikely with increasing time and thus with distance from the galaxy disc. 
Furthermore, if a GMC gets in touch with the hot surrounding medium (a few 10$^7$ K), it will evaporate on
relatively short timescales [$\tau_{\rm evap}^{\rm sphere}$ $\simeq$ 20 Myr for a cloud of $M\mathrm{(H_2)}$ = 10$^6$ 
M$_{\odot}$, radius $r$ = 100 pc, and $T_{\rm ICM}$ = 3 $\times$ 10$^7$ K as derived using the prescription of Cowie \& McKee 1977, although it is known that in low-density environments 
within a hot plasma heat conduction makes the evaporation time even more efficient (Sander \& Hensler 2021, 2023)].

\subsection{Consistency with simulations}

To compare the observed properties with theoretical expectations, we have performed a suite of idealised numerical simulations of a star-forming cloud exposed to the ram-pressure of 
the intracluster medium. We have considered two different setups, one employing the code \textsc{ramses} and the model by Calura et al. (2020) for the `SECCO' cloud, and one with 
\textsc{gizmo}, based on a star-by-star numerical model (Lupi et al. in preparation) inspired by that in Lahén et al. (2020). The goal of these simulations is to assess the survival, 
fragmentation, and possible star-formation properties of the stripped material as it interacts with the hot ICM and to assess how robust the evolution is to changes in the numerical 
technique employed and the physical conditions assumed. 

Both sets of simulations consider a uniform spherical gas cloud of mass $M_{\mathrm{cloud}} \simeq 10^6\,\rm{M}_\odot$ and radius $R_{\mathrm{cloud}} = 100\,\mathrm{pc}$ (corresponding to
an initial density of $\rho_{\rm cloud}$ $\sim 10^{-23} \mathrm{g~cm^{-3}}$), embedded in a hot ICM wind. The temperature is initialised to ensure approximate thermal pressure 
equilibrium with the ICM. The ICM wind is characterised by a velocity $v_{\mathrm{ICM}}$= 20 km s$^{-1}$ (slow wind, SW)
and $v_{\mathrm{ICM}}$= 2000 km s$^{-1}$ (fast wind, FW), density $\rho_{\mathrm{ICM}} = 10^{-28}\,\mathrm{g\,cm^{-3}}$, and temperature $T_{\mathrm{ICM}}$ = 10$^7$ K.
The detailed description of the simulations is given in Appendix~\ref{app:simulations}. 

\begin{figure}
\centering
\includegraphics[width=0.24\textwidth]{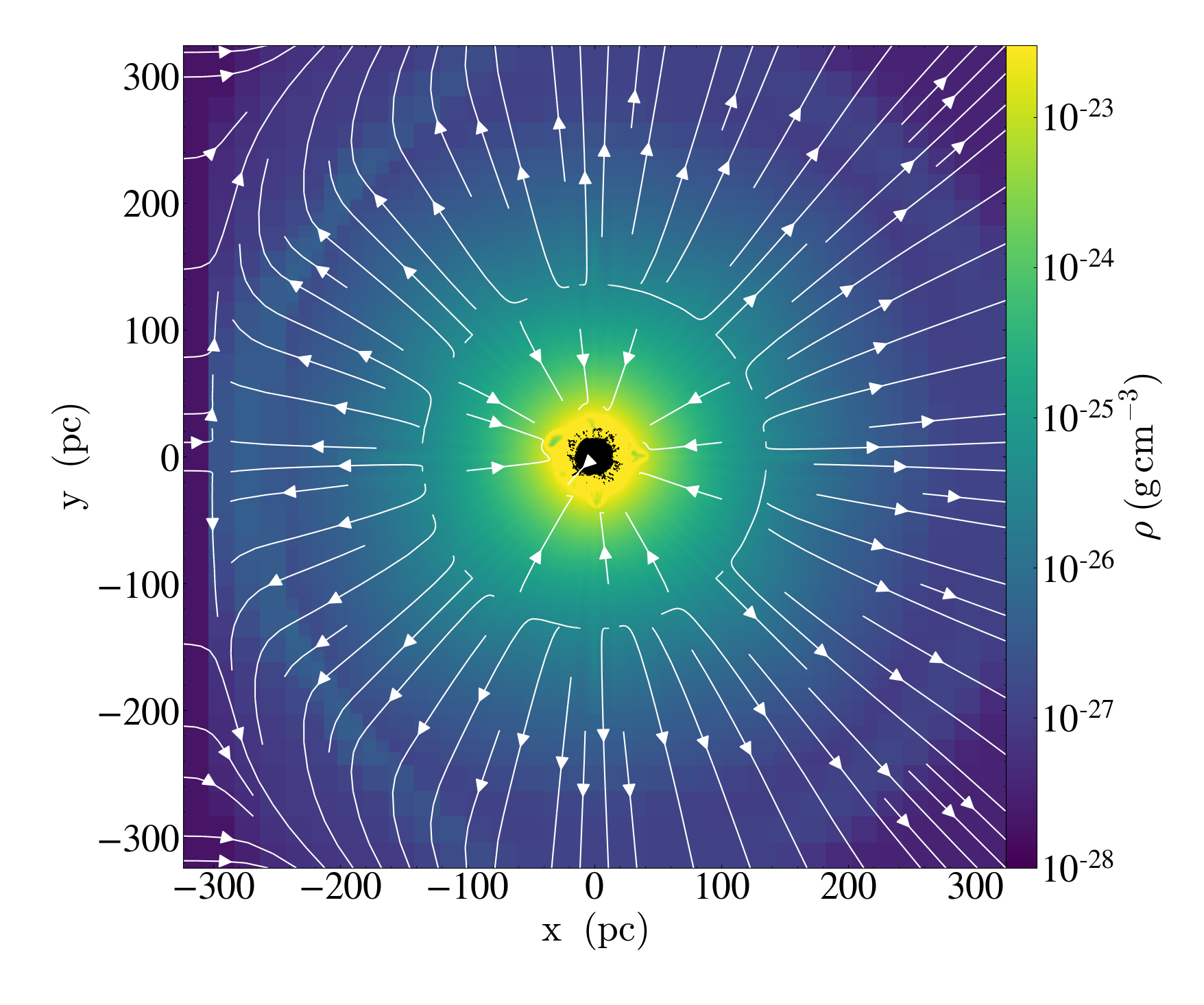}
\includegraphics[width=0.24\textwidth]{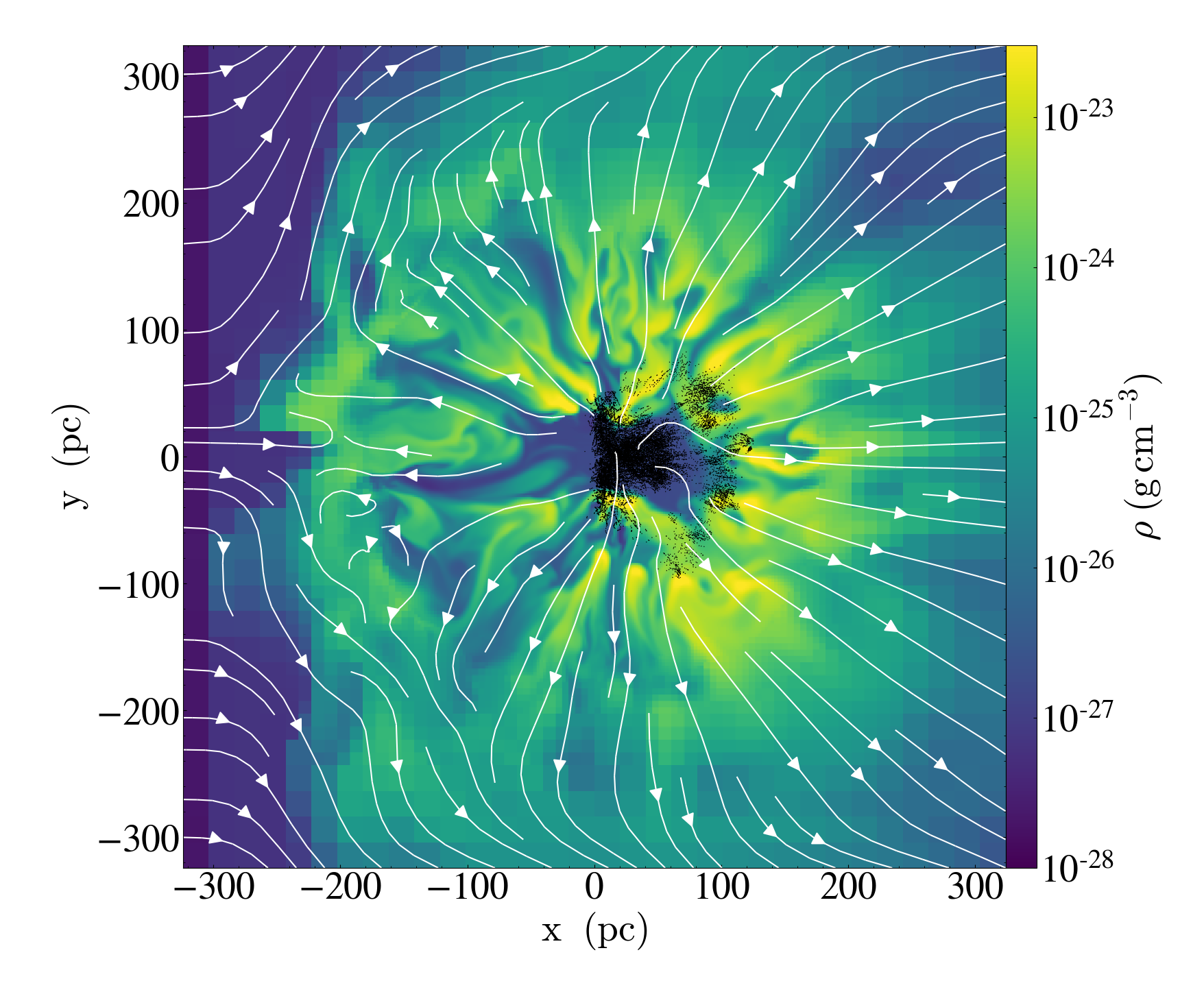}
\\
\includegraphics[width=0.24\textwidth]{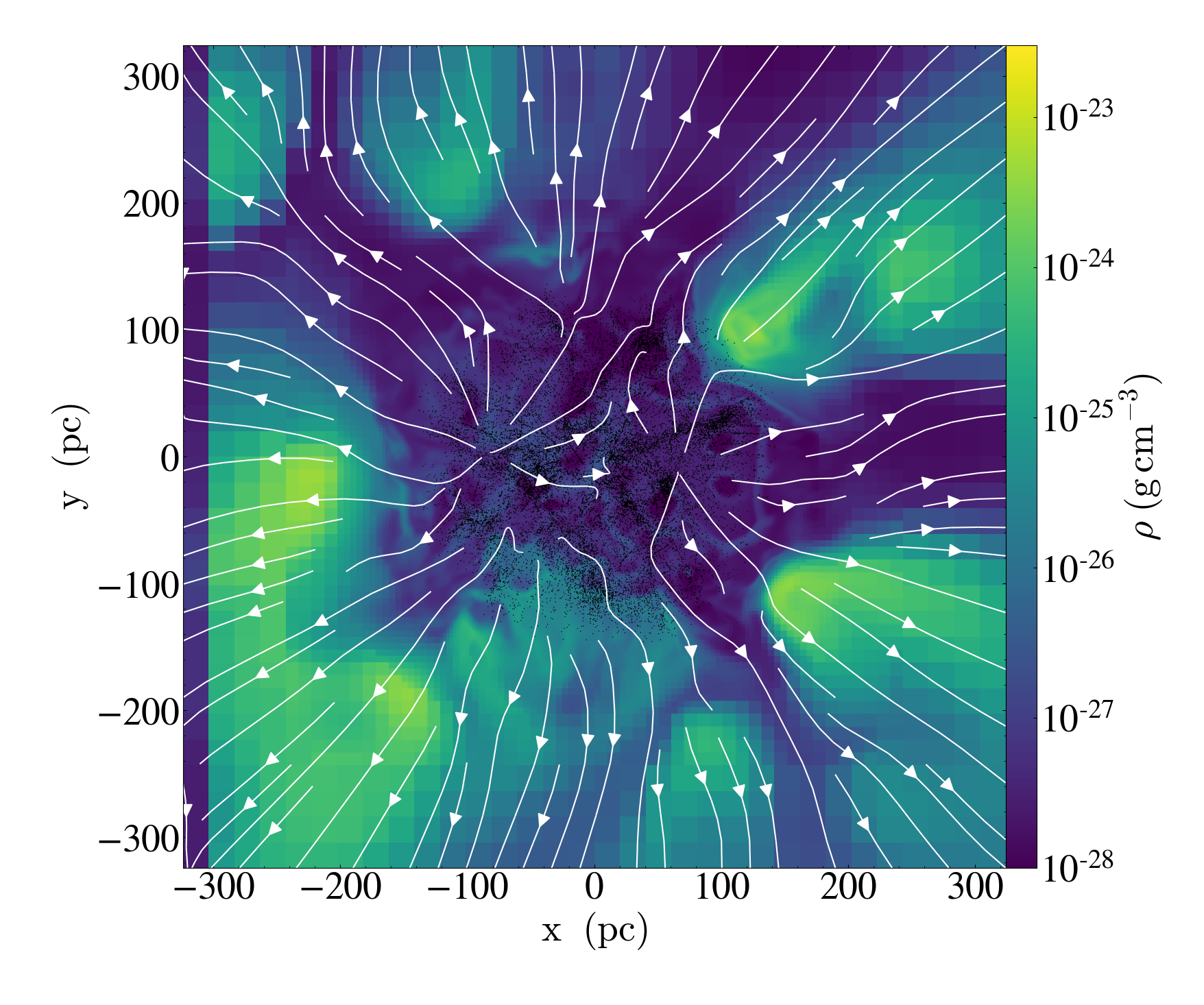}
\includegraphics[width=0.24\textwidth]{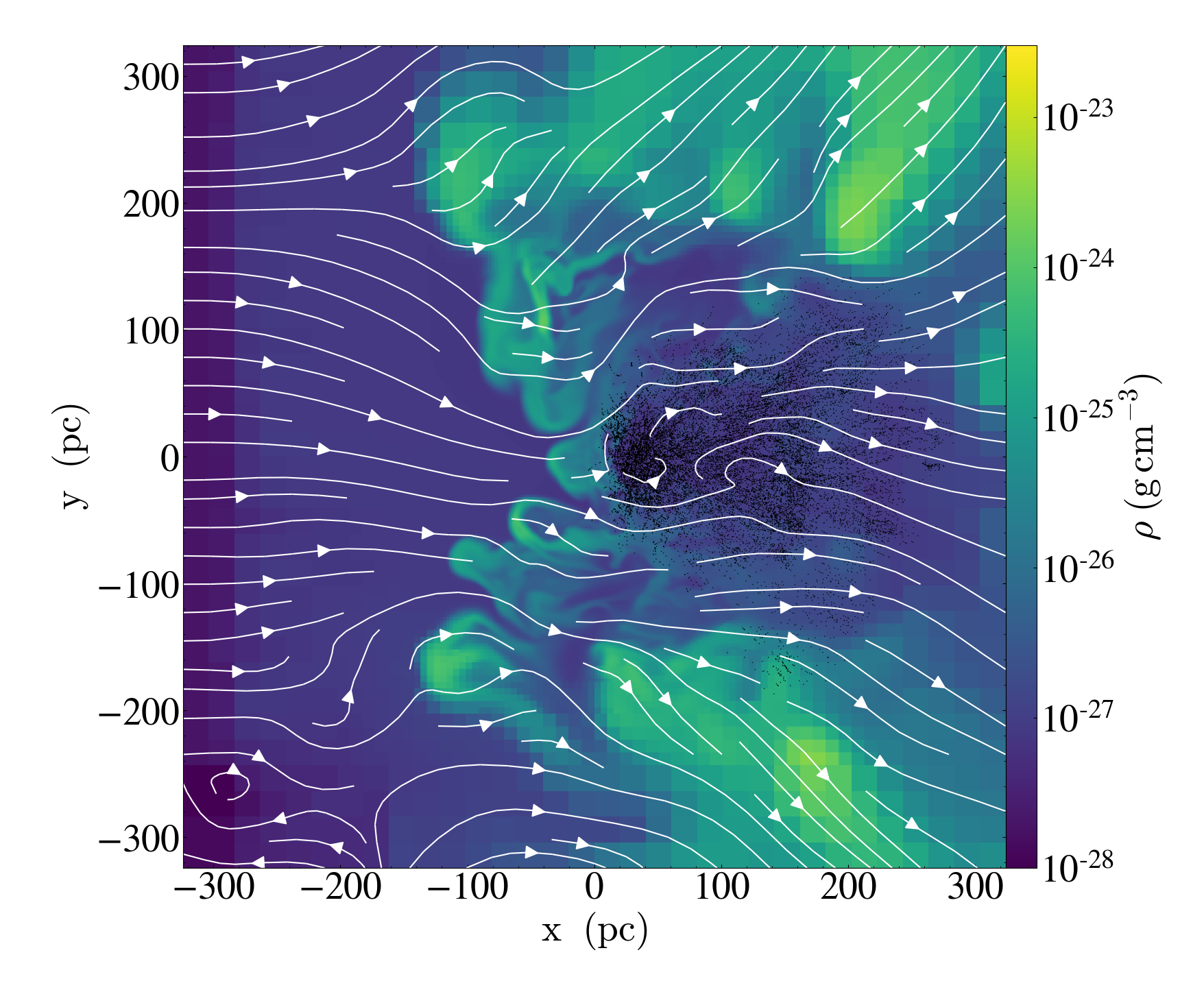}
\\
\caption{Time evolution of gas density and star formation in the slow wind (SW; \textit{left panels}) and fast wind (FW; \textit{right panels}) \textsc{ramses} models.
  \textit{Top left}: Slice density map through the central plane of the SW model at 16 Myr. The colour scale and streamlines are as in Fig.~\ref{fig_ic},
  whereas the black points indicate the stars.\textit{ Bottom left}: Same as in the top left but showing the model results at 25.6 Myr. 
  \textit{Top right and bottom right}: Density maps of the FW model at 16 Myr and 26 Myr, respectively, with the same colour-scale and symbols as above. }
\label{fig_evo}
\end{figure}

\begin{figure}
  \includegraphics[width=9.cm,height=6.cm]{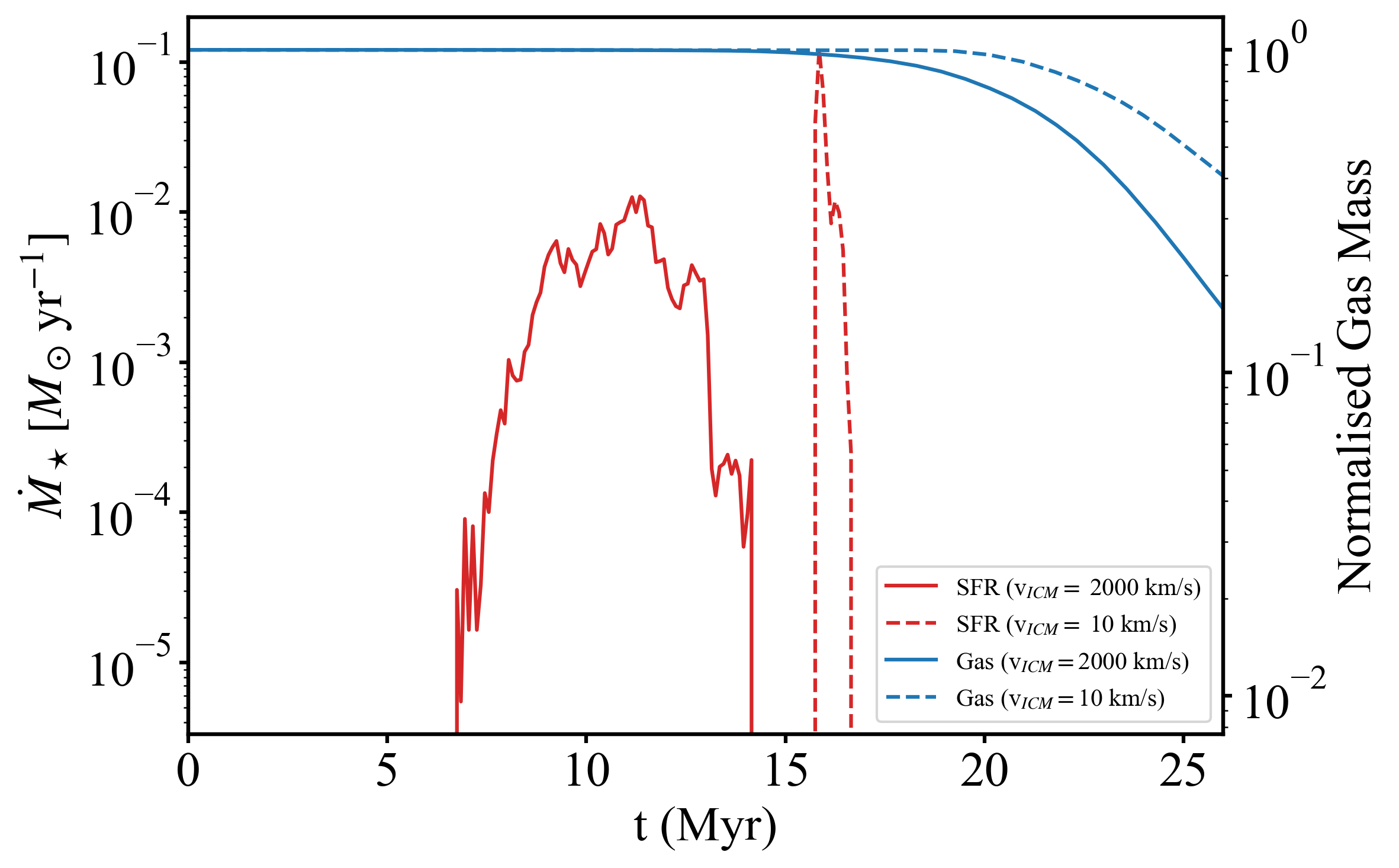}
\caption{Star-formation history and gas evolution in the two wind tunnel \textsc{ramses} simulations. 
  Red lines show the SFR evolution for the FW (solid) and SW (dashed) models.
  Blue lines indicate the gas mass fraction inside the domain, normalised to the initial value, for the FW (solid) and SW (dashed) models. For comparison, 
  the typical SFR observed in the studied regions is SFR $\simeq$ 10$^{-3}$ M$_{\odot}$ yr$^{-1}$.}
\label{fig_sfh}
\end{figure}

\begin{figure}
\centering
\includegraphics[width=0.95\columnwidth]{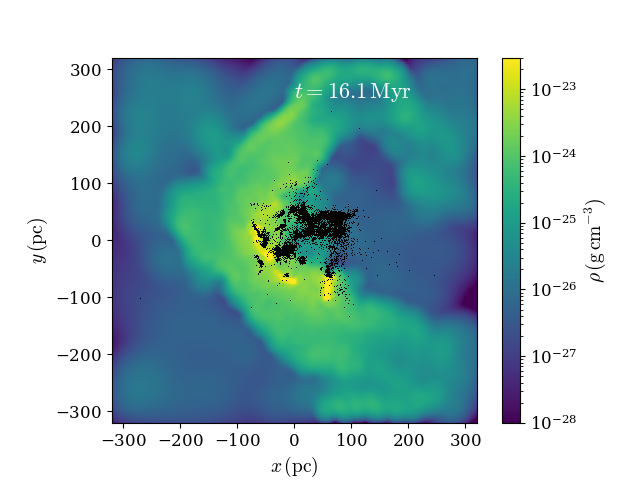}
\caption{Slice through the centre of the gas cloud from the \textsc{gizmo} HDRest run (with a static ICM) after 16 Myr. The black dots represent single stars formed during the simulation.}
\label{fig:gizmo_lt_rest}%
\end{figure}

\begin{figure*}
\centering
\includegraphics[width=0.99\columnwidth,trim=0 1.5cm 0 2cm,clip]{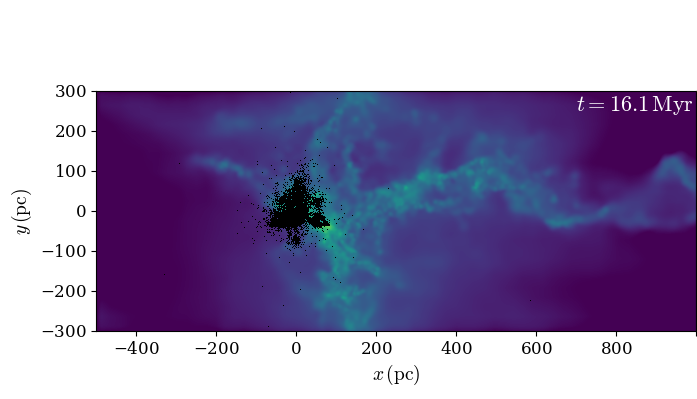}
\includegraphics[width=\columnwidth,trim=0 1.5cm 0 2cm,clip]{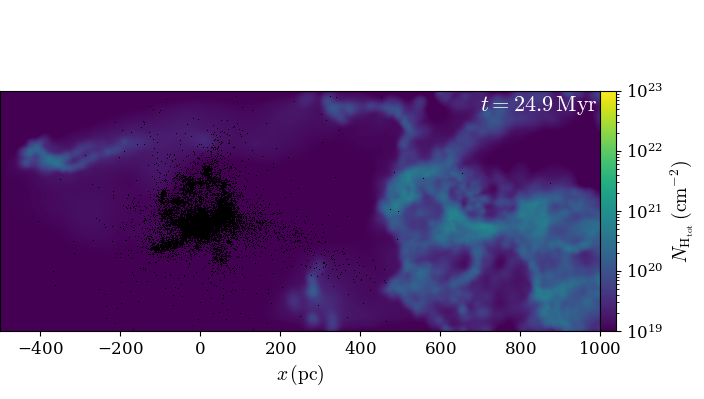}\\
\includegraphics[width=0.99\columnwidth,trim=0 0 0 2cm,clip]{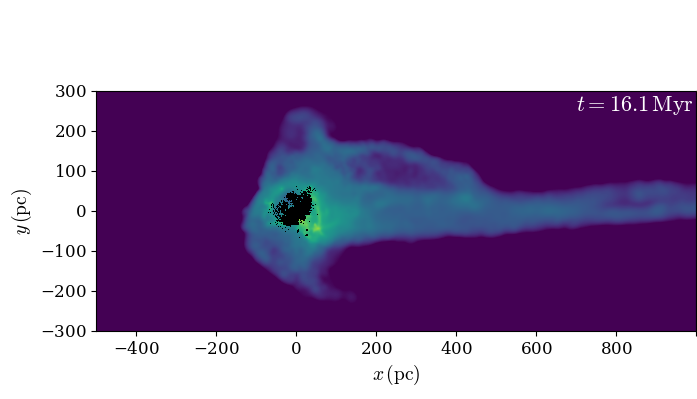}
\includegraphics[width=\columnwidth,trim=0 0 0 2cm,clip]{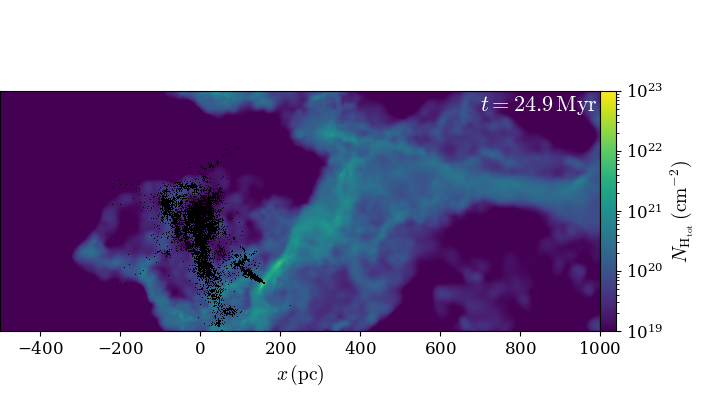}\\
\caption{Column density maps of the gas clouds from the fast wind \textsc{gizmo} simulations after 16 (\textit{left panels}) and 25 (\textit{right panels}) Myr. 
The \textit{upper and lower rows} are for simulations without and with ideal magnetohydrodynamics (HD, MHD), respectively. The wind flows from left to right. 
The black dots represent single stars formed during the simulation.}
\label{fig:gizmo_lt}%
\end{figure*}

\begin{figure}
\centering   
\includegraphics[width=\columnwidth]{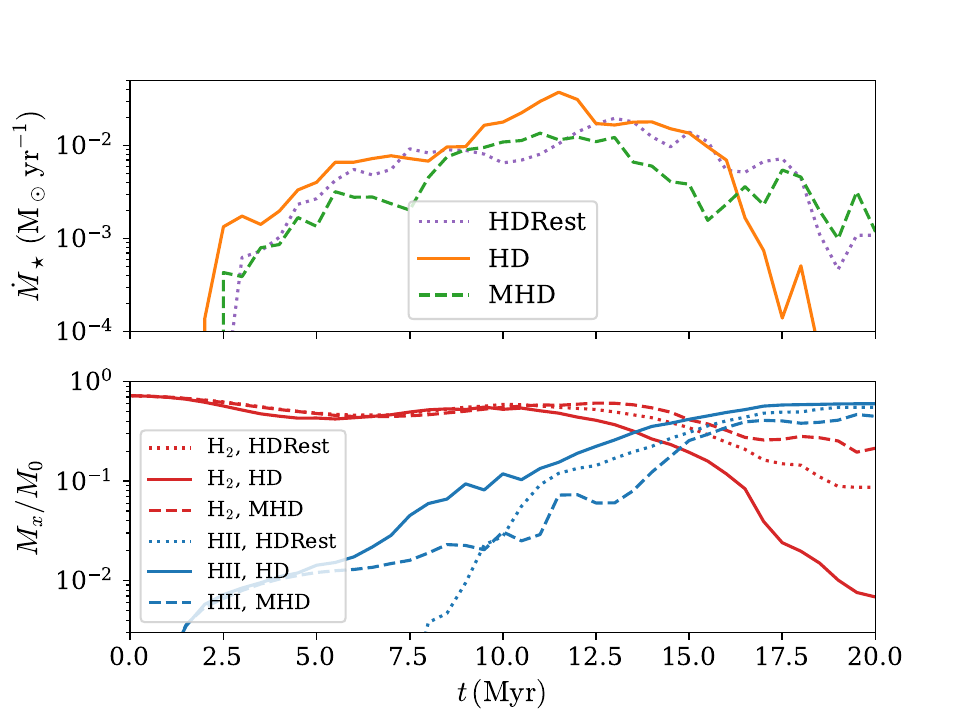}\\
\caption{\textit{Top}: Star formation history in the \textsc{gizmo} simulations. The HDRest run (static ICM) is shown as a dotted purple line, whereas the fast wind ones 
are shown as a solid orange line (HD) and a dashed green line (MHD). \textit{Bottom}: Gas mass evolution in the three \textsc{gizmo} runs, normalised to the initial cloud mass. 
We show molecular gas in red and the ionised gas in blue, respectively, following the line style of the top panel. }
\label{fig:gizmo_sfr}%
\end{figure}

By comparing the survival time-scales inferred from the observations with our numerical simulations (see Figs. \ref{fig_evo} and \ref{fig_sfh} for the \textsc{ramses} simulations and 
Figs. \ref{fig:gizmo_lt_rest}, \ref{fig:gizmo_lt}, and \ref{fig:gizmo_sfr} for the \textsc{gizmo} simulations), we tentatively explain the origin of the star-forming clouds. The two 
simulations differ both in numerical technique and initial setup. This produces some differences in the star-formation history of the cloud, which are evident in the slow wind/static 
ICM case. In particular, the \textsc{ramses} simulation needs more than 10~Myr to allow the gas to contract enough to produce the first stars. As soon as the star formation density 
is reached, however, many stars form at the same time, producing enough feedback to shut-off star formation a few million years later. In the \textsc{gizmo} run, instead, the 
initial turbulence accelerates the formation of self-gravitating clumps, producing a prolonged star formation event which lasts 15--20~Myr before stellar feedback completely 
destroys the cloud. The differences between the two runs become less severe when we include the effect of ram pressure due to the cloud motion within the ICM. Indeed, because 
of the initial compression by the wind, gas collapse is accelerated, leading to an earlier formation of stars in the \textsc{ramses} run, whereas in the \textsc{gizmo} case, 
because of turbulence, only a mild enhancement is produced. Over time, ram pressure starts depleting the cloud in both runs, resulting in the complete shut-off of star formation 
in a slightly shorter time, but still around 15--20~Myr. The presence of magnetic field in one of the \textsc{gizmo} runs slightly prolongs star formation compared to the hydrodynamic 
case, but still of the order of the time resulting from the other runs. 

In general, these results seem in line with the values inferred from observations, and seem to be consistent with a picture in which a gas cloud is stripped from the galactic 
disc and becomes part of the tail. Within the tail, the cloud is partially shielded from the ICM, and forms stars over time-scales comparable to those of isolated molecular clouds. 
Indeed, the critical density is reached only in isolated regions, preventing self-propagating star formation from taking place. Consistently with this picture, clouds farther away 
from the galaxy should not exhibit any evidence of recent star formation, due to the longer time needed to move by tens of kpc, which results in young stars dying and the star-forming 
gas being completely depleted. Any evidence of young stars in such systems would point instead towards a different evolutionary path, in which the cloud might form out of the moderately 
dense gas in the tail via thermal instabilities, which allow the gas to contract and become self-gravitating.

\subsection{GMC and star formation in the tails of other galaxies}

Interesting is understanding how this evolutionary picture can be extended to explain the large variety of properties observed in other perturbed cluster systems.
The simple calculations done in the previous sections suggest two possible different scenarios for the gas stripped during hydrodynamic interactions (e.g. ram pressure 
stripping) and gravitational perturbations. In gravitational perturbations, such as the one observed in NGC 4254, all gas phases can be removed during the perturbation. If 
the perturbation is sufficiently strong, the dense molecular gas located within GMCs can be removed along with all the other baryonic components. 
Hydrodynamic simulations of a cold gas cloud within a turbulent multi-phase medium suggest that the cloud can survive and continuously grow in dimension whenever its original 
size is larger than a critical size which depends on the properties of the surrounding medium. The growth in mass, however, occurs via the formation of several smaller clouds 
(Gronke et al. 2022; see also Scannapieco \& Bruggen 2015). The stripped GMCs can thus fragment into smaller clouds where star formation occurs. Other molecular clouds can be 
formed within the tail of stripped material whenever the density of the gas is sufficiently high to allow efficient cooling, and the clouds evolve similarly to those removed 
during the interaction.  
Once the density of the gas in the tail drops below a critical value and the temperature of the surrounding ICM is too high, cooling becomes not efficient and the diffuse principally 
atomic gas evaporates, becoming first ionised, than hot gas emitting in X-rays.

Ram pressure stripping is an hydrodynamic process acting only on the gaseous component. Simulations indicate that it mostly removes the diffuse gas component, which is principally atomic 
gas. GMCs can be directly stripped, although in a less efficient way than the diffuse gas given that their gas density is a factor of $\sim$ 100 higher. Furthermore, gas compression can 
produce density inhomogeneities in the ISM, reaching the conditions for GMC formation. It is thus likely that star formation can occur in ram pressure stripped tails only within these 
high density regions (Steyrleithner et al. 2020). This has been observed in several local ram pressure stripped candidates, such as ESO 137-001 and D100, where clumpy molecular gas 
complexes have been observed within the extended tail of stripped material (Jachym et al. 2014, 2017, 2019). The fragmentation of the growing GMC in the tail induces the formation of 
a shower of compact molecular gas clouds where star formation can occur, as observed in the ram pressure stripped galaxy ESO137-001 (Jachym et al. 2014, 2019). There are a few examples 
where the tail contains a large amount of molecular gas. D100 in the Coma cluster, for instance, has more molecular gas in the tail than on the stellar disc (Jachym et al. 2017). D100 
is a dwarf system ($M_{\rm star}$ = 2.1 $\times$ 10$^9$ M$_{\odot}$, Jachym et al. 2017) orbiting close to the core of a very massive cluster (Coma, 
$M_{\rm halo}$ $\simeq$ 8.9 $\times$ 10$^{14}$ M$_{\odot}$, Boselli et al. 2022). The narrow-band H$\alpha$ image of the galaxy shows that the stripping process, which is acting 
outside-in, has reached the very inner regions. It is likely that part of the diffuse molecular gas component, mostly located in the inner disc, with its associated dust, has been 
removed during the interaction. Dust acts as a catalyst in the process of molecular gas formation (Hollenbach \& Salpeter 1971; Wakelam et al. 2017). Thus, it can contribute to make 
the molecular gas formation more efficient. We recall that the galaxy D100 has a molecular gas mass of $M\mathrm{(H_2)}$ = 4.8 $\times$ 10$^8$ M$_{\odot}$ on the disc, 
$M\mathrm{(H_2)}$ $\simeq$ 10$^9$ M$_{\odot}$ in the tail, and it is undetected in \hi\ in the recent survey of Molnar et al. (2022) which is sensitive to \mhi\ $\simeq$ 9.6 $\times$ 
10$^7$ M$_{\odot}$ (5$\sigma$ for 100 km s$^{-1}$ line width), which gives a molecular gas fraction in the tail $R_{\rm mol}$ $>$ 10. The jellyfish galaxies such as JW100, which has 
$\simeq$ 30\%\ of its molecular gas in the tail (Moretti et al. 2020b), might be extreme cases where the density of the stripped \hi\ gas is sufficiently high to allow GMC formation via gas cooling.

\section{Conclusion}

During the VESTIGE survey, we discovered 60 star-forming regions located outside the stellar disc of the spiral galaxy NGC 4254 infalling into the Virgo cluster.
These star-forming regions are all located within a \hi\ gas tail formed during the gravitational interaction of the galaxy with another cluster member that occurred 
$\simeq$ 280--750 Myr ago, as suggested by tuned simulations. We observed 42 of these regions in the $^{12}$CO(1-0) line using ALMA and detected molecular gas in four of them, which are 
located in ten individual and resolved GMCs with masses $M\mathrm{(H_2)}$ = 0.75--2.08 $\times$ 10$^6$ M$_{\odot}$, velocity dispersions $\sigma_{v}(\mathrm{CO})$ $\simeq$ 3-12 km s$^{-1}$, 
sizes $>$ 160 pc, and column densities $\Sigma\mathrm{(H_2)}$ $\simeq$ 10 M$_{\odot}$ pc$^{-2}$. These clouds follow the Schmidt relation between the molecular gas column density and the star 
formation surface density observed within the disc of the galaxy, in other local objects, and in similar galaxies in the Virgo cluster. Analytic calculations and tuned simulations consistently 
suggest that these clouds are unstable and expected to dissolve under the energy input of stellar feedback on short timescales ($\simeq$ 10--30 Myr). They formed in the densest regions of the 
stripped gas, where self-shielding from the external hot ICM allowed gas cooling and the formation of GMCs. These results are important in the study of the fate of the gas stripped from their 
parent galaxies in dense environments and show that the formation of dense clouds where star formation can occur only happens under specific conditions, and these conditions are not always 
encountered in gravitational or hydrodynamic perturbations. The results also suggest that a coherent and complete understanding of the evolution of the stripped material in harsh environments 
requires more multi-frequency data of a wide sampling of the physical conditions of the perturbed galaxies (total mass, gas fraction), of the stripped gas and of the surrounding ICM (gas density 
and temperature), and of the impact parameters characterising the perturbation responsible for the gas stripping event.

\begin{acknowledgements}

Based on observations obtained with MegaPrime/MegaCam, a joint project of CFHT and CEA/DAPNIA, at the Canada-French-Hawaii Telescope
(CFHT) which is operated by the National Research Council (NRC) of Canada, the Institut National des Sciences de l'Univers of the Centre National de la Recherche Scientifique (CNRS) of France 
and the University of Hawaii. We thank the anonymous referee for constructive comments which helped improving the quality of the manuscript. 
This paper makes use of the following ALMA data: ADS/JAO.ALMA2011.0.01234.S. ALMA is a partnership of ESO (representing its member states), NSF (USA) and NINS (Japan), together with NRC (Canada), 
NSTC and ASIAA (Taiwan), and KASI (Republic of Korea), in cooperation with the Republic of Chile. The Joint ALMA Observatory is operated by ESO, AUI/NRAO and NAOJ. We are grateful to the whole CFHT 
team who assisted us in the preparation and in the execution of the observations and in the calibration and data reduction.
We acknowledge financial support from ``Programme National de Cosmologie and Galaxies" (PNCG) funded by CNRS/INSU-IN2P3-INP, CEA and CNES, France.
The MeerKAT telescope is operated by the South African Radio Astronomy Observatory, which is a facility
of the National Research Foundation, an agency of the Department of Science and Innovation. The research activities
described in this paper have been co-funded by the European Union - Next Generation EU within PRIN 2022 project no. 20229YBSAN - Globular clusters in cosmological simulations and in lensed 
fields: from their birth to the present epoch. We acknowledge support from the INAF Minigrant `Clumps at cosmological distance: revealing their formation, nature, and evolution' (Ob. Fu.
1.05.23.04.01) and from  the INAF Theory Grant ``Magnetohydrodynamic Simulations of Galactic Molecular Clouds: Resolving Stellar Birth and Proto-planetary Discs with an Enhanced Chemical Network". 
We acknowledge PRACE for awarding us access to Discoverer at
Sofia Tech Park, Bulgaria. This research was supported in part by Lilly Endowment, Inc., through its support for the Indiana University Pervasive Technology
Institute. We acknowledge EuroHPC JU for awarding the project IDs EHPC-REG-2021R0052 and EHPC-REG-2024R01-042 access to DISCOVERER at
the Sofia Tech Park, Bulgaria. M.S. thanks the support by the NSF grant 2407821. HP acknowledges financial support from the Brazilian National Council for Scientific and Technological Development 
(CNPq) under Grant No. 404160/2025-5 (CNPq/MCTI Call No. 44/2024 - Universal).
The INAF - OAC computer cluster used in this work has been acquired within a project aimed to enhance the Sardinia Radio Telescope (SRT). The Enhancement of the SRT for the study of the Universe 
at high radio frequencies is financially supported by the National Operative Program (Programma Operativo Nazionale - PON) of the Italian Ministry of University and Research "Research and 
Innovation 2014-2020", Notice D.D. 424 of 28/02/2018 for the granting of funding aimed at strengthening research infrastructures, in implementation of the Action II.1 - Project Proposal 
PIR01\_00010.
\end{acknowledgements}

\clearpage
\newpage
\begin{appendix}

\section{ALMA continuum subtraction improvements}

\begin{figure*}
\centering
\includegraphics[width=0.9\textwidth]{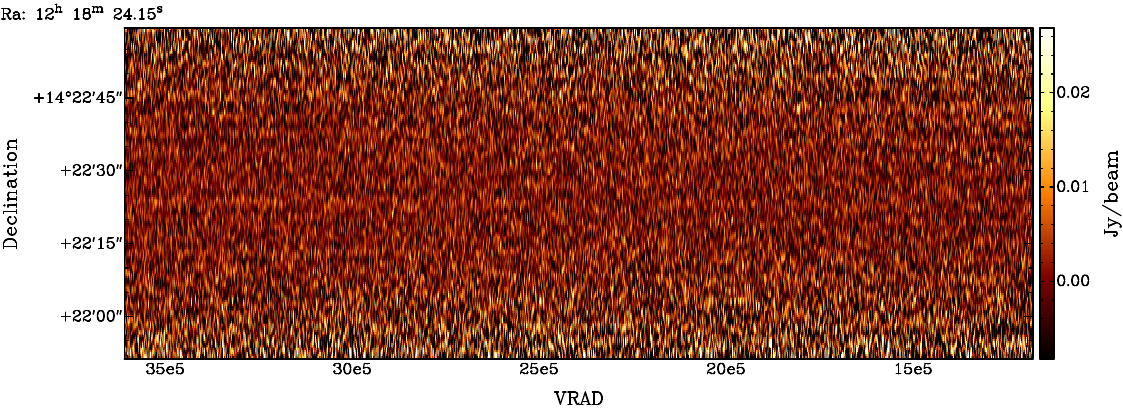}
\\
\includegraphics[width=0.9\textwidth]{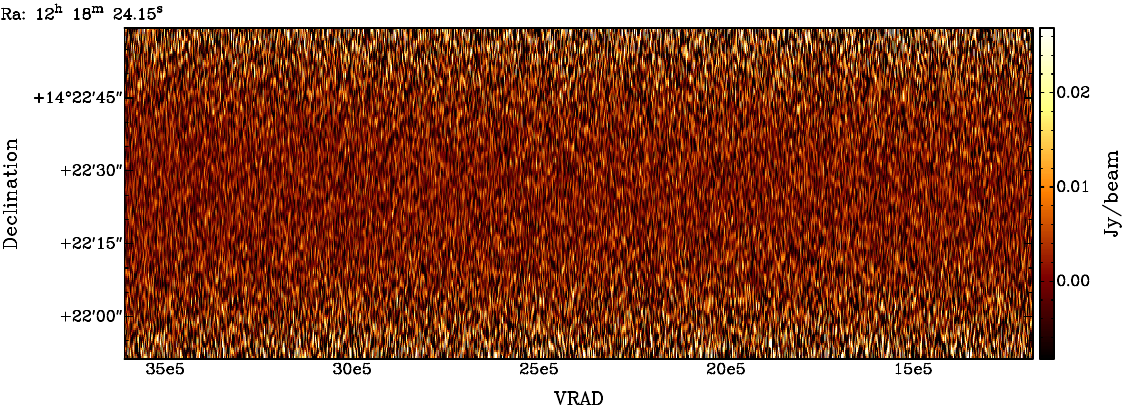}
\caption{Velocity - declination slice through one of our ALMA cubes showing the artefacts described in Sect.~\ref{sec:almadata} (\textit{top}), and the improvement 
obtained by running \texttt{imcontsub} as described in that same section (\textit{bottom}). The units of the velocity axis are m/s.}
\label{fig:contsub}%
\end{figure*}

Figure \ref{fig:contsub} shows an example of the improvements in the continuum subtraction of the ALMA cubes. See figure caption and Sect.~\ref{sec:almadata} for details.

\section{Molecular gas masses and upper limits}

Table \ref{TabCO} gives the physical properties of the molecular clouds. Interesting is the significant difference in recessional velocity of the GMC E with respect to the others 
detected clouds. This difference could be due to a large scale turbulent motion resulting from the gravitational perturbation which gave birth to the tail.

\begin{table*}
\caption{Physical properties of the molecular clouds.}
\label{TabCO}
{\tiny
\[
\begin{tabular}{cccccccccccc}
\hline
\noalign{\smallskip}
\hline
Id  &Region & R.A.(J2000) & Dec.      &  $S(\rm CO)$ & ${\rm err}_{S(\rm CO)}$ & $rms$ & $S/N$ & $\sigma_{v}({\rm CO})$ & $v({\rm CO})$ & $R({\rm CO})$ & $M\mathrm{(H_2)}$ \\      
\hline
 &       & deg       & deg  & mJy km s$^{-1}$&  mJy km s$^{-1}$& mJy beam$^{-1}$ &   &  km s$^{-1}$ & km s$^{-1}$ & pc & $\times$10$^6$M$_{\odot}$ \\
\hline
A  & J121831.64+142404.4  & 184.63185  &  14.40122   &   400.02  &     62.84    &    5.19   &     6.37    &     7.2   & 2356.0  &178.21 &   1.20\\
B1 & J121837.05+142407.8  & 184.65439  &  14.40216   &   385.78  &     53.92    &    5.33   &     7.15    &     8.2   & 2286.5  &202.37 &   1.16\\
B2 & J121837.13+142423.6  & 184.65472  &  14.40656   &   415.81  &     59.01    &    5.24   &     7.05    &    11.7   & 2286.8  &197.93 &   1.25\\
B3 & J121837.36+142431.0  & 184.65567  &  14.40862   &   383.82  &     59.60    &    5.94   &     6.44    &     4.3   & 2289.0  &174.28 &   1.15\\
B4 & J121837.47+142416.7  & 184.65614  &  14.40464   &   626.45  &     62.11    &    4.72   &    10.09    &     7.0   & 2303.3  &242.47 &   1.88\\
B5 & J121838.15+142412.9  & 184.65895  &  14.40358   &   306.50  &     48.65    &    4.92   &     6.30    &     4.3   & 2305.9  &170.59 &   0.92\\
C  & J121838.79+142149.3  & 184.66165  &  14.36371   &   251.18  &     43.07    &    5.73   &     5.83    &     3.7   & 2294.9  &175.78 &   0.75\\
D1 & J121843.51+142324.6  & 184.68134  &  14.39017   &   266.71  &     42.99    &    5.10   &     6.20    &     3.5   & 2312.9  &167.09 &   0.80\\
D2 & J121844.11+142324.4  & 184.68381  &  14.39012   &   320.89  &     48.72    &    5.61   &     6.59    &     3.2   & 2330.6  &175.75 &   0.96\\
E  & J121844.60+142221.8  & 184.68584  &  14.37272   &   695.47  &    106.13    &   12.30   &     6.55    &     9.5   & 2482.2  &120.45 &   2.08\\
\noalign{\smallskip}
\hline
\end{tabular}
\]
Column 1: Region Id.
Column 2: Region name.
Columns 3 and 4: Right ascension (J2000) and declination of the molecular cloud.
Columns 5 and 6: Integrated $^{12}$CO(1-0) flux and associated error, in mJy km s$^{-1}$.
Column 7: rms, in mJy beam$^{-1}$.
Column 8: Integrated signal-to-noise ratio within the 3D detection mask of the cloud.
Column 9: CO line velocity dispersion, in km s$^{-1}$, calculated as described in Sect. \ref{sec:analysis}.
Column 10: Recessional velocity (optical definition) of the cloud, in km s$^{-1}$.
Column 11: Radius of the equivalent circular area covered by the GMC, in pc, calculated as described in Sect. \ref{sec:analysis}.
Column 12: Molecular hydrogen mass of the cloud, in M$_{\odot}$, derived assuming a conversion factor $X_{\rm CO}$ = 2 $\times$ 10$^{20}$ cm$^{-2}$ [K km s$^{-1}$]$^{-1}$ multiplied 
by a factor of 1.36 to take into account helium and other elements, corresponding to $\alpha_{10}$ = 4.3 M$_{\odot}$ pc$^{-2}$.
}
\end{table*}

\section{Atomic gas distribution}

\begin{figure}
\centering
\includegraphics[width=0.49\textwidth]{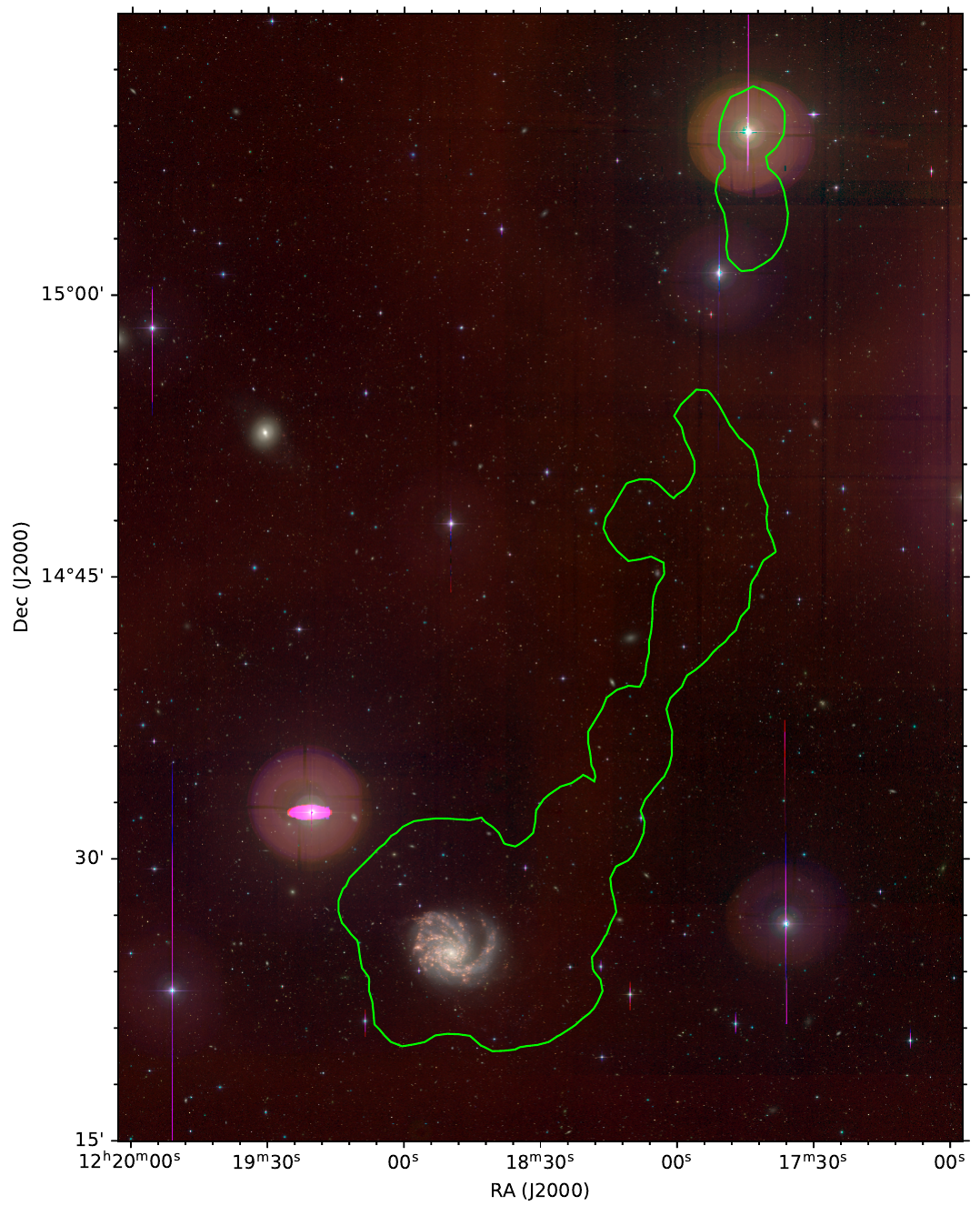}\\
\caption{\hi\ gas distribution in NGC 4254 and in its tail is compared to that of the stellar distribution. The colour image of the galaxy is obtained combining the NGVS (Ferrarese et al. 2012) 
optical $u$ and $g$ in the blue channel, the $r$ and the NB in the green, and the $i$ and the continuum-subtracted H$\alpha$ in the red. The green contour gives the \hi\ column 
density \nhi\ = 2.6$\times$ 10$^{18}$ cm$^{-2}$.}
\label{MeerKAT_color}%
\end{figure}

Figure \ref{MeerKAT_color} shows the low resolution (120\arcsec) \hi\ gas distribution compared to that of the stellar distribution.

\section{Comparison of FUV UVIT and GALEX magnitudes}

\begin{figure*}
\centering
\includegraphics[width=0.99\textwidth]{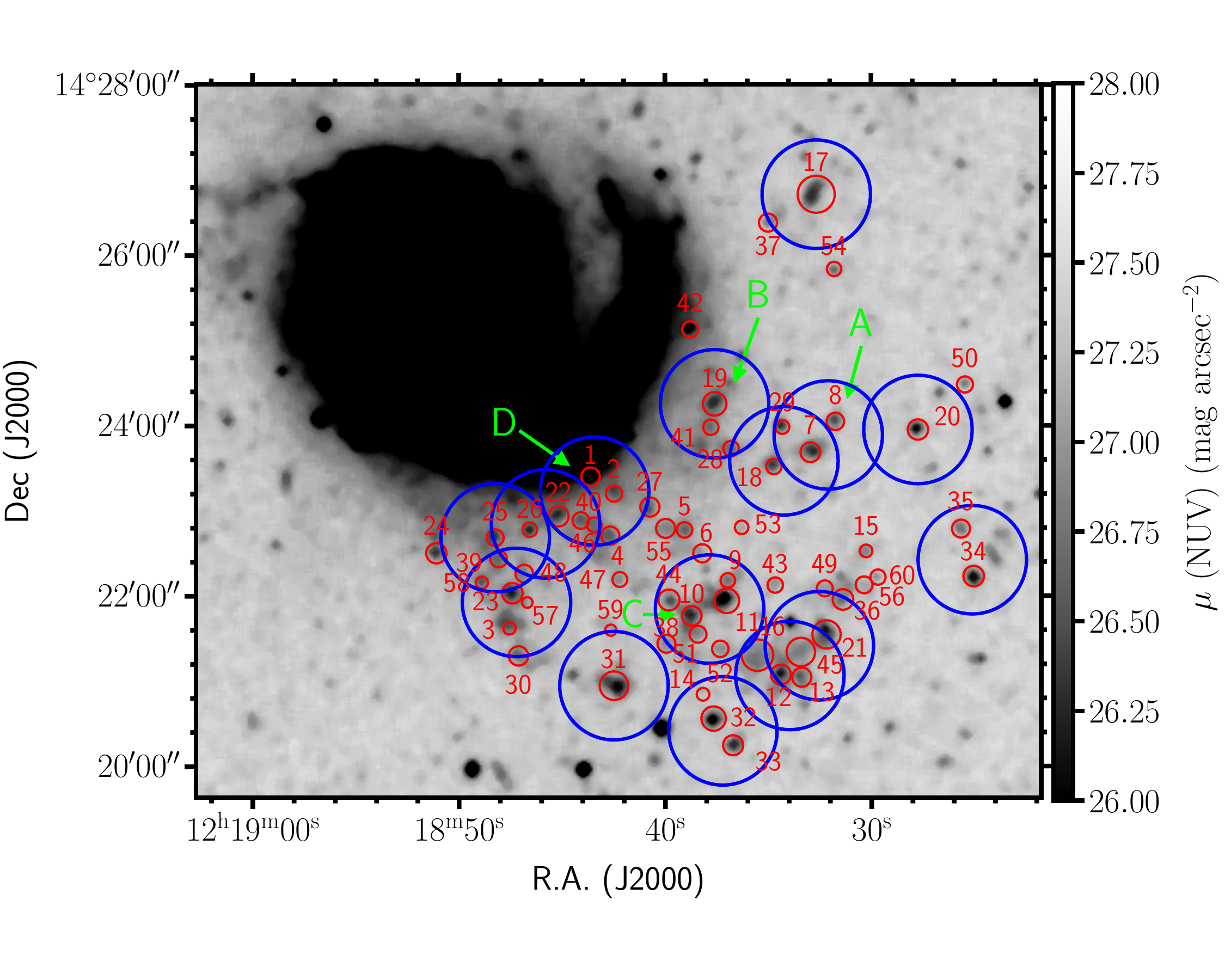}\\
\caption{GALEX NUV images of the galaxy NGC 4254, with the star-forming regions located outside the stellar disc and identified as in Boselli et al. (2018b) The coverage of the 15 different 
ALMA pointings are indicated with blue circles of diameter 75\arcsec, the size of the cube, while green arrows indicate the corresponding detected molecular cloud complexes.} 
\label{idregions}%
\end{figure*}

\begin{figure}
\centering
\includegraphics[width=0.48\textwidth]{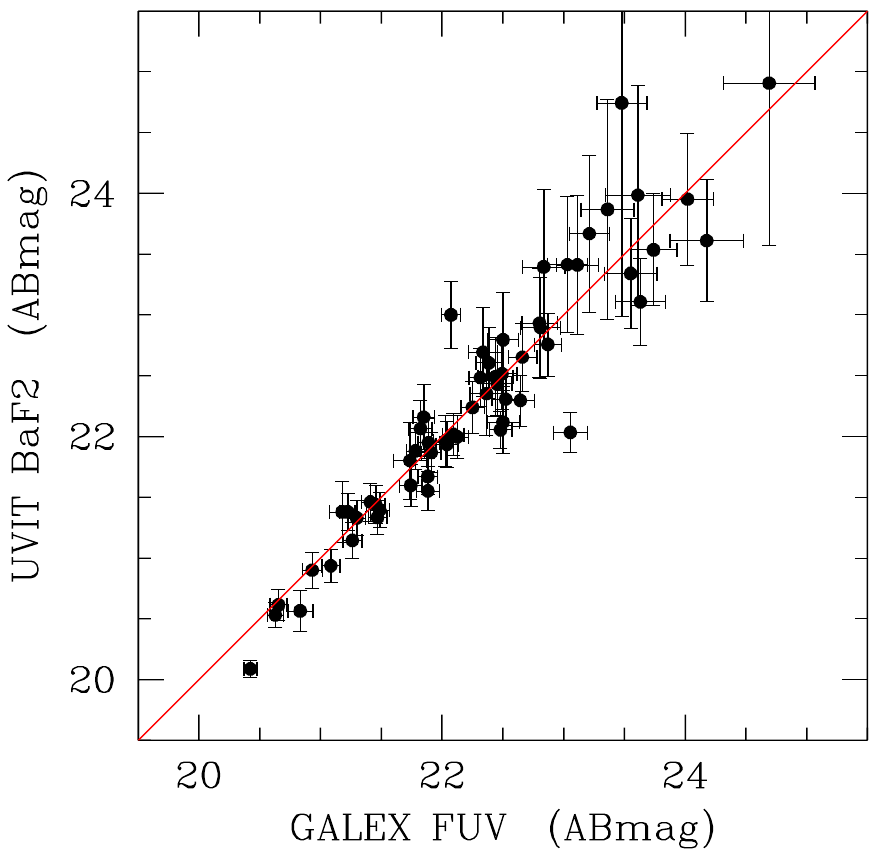}\\
\caption{Comparison of the FUV magnitudes of the external star-forming regions of NGC 4254 measured by GALEX
and UVIT in the BaF2 band. The solid red line shows the 1:1 relation.
}
\label{UVITGALEX}%
\end{figure}

We derived FUV magnitudes of each single region and their associated uncertainties using the same flux extraction procedure described in Fossati et al. (2018) and Boselli et al. (2018b). 
Magnitudes are measured within the same apertures adopted for the other bands in Boselli et al. (2018b). 
Table \ref{HIIUVIT} gives the FUV magnitudes and their associated uncertainties of the star-forming regions located outside the stellar disc of NGC 4254 and identified in Fig. \ref{idregions}. 
These magnitudes are consistent with those derived from the GALEX data (Boselli et al. 2011) when extracted using the same procedure, as shown in Fig. \ref{UVITGALEX}. The larger uncertainty in 
the UVIT data is due to
the shorter integration time in UVIT (8055\,s) versus to GALEX (18\,131\,s) and to the use of large apertures necessary 
to include all the flux of the unresolved star-forming regions in the GALEX band, where the PSF is $\sim$ 5\arcsec\
versus $\sim$ 1.5\arcsec\ in the UVIT image.  

\begin{table*}
\caption{UV, {\hi}, and H$_2$ properties of the external star-forming regions. }
\label{HIIUVIT}
{
\[
\begin{tabular}{cccccccccccc}
\hline
\noalign{\smallskip}
\hline
Region & R.A.(J2000) & Dec.      & $R$     & BaF2 mag   & err           & $S/N$  & \mhi\  & vel   & $W_{20}(\ion{H}{i})$  &  $M\mathrm{(H_2)}$ & GMC\\      
\hline
       & h m s       &$^{\deg}$ \arcmin \arcsec &   \arcsec     & AB mag & AB mag &  & $\times$10$^6$M$_{\odot}$& km s$^{-1}$ & km s$^{-1}$ & $\times$10$^6$M$_{\odot}$ & \\
\hline
 1 & 12:18:43.617 & +14:23:24.13 & 6.527 &  20.087 &  0.0712       & 15.248 &     $<$1     &  -   &  -   & $<$0.85 & D \\
 2 & 12:18:42.478 & +14:23:12.26 & 5.769 &  21.867 &  0.1666       & 6.517  &     $<$1     &  -   &  -   & $<$0.81 & - \\
 3 & 12:18:47.554 & +14:21:37.30 & 4.367 &  22.055 &  0.1527       & 7.109  & 1.7$\pm$0.4  & 2285 & 18   & $<$0.76 & - \\
 4 & 12:18:42.664 & +14:22:42.60 & 6.383 &  21.949 &  0.1987       & 5.464  &     $<$1     &  -   &  -   & $<$2.27 & - \\
 5 & 12:18:39.060 & +14:22:46.56 & 5.464 &  21.996 &  0.1735       & 6.257  &     $<$1     &  -   &  -   & -       & - \\
 6 & 12:18:38.200 & +14:22:30.09 & 6.377 &  23.394 &  0.6389       & 1.699  &     $<$1     &  -   &  -   & $<$1.94 & - \\
 7 & 12:18:32.956 & +14:23:41.26 & 6.949 &  21.394 &  0.1437       & 7.558  &     $<$1     &  -   &  -   & $<$1.12 & - \\
 8 & 12:18:31.743 & +14:24:02.94 & 6.140 &  21.933 &  0.1914       & 5.673  &     $<$1     &  -   &  -   & $<$0.83 & A \\
 9 & 12:18:36.965 & +14:22:10.97 & 5.166 &  22.296 &  0.2099       & 5.172  &     $<$1     &  -   &  -   & $<$1.19 & - \\
10 & 12:18:38.714 & +14:21:46.03 & 6.870 &  21.334 &  0.1409    & 7.705  &     $<$1     &  -   &  -   & $<$1.00 & C \\
11 & 12:18:37.016 & +14:21:56.62 & 8.672 &  20.531 &  0.1022    & 10.624 &     $<$1     &  -   &  -   & $<$1.14 & - \\
12 & 12:18:34.395 & +14:21:04.64 & 7.032 &  21.380 &  0.1493    & 7.273  &     $<$1     &  -   &  -   & $<$0.80 & - \\
13 & 12:18:33.383 & +14:21:02.78 & 6.752 &  22.159 &  0.2653    & 4.092  &     $<$1     &  -   &  -   & $<$0.81 & - \\
14 & 12:18:38.171 & +14:20:50.76 & 4.500 &  23.611 &  0.5004    & 2.170  &     $<$1     &  -   &  -   & $<$1.37 & - \\
15 & 12:18:30.260 & +14:22:31.69 & 4.455 &  22.756 &  0.2584    & 4.202  &     $<$1     &  -   &  -   & -       & - \\
16 & 12:18:35.531 & +14:21:18.33 & 11.138&  21.379 &  0.2476    & 4.385  &     $<$1     &  -   &  -   & $<$2.51 & - \\
17 & 12:18:32.664 & +14:26:42.99 & 13.045&  20.564 &  0.1656    & 6.555  &     $<$1     &  -   &  -   & $<$1.28 & - \\
18 & 12:18:34.726 & +14:23:31.23 & 5.461 &  22.016 &  0.1735    & 6.259  &     $<$1     &  -   &  -   & $<$0.69 & - \\
19 & 12:18:37.598 & +14:24:15.38 & 8.315 &  21.144 &  0.1447    & 7.505  &     $<$1     &  -   &  -   & $<$0.88 & B \\
20 & 12:18:27.741 & +14:23:57.16 & 7.181 &  21.439 &  0.1557    & 6.971  &     $<$1     &  -   &  -   & $<$0.80 & - \\
21 & 12:18:32.182 & +14:21:32.91 & 9.948 &  20.900 &  0.1483    & 7.323  &     $<$1     &  -   &  -   & $<$1.12 & - \\
22 & 12:18:45.170 & +14:22:56.25 & 7.136 &  21.460 &  0.1570    & 6.917  &     $<$1     &  -   &  -   & $<$0.88 & - \\
23 & 12:18:47.413 & +14:22:02.02 & 7.206 &  21.329 &  0.1417    & 7.661  &     $<$1     &  -   &  -   & $<$0.82 & - \\
24 & 12:18:51.095 & +14:22:30.27 & 7.155 &  22.064 &  0.2290    & 4.741  &     $<$1     &  -   &  -   & $<$1.43 & - \\
25 & 12:18:48.241 & +14:22:41.21 & 5.908 &  21.670 &  0.1472    & 7.378  &     $<$1     &  -   &  -   & $<$0.68 & - \\
26 & 12:18:46.564 & +14:22:46.86 & 5.153 &  21.880 &  0.1540    & 7.052  &     $<$1     &  -   &  -   & $<$0.71 & - \\
27 & 12:18:40.739 & +14:23:02.72 & 6.745 &  21.963 &  0.2087    & 5.201  &     $<$1     &  -   &  -   & $<$1.86 & - \\
28 & 12:18:36.803 & +14:23:43.90 & 5.441 &  22.306 &  0.2152    & 5.046  &     $<$1     &  -   &  -   & $<$2.12 & - \\
29 & 12:18:34.313 & +14:23:59.06 & 4.941 &  22.032 &  0.1631    & 6.657  &     $<$1     &  -   &  -   & $<$1.15 & - \\
30 & 12:18:47.116 & +14:21:17.90 & 6.810 &  22.490 &  0.3258    & 3.333  &     $<$1     &  -   &  -   & $<$2.12 & - \\
31 & 12:18:42.488 & +14:20:56.87 & 10.091&  20.614 &  0.1271    & 8.542  &     $<$1     &  -   &  -   & $<$1.00 & - \\
32 & 12:18:37.661 & +14:20:33.67 & 8.579 &  20.936 &  0.1335    & 8.134  &     $<$1     &  -   &  -   & $<$1.00 & - \\
33 & 12:18:36.716 & +14:20:14.85 & 7.053 &  21.551 &  0.1581    & 6.867  &     $<$1     &  -   &  -   & $<$0.94 & - \\
34 & 12:18:25.045 & +14:22:13.88 & 7.252 &  21.597 &  0.1736    & 6.255  &     $<$1     &  -   &  -   & $<$0.94 & - \\
35 & 12:18:25.655 & +14:22:47.32 & 6.343 &  22.932 &  0.4478    & 2.425  &     $<$1     &  -   &  -   & $<$1.28 & - \\
36 & 12:18:31.383 & +14:21:57.49 & 7.265 &  22.117 &  0.2591    & 4.190  &     $<$1     &  -   &  -   & $<$2.31 & - \\
37 & 12:18:35.001 & +14:26:23.02 & 6.362 &  22.517 &  0.2982    & 3.641  &     $<$1     &  -   &  -   & $<$1.80 & - \\
38 & 12:18:39.944 & +14:21:26.40 & 6.322 &  22.894 &  0.4134    & 2.626  &     $<$1     &  -   &  -   & $<$1.94 & - \\
39 & 12:18:48.075 & +14:22:26.20 & 5.875 &  22.605 &  0.2927    & 3.709  &     $<$1     &  -   &  -   & $<$1.49 & - \\
40 & 12:18:44.091 & +14:22:53.29 & 5.821 &  22.239 &  0.2168    & 5.008  &     $<$1     &  -   &  -   & $<$1.26 & - \\
41 & 12:18:37.781 & +14:23:58.92 & 5.386 &  22.653 &  0.2792    & 3.889  &     $<$1     &  -   &  -   & $<$0.90 & - \\
42 & 12:18:38.773 & +14:25:07.86 & 5.685 &  23.000 &  0.2719    & 3.993  &     $<$1     &  -   &  -   & -       & - \\
43 & 12:18:34.668 & +14:22:07.67 & 5.418 &  24.743 &  1.7534    & 0.619  &     $<$1     &  -   &  -   & -       & - \\
44 & 12:18:39.821 & +14:21:57.23 & 7.392 &  22.352 &  0.3427    & 3.168  &     $<$1     &  -   &  -   & $<$2.32 & - \\
45 & 12:18:33.428 & +14:21:20.48 & 10.047&  21.800 &  0.3172    & 3.423  &     $<$1     &  -   &  -   & $<$1.26 & - \\
46 & 12:18:43.437 & +14:22:50.23 & 5.227 &  22.484 &  0.2389    & 4.545  &     $<$1     &  -   &  -   & $<$1.11 & - \\
47 & 12:18:42.200 & +14:22:11.70 & 5.283 &  23.668 &  0.6444    & 1.685  &     $<$1     &  -   &  -   & -       & - \\
48 & 12:18:46.838 & +14:22:15.93 & 6.176 &  22.795 &  0.3858    & 2.814  &     $<$1     &  -   &  -   & $<$1.16 & - \\
49 & 12:18:32.262 & +14:22:05.30 & 5.609 &  23.410 &  0.5747    & 1.889  &     $<$1     &  -   &  -   & $<$1.61 & - \\
50 & 12:18:25.457 & +14:24:28.89 & 5.674 &  23.985 &  0.9066    & 1.198  &     $<$1     &  -   &  -   & -       & - \\
51 & 12:18:38.412 & +14:21:33.05 & 5.975 &  22.433 &  0.2588    & 4.196  &     $<$1     &  -   &  -   & $<$1.08 & - \\
52 & 12:18:37.334 & +14:21:22.79 & 5.789 &  23.413 &  0.5597    & 1.940  &     $<$1     &  -   &  -   & $<$1.87 & - \\
53 & 12:18:36.292 & +14:22:48.38 & 4.793 &  23.107 &  0.3606    & 3.011  &     $<$1     &  -   &  -   & -       & - \\
54 & 12:18:31.799 & +14:25:50.36 & 5.106 &  23.340 &  0.4510    & 2.407  &     $<$1     &  -   &  -   & -       & - \\
55 & 12:18:39.979 & +14:22:47.69 & 6.684 &  22.693 &  0.3669    & 2.959  &     $<$1     &  -   &  -   & -       & - \\
56 & 12:18:30.357 & +14:22:07.78 & 5.951 &  23.867 &  0.9026    & 1.203  &     $<$1     &  -   &  -   & -       & - \\
57 & 12:18:46.687 & +14:21:55.55 & 3.738 &  23.952 &  0.5448    & 1.993  &     $<$1     &  -   &  -   & $<$0.49 & - \\
58 & 12:18:48.887 & +14:22:09.52 & 4.284 &  23.535 &  0.4609    & 2.356  &     $<$1     &  -   &  -   & $<$1.22 & - \\
59 & 12:18:42.639 & +14:21:35.98 & 3.983 &  24.906 &  1.3317    & 0.815  &     $<$1     &  -   &  -   & $<$1.34 & - \\
60 & 12:18:29.684 & +14:22:13.22 & 5.351 &  -      &  -             &   -    &     $<$1     &  -   &  -   & -       & - \\
\noalign{\smallskip}
\hline
\end{tabular}
\]
FUV magnitudes of the 60 individual regions identified in Boselli et al. (2018b) are extracted within circular apertures of radius $R$, in arcsec. \hi\ and H$_2$ upper 
limits are 3$\sigma$ assuming a line width of $W_{20}(\ion{H}{i})$ = $W_{20}({\rm CO})$ = 25 km s$^{-1}$. Column GMC indicates the corresponding detected molecular gas complex. Some 
complexes have detected individual GMCs in Table \ref{TabCO}, but upper limits here.  This is due to the fact that the detected GMCs do not necessary fall within the regions defined in 
columns 2, 3 and 4. A few regions are at the very edge of the ALMA beam, but they are partly covered by the ALMA cube. We measured an upper limit of these regions using only the available 
channels. Detections or upper limits are thus available for 49 regions.
}
\end{table*}

\section{CO line profiles}

\begin{figure*}
\centering
\includegraphics[width=0.19\textwidth]{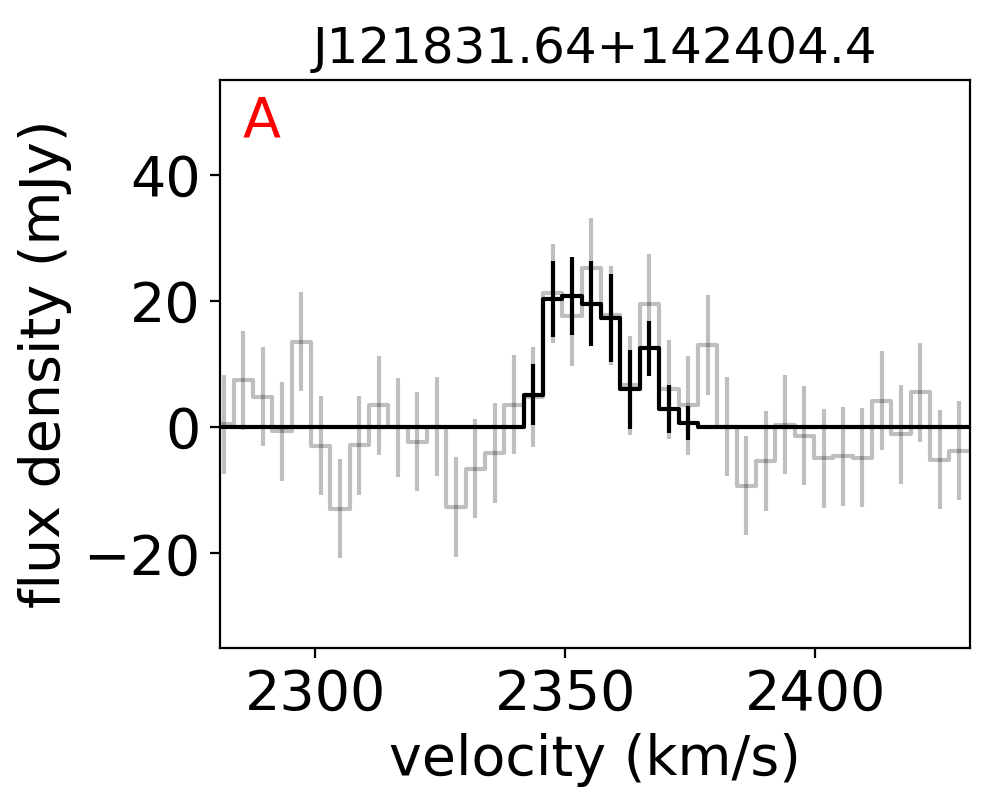}
\includegraphics[width=0.19\textwidth]{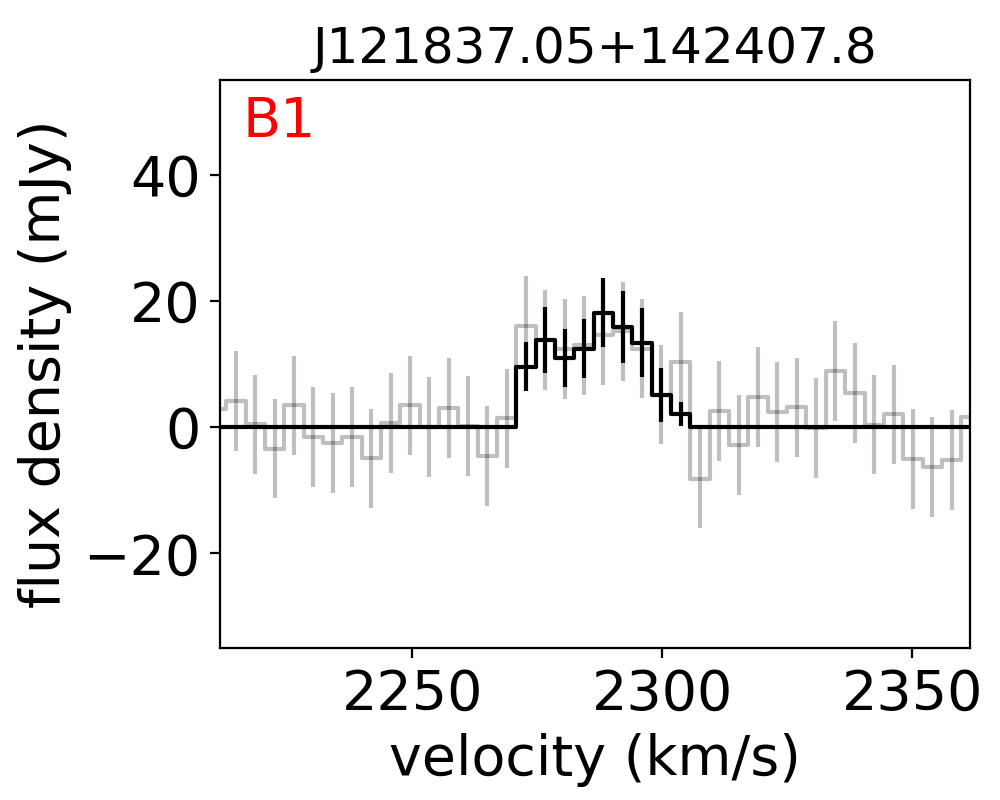}
\includegraphics[width=0.19\textwidth]{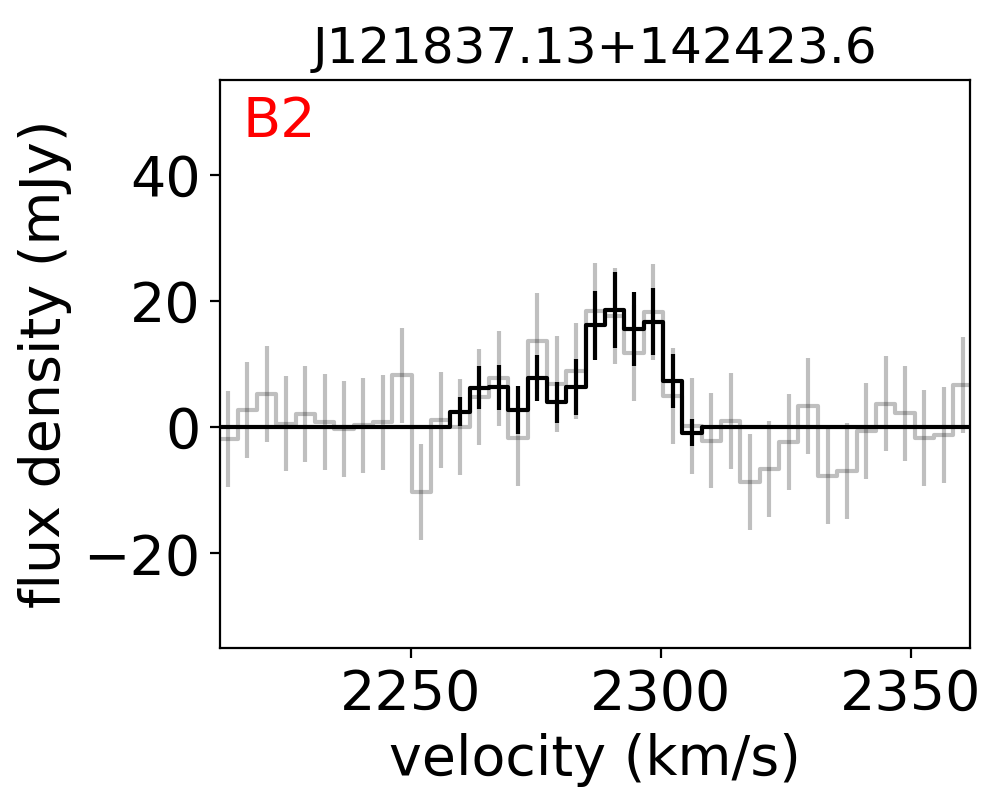}
\includegraphics[width=0.19\textwidth]{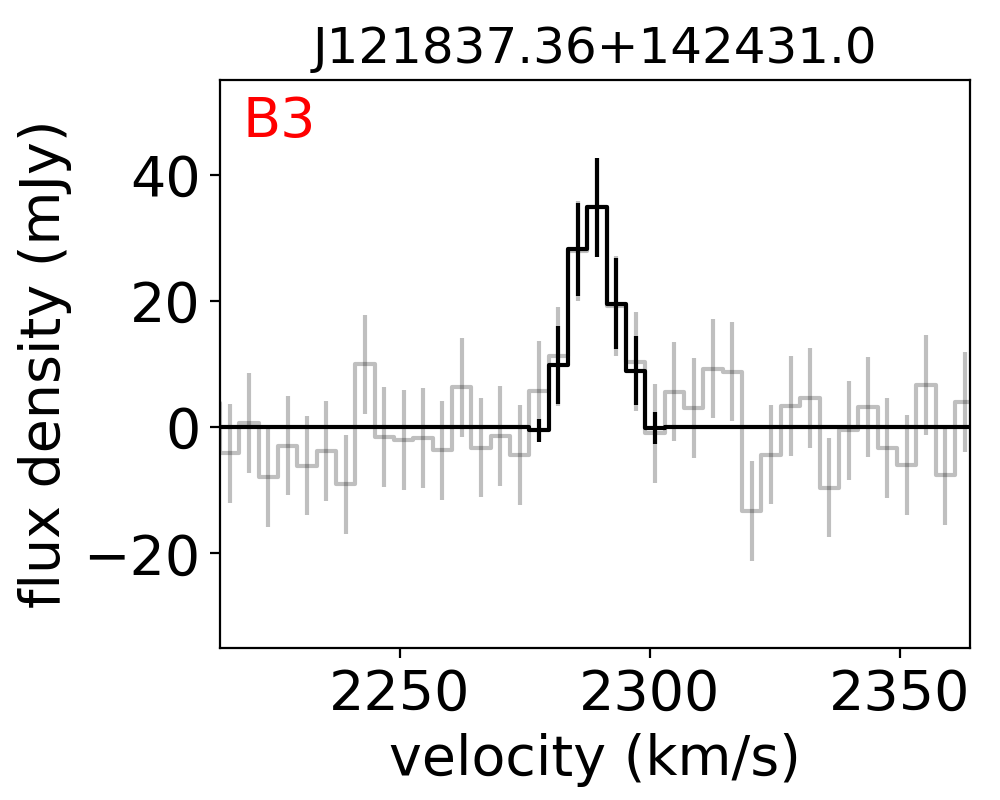}
\includegraphics[width=0.19\textwidth]{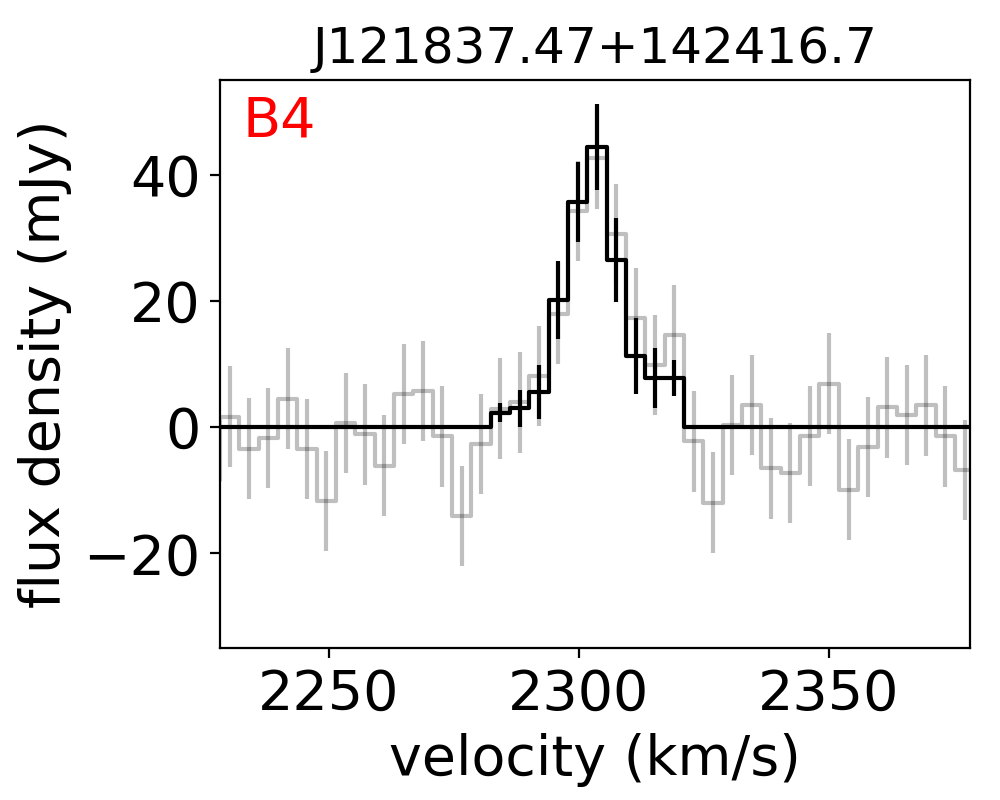}
\\
\includegraphics[width=0.19\textwidth]{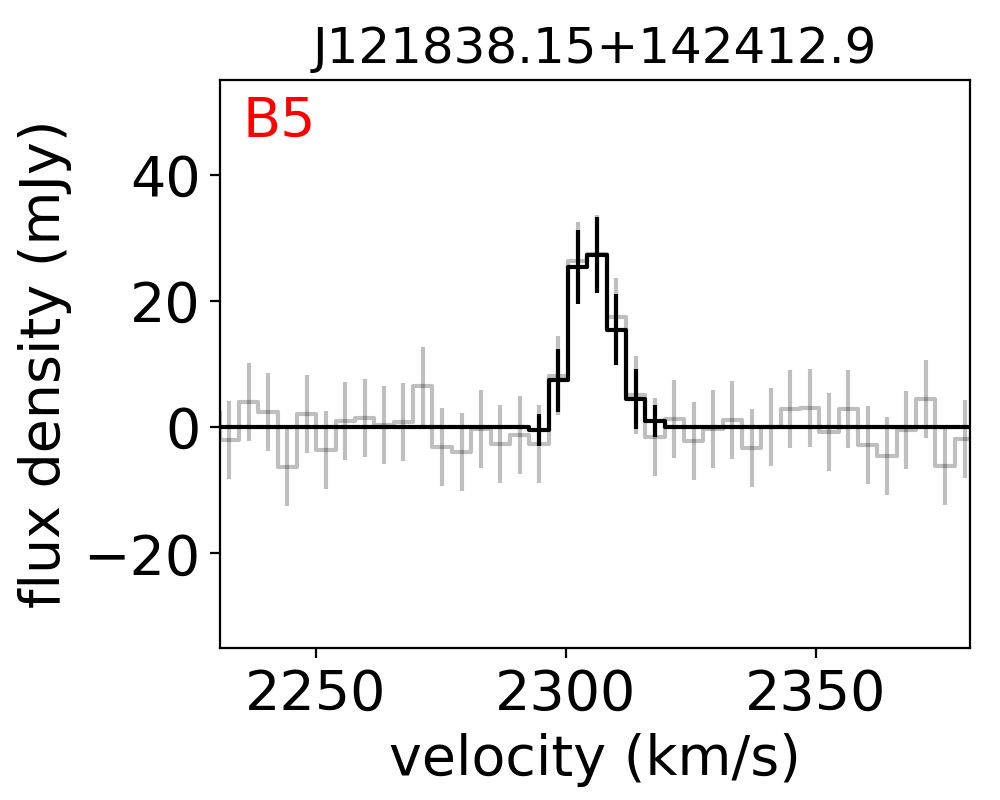}
\includegraphics[width=0.19\textwidth]{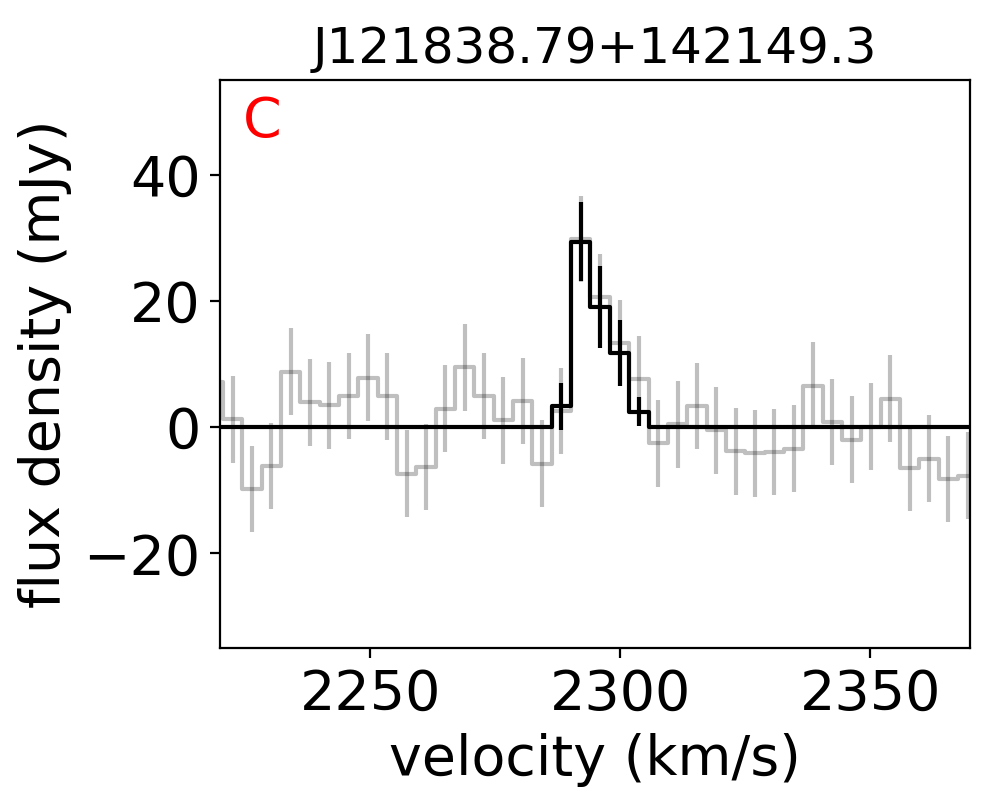}
\includegraphics[width=0.19\textwidth]{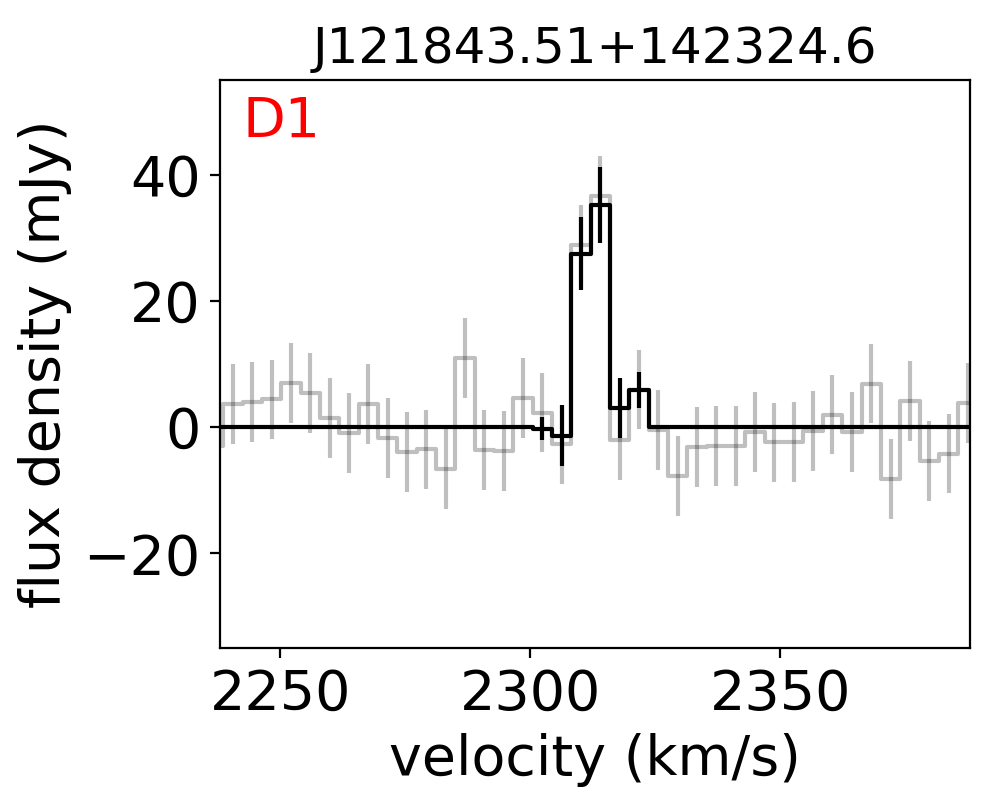}
\includegraphics[width=0.19\textwidth]{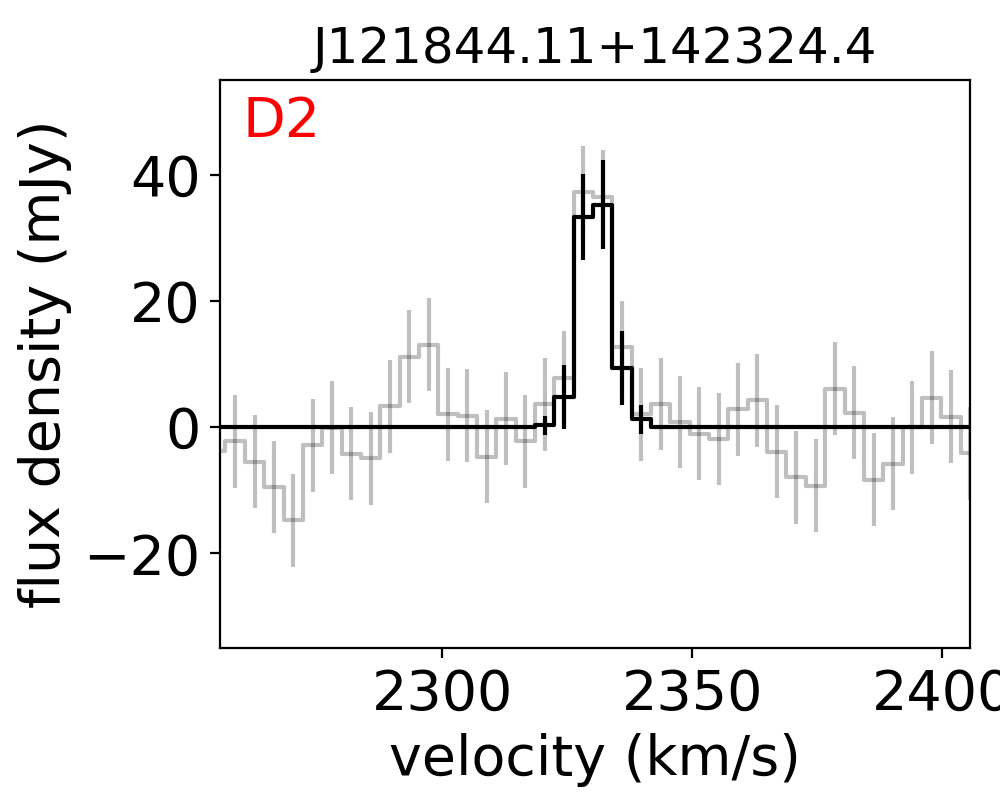}
\includegraphics[width=0.19\textwidth]{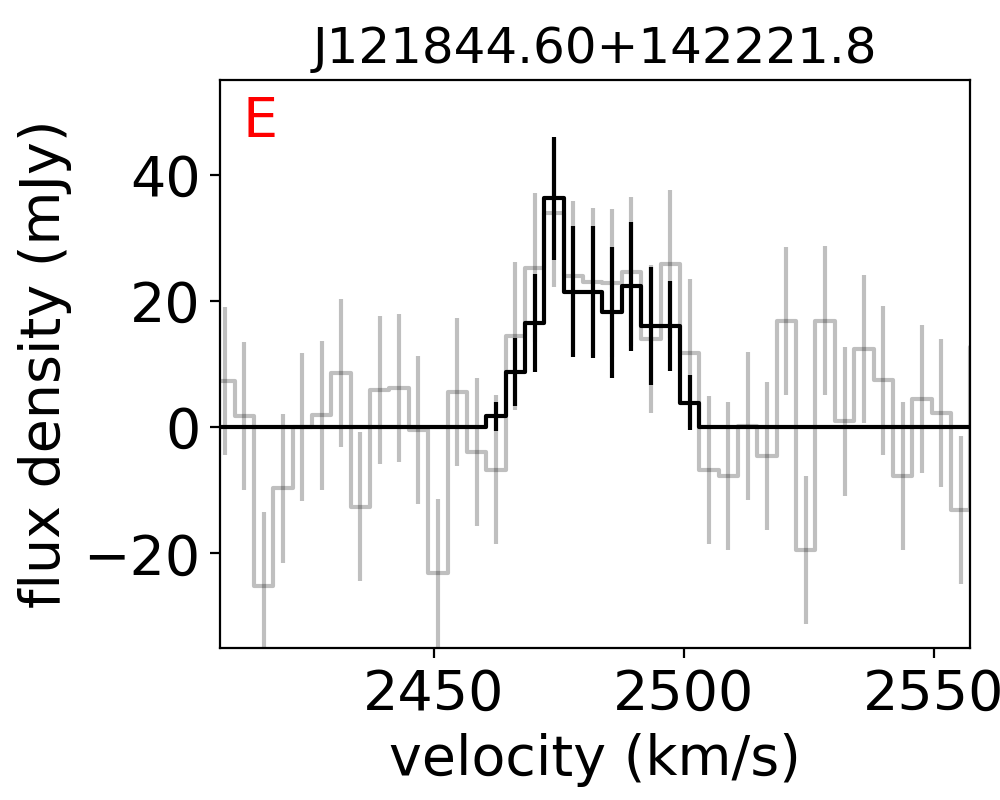}
\\
\caption{ CO(1-0) line velocity profile of the detected GMCs shown in Fig. \ref{zoom_region}. 
Velocities are in the barycentric reference frame and follow the optical definition. The spectra are binned to a channel width of 3.8 \kms. The black line shows the spectrum obtained 
integrating the emission within the 3D detection mask. The grey line shows the spectrum obtained integrating within a 2D aperture obtained collapsing the 3D detection mask along the 
velocity axis. We show statistical error bars calculated using the local noise level around the emission and the number of spatial pixels included in each channel.}
\label{COvelprofile}%
\end{figure*}

Figure \ref{COvelprofile} shows the CO line profile of the detected GMCs.

\section{Simulations}
\label{app:simulations}
\subsection{\textsc{ramses} simulations}

The simulations were carried out using the \textsc{ramses} hydrodynamic code (Teyssier 2002), featuring adaptive mesh refinement (AMR) to achieve high spatial resolution in dense regions 
or in regions with strong density gradients. 

\subsubsection{Simulation setup}

The simulations incorporate updated individual stellar feedback prescriptions from the SIEGE cosmological framework (Calura et al. 2022, 2025). 
We include radiative cooling, self-gravity, and a star formation prescription based on local gravitational collapse,
adopting a density threshold of $n > 100\,\mathrm{cm^{-3}}$.
The generation of single stars occurs via stochastic, discrete sampling (Sormani et al. 2017, Calura et al. 2022) from a Kroupa (2001) stellar initial mass function. 
Star formation is regulated by a stellar feedback module which includes time-resolved injection of thermal energy and mass
from individual massive stars in the pre-supernova phase through stellar winds and, at the end of their life, through supernova explosions. 

\begin{table}
  \caption{Initial conditions parameters of \textsc{ramses} and \textsc{gizmo} hydrodynamic simulations.}
\label{models}
{
\[
\begin{tabular}{cccccc}
\hline
\noalign{\smallskip}
\hline
Model      & $\rho_{\rm cloud}$ & $\rho_{\rm ICM}$ & $T_{\rm ICM}$ & $ v_{\rm ICM}$   & $B_z$        \\
\hline
           & g cm$^{-3}$    & g cm$^{-3}$ &  K      &   km s$^{-1}$      & {\rm $\mu$ gauss}  \\
\hline
\textsc{ramses} & & & & & \\
\hline
SW         &  $\sim 10^{-23}$  & $ 10^{-28}$  & $10^7$  &  20  & -\\
FW         &  $\sim 10^{-23}$ & $  10^{-28}$  & $10^7$  & 2000 & -\\
\hline
\textsc{gizmo} & & & & & \\
\hline
HDrest  & $\sim 10^{-23}$ & $3.4\times 10^{-28}$ & $10^7$ & 0 & -\\
HD        & $\sim 10^{-23}$ & $3.4\times 10^{-28}$ & $10^7$ & 2000 & -\\
MHD     & $\sim 10^{-23}$ & $3.4\times 10^{-28}$ & $10^7$ & 2000 & 2\\
\noalign{\smallskip}
\hline
\noalign{\smallskip} 
\end{tabular}
\]
}
\end{table}

In the stellar wind phase, each massive star contributes to stellar feedback by injecting mass and energy continuously  (Calura et al. 2025).  
This feedback is coupled to the surrounding ISM through a delayed cooling scheme, in which radiative cooling
is temporarily switched off based on the local concentration of an appropriate tracer, accounted for by means of a
passive scalar and aimed at modelling crudely the non-thermal stellar energy deposition (Teyssier et al. 2013, Calura et al. 2015, 2025).

We employ a Cartesian box of size 2$\times$$L_x = 325~$pc, 
with a base resolution of $20$~pc and refinement down to $0.6$~pc in maximum resolution cells (see Table \ref{models}). 
The ICM wind is imposed as a steady inflow along one boundary of the domain. 
We consider two runs that differ in the relative velocity of the impinging ICM wind $v_{\mathrm{ICM}}$ and temperature $T_{\mathrm{ICM}}$. 

In the first case (`SLOW WIND', SW), the ICM wind has a velocity $v_{\mathrm{ICM}} = 20\,\mathrm{km\,s^{-1}}$,
chosen to approximate the physical conditions of a quiescent average Galactic environment, or representing a cloud
moving through a diffused, ionised gas distribution with velocity comparable to the local sound speed. 
In the second case (`FAST WIND', FW), the relative velocity is set to $v_{\mathrm{ICM}} = 2000\,\mathrm{km\,s^{-1}}$,
representative of strong ram-pressure conditions in the Virgo cluster and consistent with Calura et al. (2020). 
In both cases, the assumed ICM gas temperature is $T_{\mathrm{ICM}} \sim 10^7\,\mathrm{K}$.

The wind is imposed as a steady inflow along one boundary of the domain, with outflow conditions on the opposite face and on the remaining sides.
Both simulations are evolved for $\sim 25$\,Myr, of the order of the typical survival value for cold gas clouds in local star-forming regions (Jeffreson et al. 2024). 

\begin{figure}
  \includegraphics[width=7.8cm,height=6.8cm]{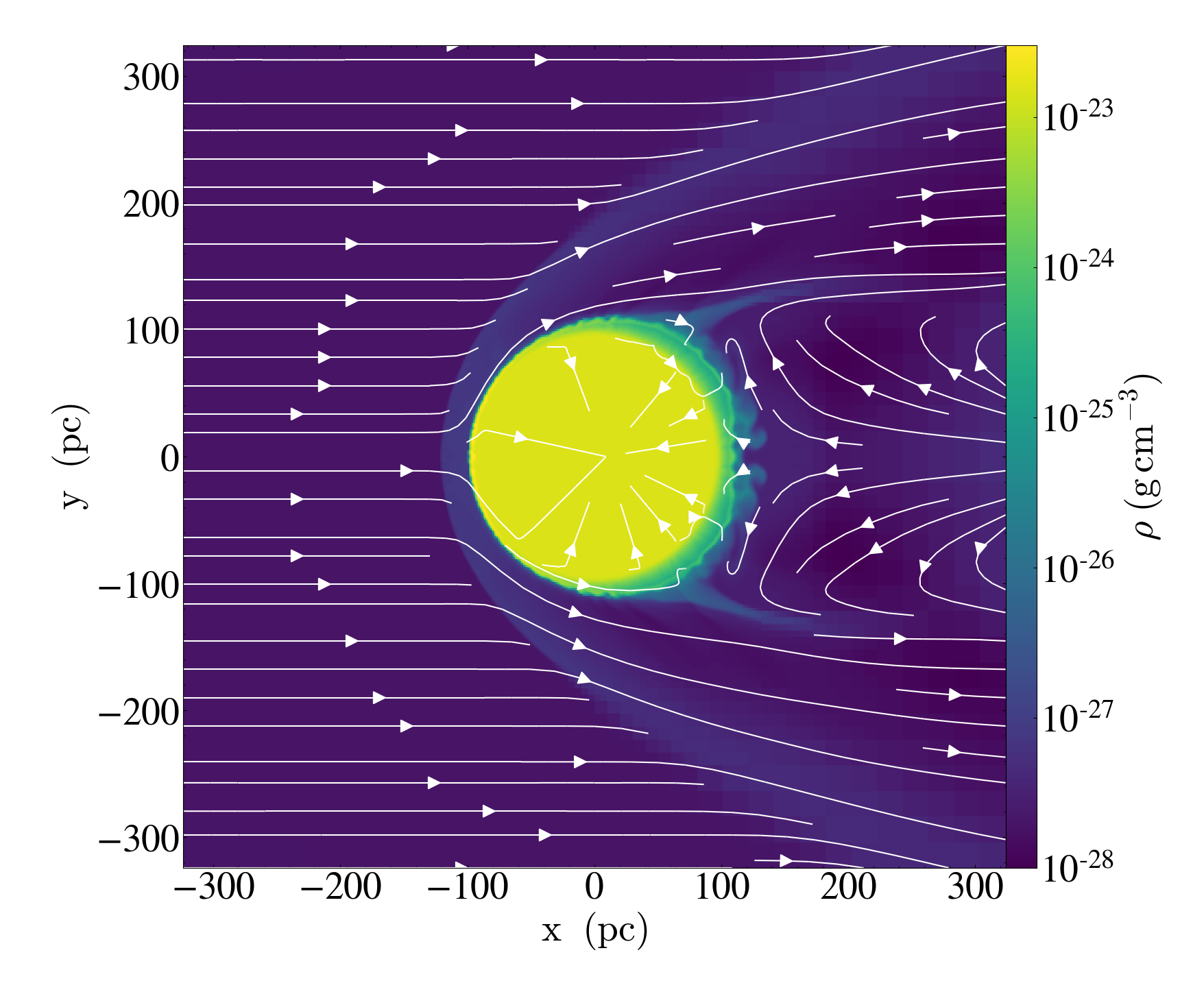}
\caption{Initial density map of the cold gas cloud in the fast-wind model. 
  Slice density map through the central plane of the simulation box at 2 Myr,
  showing the early interaction between the dense cloud and the hot ICM wind, which flows from left to right.
  The colour scale represents gas density, whereas the streamlines trace the velocity field of the ICM.}
\label{fig_ic}
\end{figure}

The initial configuration, consisting of a cold gas cloud at the centre of the box bathed in a tenuous ICM wind, is illustrated in Fig. ~\ref{fig_ic},
representing a slice density map of the simulated system at an early evolutionary stage.  
The figure shows the cloud immersed in the fast wind after 2 Myr of evolution.   
The direction of the flow is shown by the velocity streamlines, with the hot ICM gas moving from left to right.  
A bow shock is visible upstream of the initially spherical cold gas sphere, which at this time has already undergone a slight compression.   
The properties of the system at different times are described in the next subsection.

\subsubsection{Results} 

The system at later evolutionary stages is shown in Fig.~\ref{fig_evo},
representing slice density maps of the two simulations at an intermediate time ($\sim 16$ Myr) and at an advanced time
close to the end of the runs ($\sim 26$ Myr).  
In the case of the SW, at $\sim 16$ Myr the dense gas has undergone significant collapse, 
reaching densities orders of magnitude higher than the initial value of $\rho_{\rm cloud}$ $\sim 10^{-23} \mathrm{g~cm^{-3}}$.  
The collapse led to the birth of $\sim 5 \times 10^4$ M$_{\odot}$ of stars (black dots in Fig.~\ref{fig_evo}).   
After the onset of the star formation, all the stars are grouped in the centre. 
The central gas distribution presents a structure significantly perturbed by the stellar feedback,
which gives place to strong compression and cavities around the stellar aggregate. 
In the central region, part of the gas is still collapsing, but  
the starburst generated an outflow, outlined by the velocity streamlines that indicate
clearly an outwards-directed bulk motion, perturbed only on the
left-side boundary by the steady presence of the wind. 
At a later time ($\sim 26$ Myr), a significant fraction of the initially cold cloud gas 
has been pushed away from the centre, with moderately dense regions 
with maximum density $\rho_{\rm cloud}$ $\sim 10^{-24}\, \mathrm{g~cm^{-3}}$ visible around the stars,
at distances $>100$ pc. In response to the outflow, the stars have
diffused significantly and, at this time, are distributed across a vast, rarefied region.

From a morphological point of view, at $\sim 16$ Myr the stellar distribution of the FW model (top-right panel of Fig.~\ref{fig_evo}) 
shows remarkable differences with the one of the SW. 
Even though most stars are concentrated in the centre, the stellar component 
is more diffuse and is partially reminiscent of the typical shape of 
a cloud subject to strong ram pressure, with a visible compression
at the upstream front edge and an elongated, fan-shaped distribution downstream.

At variance with the SW, at this epoch the collapse of the gas has already ceased,
and several `fingers' of dense gas are scattered around the centre, with the densest parts pointing towards it and 
in most cases surrounded by elongated cavities.
This indicates that the dense gas that composed the original
cloud is being pushed away from the centre by the combined effects of the
ram pressure of the hot ICM and stellar feedback. 
The stellar feedback pushes some of 
the dense gas leftwards toward the ICM wind, which is
significantly affected, as visible from the complex pattern of
upstream velocity streamlines, indicating 
the deviated trajectories of some hot wind gas on the left border.

At a later time ($\sim 26$ Myr),
the stellar distribution has become more diffuse and, as in the SW
case, it has mainly left the central regions, in which
the gas is rarefied and presents low density, 
$\rho_{\rm cloud}$ $< 10^{-26}\, \mathrm{g~cm^{-3}}$.
Also, the moderately dense fingers have been pushed apart or 
disrupted by these processes.

The evolution of the global gas content in the two
models is shown in Fig. \ref{fig_sfh}, along with the star formation history.
The compression of the fast wind leads to positive feedback, shown by the fact that
the star formation begins earlier in the FW model (solid red curve). 
However, the ram pressure causes also stronger regulation,
as witnessed by the lower SFR values and final stellar mass $\sim 4.8 \times 10^4$ M$_{\odot}$.
In the SW model, the SFR can reach maximum values that are 10 times higher,
but the SF period lasts a significantly shorter time.
The final stellar mass in this case is $\sim 5.1 \times 10^4$ M$_{\odot}$, slightly higher than in the other model. 
The blue curves show the evolution of the gas fraction inside the domain,
normalised to the initial value. 
In both models, the gas fraction starts to decrease significantly
soon after the beginning of the star formation and reaches a lower value at the final time of the simulations.

\subsection{\textsc{gizmo} simulations}

Unlike in the \textsc{ramses} runs, in this case the cloud is initially turbulent, with the turbulence cascade following a Burgers-like power spectrum (see Bovino et al. 2019). We 
employed $\sim 50\,000$ particles to map the cloud, corresponding to $\sim 20\rm\, M_\odot$ per particle. We embedded the cloud in a cuboidal box of $600\times 600\times 60\,000$ pc, 
which ensures that the wind material and the stripped gas do not interact with the cloud twice during the simulated time because of the periodic boundary conditions. For the wind, we 
assume $v_{\rm ICM}=2000\rm\, km\ s^{-1}$ and $T_{\rm ICM}=10^7$~K.
We performed three simulations, one with a static ICM (`HDRest' run) and two with $v_{\rm ICM}=2000\rm\, km~s^{-1}$, with (`MHD' run) and without (`HD' run) ideal magnetohydrodynamics 
(MHD; Hopkins \& Raives 2016) respectively. Gas thermochemistry is incorporated as described in Lupi et al. (2024). Star formation is based on the Jeans criterion; i.e. we allow star 
formation to occur when the Jeans mass within the kernel of each particle (encompassing 32 neighbours) becomes lower than $1000\rm\, M_\odot$. In addition, we require the gas to be 
denser than $n$ $>$ 1000~$\rm cm^{-3}$, converging ($\nabla\cdot \mathbf{v}$), and that the gas cells do not overlap with newly formed stars.

\begin{figure}
  \includegraphics[width=\columnwidth]{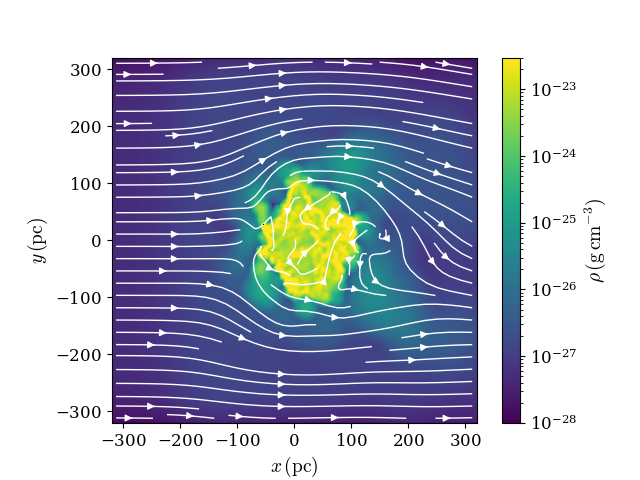}
\caption{Same as Fig.~\ref{fig_ic} but for the fast wind `HD' \textsc{gizmo} run at $t=2\rm\, Myr$.}
\label{fig:ic_gizmo}
\end{figure}

In Fig.~\ref{fig:ic_gizmo}, we report the cloud from the HD run after 2~Myr of evolution. The stream lines highlight the wind direction and the interaction with the cloud. Despite the 
large Mach number ($\mathcal{M}\sim 2000$), the bow shock surface is not clearly visible in this run, due to the irregular shape of the cloud resulting from the initial turbulence. We 
can nonetheless observe the compression of the gas on the upwind side of the cloud, that initially boosts star formation. 

\subsubsection{Long-term evolution}

In Fig.~\ref{fig:gizmo_lt_rest}, we report the results from the HDRest simulation after 16~Myr. The pressurised environment around the cloud, being at rest, does not significantly affect 
the cloud evolution, which cools down, fragments, and form stars, as can be seen from the black dots in the figure. Stellar feedback, in the form of radiation, winds, and SNe (the first 
one occurring at $t\sim 8$~Myr), then starts expelling the gas, destroying the cloud after about 25 Myr (when the total energy injected by SNe is about $10^{53}\rm erg$, much larger than 
the work done by ram pressure). At $t\sim 16$~Myr, the right side of the cloud has been already dispersed, whereas the left one still shows dense clumps on the verge of collapsing to form new stars. 

In order to assess the impact of ram pressure, in Fig.~\ref{fig:gizmo_lt} we show instead the total hydrogen column density maps for the HD and MHD runs after 16~Myr and 25~Myr. After 16~Myr, 
the gas in the HD run has already been stripped away from the star-forming region by ram pressure, almost completely dissolving the cloud. In the MHD run, instead, the additional magnetic 
pressure slows down the initial fragmentation, leaving a dense core in the centre that is more resilient to ram pressure, and forms stars in a more compact configuration. The cloud outskirts, 
on the other hand, have been stripped more easily, and have formed a thick stream behind the cloud. At 25~Myr, the star cluster appears slightly more massive and extended than at 16~Myr. This 
is partially due to some additional SF occurring in the still present dense clumps, but mostly to the potential well of the cloud becoming shallower and the cluster relaxation. The remaining 
gas at the end of the simulation has been pushed to several hundred pc away from the cluster location, and has mixed with the surrounding ICM, resulting in a complete shut-off of further star 
formation. 

In the MHD run, instead, we see a somewhat different behaviour. At 25~Myr, the cluster appears more irregular and extended, due to the dense clumps in the cloud centre surviving for a longer 
time. We also note that, while significant mass was already removed from the cloud in the first 16~Myr by ram pressure, some dense gas still remained and was evacuated at later times by SN 
feedback. Interestingly, cooling at the edge of the SN bubble was able to produce relatively dense star-forming clumps - see, for instance, the region at $(x,y)=(-200,200)$~pc, further 
prolonging the period of active star formation.

To more quantitatively highlight these results, we report in Fig.~\ref{fig:gizmo_sfr} the SFR (top panel) and the H$_2$ and \hii\ masses (bottom panel) in the cloud - only the gas that originally 
belonged to the cloud, without accounting for the wind material interspersed within the cloud - as a function of time for the three \textsc{gizmo} runs. Because of the initial turbulence that 
rapidly triggers the formation of dense clumps, we do not see any significant differences in the onset time of star formation among the different runs. We note, however, that due to the initial 
compression in the HD run, the peak of SF is a few times higher than that of the other runs. Such a higher SFR is maintained for about 10~Myr, and is followed by a fast decline which shuts off 
SF within the first 20~Myr. At late times, the difference with the HD run is significant, as the HDRest case shows prolonged SF well beyond 20~Myr. As already described, the MHD run exhibits a 
SFR evolution which is mildly lower than the HD case at early times (due to the additional magnetic pressure), and a prolonged star-forming phase which resembles the HDRest run in the last 5~Myr 
of the run. All runs show a decline in molecular gas mass after 10~Myr, steeper in the HD case than in the other two runs (due to ram pressure and enhanced stellar feedback). By looking at the 
\hii\ mass evolution, we can clearly see that the presence of ram pressure enhances the abundance of ionised gas at early times. This is due to ram pressure removing the low density gas in the 
cloud outskirts and clearing out channels through which stellar radiation can propagate, more effectively ionising the gas. At late times, these differences are attenuated, as most of the gas 
is dispersed and ionised by both stellar feedback and fluid mixing with the ICM.

\end{appendix}
\end{document}